\documentclass[draft]{agujournal2019}
\usepackage{url} 
\usepackage{lineno}
\usepackage[inline]{trackchanges} 
\usepackage{soul}
\usepackage{flafter}
\usepackage{subcaption}
\usepackage{caption}
\usepackage{graphicx}
\usepackage{booktabs} 
\usepackage{multirow}
\usepackage{array}
\usepackage{amsfonts}
\usepackage{amsmath}
\usepackage{nameref}
\usepackage{amssymb}
\usepackage{siunitx}
\usepackage{algorithm}
\usepackage{algpseudocode}
\usepackage{pifont}
\usepackage{xcolor}
\usepackage{morefloats}
\usepackage{xr}
\newcommand{\cmark}{\ding{51}} 
\newcommand{\xmark}{\ding{55}} 

\draftfalse

\journalname{JGR: Machine Learning and Computation}

 \usepackage{amsmath,amsfonts,bm}

\def\eqref#1{equation~\ref{#1}}

\def\1{\bm{1}}

\def\vzero{{\bm{0}}}

\def\vmu{{\bm{\mu}}}
\def\vtheta{{\bm{\theta}}}
\def\vphi{{\bm{\phi}}}
\def\vsigma{{\bm{\sigma}}}
\def\vpsi{{\bm{\psi}}}

\def\vc{{\bm{c}}}

\def\vw{{\bm{w}}}
\def\vx{{\bm{x}}}

\def\vz{{\bm{z}}}

\def\mSigma{{\bm{\Sigma}}}

\DeclareMathAlphabet{\mathsfit}{\encodingdefault}{\sfdefault}{m}{sl}
\SetMathAlphabet{\mathsfit}{bold}{\encodingdefault}{\sfdefault}{bx}{n}

\DeclareMathOperator*{\E}{\mathbb{E}} 

\newcommand{\R}{\mathbb{R}}

\DeclareMathOperator{\diag}{diag}

\def\dd{{\mathrm{d}}}

\begin{document}

\title{High Resolution Seismic Waveform Generation using Denoising Diffusion}

%
%

\authors{Kadek Hendrawan Palgunadi\affil{1,*}, Andreas Bergmeister\affil{2,\dagger,*}, Andrea Bosisio\affil{3}, Laura Ermert\affil{1, \ddagger}, Maria Koroni\affil{1}, Nathanaël Perraudin\affil{2,\dagger}, Simon Dirmeier\affil{2}, Men-Andrin Meier\affil{4}}

\affiliation{1}{Swiss Seismological Service (SED), ETH Zürich, Switzerland}
\affiliation{2}{Swiss Data Science Center (SDSC), ETH Zürich, Switzerland}
\affiliation{3}{Politecnico di Milano, Italy}
\affiliation{4}{Earth and Planetary Science Department, ETH Zürich, Switzerland}
\affiliation{\dagger}{Work conducted while being employed at SDSC}
\affiliation{\ddagger}{Now at: Univ. Grenoble Alpes, Univ. Savoie Mont Blanc, CNRS, IRD, Univ. Gustave Eiffel, ISTerre, Grenoble, France}
\affiliation{*}{Equal contributions}
\correspondingauthor{Men-Andrin Meier}{menandrin.meier@eaps.ethz.ch}

\vspace{15mm}
{\textbf{
This manuscript is an arXiv preprint that has been submitted to a peer-reviewed journal and has not yet undergone peer review.
}}


\begin{keypoints}
\item A novel generative latent denoising diffusion model generates realistic synthetic seismic waveforms with frequency content up to 50 Hz.
\item The model predicts peak amplitudes at least as accurately as local ground motion models, and with the same variability as in real data.
\item We introduce \texttt{tqdne}, an open-source Python library for using the pre-trained model, and to train new generative models.
\end{keypoints}

%
%

\begin{abstract}
Accurate prediction and synthesis of seismic waveforms are crucial for seismic-hazard assessment and earthquake-resistant infrastructure design. Existing prediction methods, such as ground-motion models and physics-based wave-field simulations, often fail to capture the full complexity of seismic wavefields, particularly at higher frequencies. This study introduces \emph{HighFEM}, a novel, computationally efficient, and scalable (i.e., capable of generating many seismograms simultaneously) generative model for high-frequency seismic-waveform generation. Our approach leverages a spectrogram representation of the seismic-waveform data, which is reduced to a lower-dimensional manifold via an autoencoder. A state-of-the-art diffusion model is trained to generate this latent representation conditioned on key input parameters: earthquake magnitude, recording distance, site conditions, hypocenter depth, and azimuthal gap. The model generates waveforms with frequency content up to 50 Hz. Any scalar ground-motion statistic, such as peak ground-motion amplitudes and spectral accelerations, can be readily derived from the synthesized waveforms. We validate our model using commonly employed seismological metrics and performance metrics from image-generation studies. Our results demonstrate that the openly available model can generate realistic high-frequency seismic waveforms across a wide range of input parameters, even in data-sparse regions. For the scalar ground-motion statistics commonly used in seismic-hazard and earthquake-engineering studies, we show that our model accurately reproduces both the median trends of the real data and their variability. To evaluate and compare the growing number of these and similar “Generative Waveform Models” (GWMs), we argue that they should be openly available and included in community ground-motion-model evaluation efforts.

\end{abstract}

\section*{Plain Language Summary}
Predicting how the ground shakes during an earthquake is crucial for understanding earthquake hazards and for designing earthquake-resistant buildings. In this study, we use a recently developed artificial intelligence (AI) method to generate realistic, synthetic earthquake seismograms. After transforming the training seismograms from a time-domain into a time-frequency representation, we use a special type of AI model called a diffusion model—originally successful in generating images—to create synthetic seismograms. Our model takes five input parameters (earthquake magnitude, recording distance, site condition, hypocenter depth, and azimuthal gap) and can produce any number of realistic synthetic seismograms for these parameter choices, with high-frequency details up to 50 Hz. Our study shows that the open-source model we present can create realistic seismograms in a wide variety of settings, even in areas with limited training data. For a suite of performance metrics commonly used in assessing earthquake risk and for designing safe buildings, the model closely matches the average trends and variations shown in real earthquake records. To help evaluate and compare the increasing number of such Generative Waveform Models, we argue that such models should generally be made publicly available and included in community efforts to assess ground motion prediction models.


%
%

\section{Introduction}
The accurate prediction of earthquake ground motions is one of the critical components to improve earthquake safety efforts and infrastructure resilience. Recent devastating earthquakes, such as the 2025 $M_w$ 7.7 Myanmar earthquake, the 2023 $M_w$ 7.8 and $M_w$ 7.6 Türkiye, and the 2020 $M_w$ 6.4 Croatia events, have demonstrated that our ability to anticipate and prepare for seismic events remains incomplete, resulting in catastrophic structural failures and loss of life. As urban populations grow and critical infrastructure ages, the need for more realistic, waveform-based models, and computationally efficient ground motion simulations becomes increasingly critical.

The study and prediction of earthquake ground motions are central to seismology. Wavefield models across scales and frequencies are required to assess seismic hazard and the response of critical infrastructure to ground motion. State-of-the-art seismic hazard models use empirical ground motion models (GMMs) to estimate the expected level of ground shaking (i.e., intensity measures) at a site given an earthquake and site properties information. Other applications require the prediction of full-time histories of ground motion at sites of interest. An important example is nonlinear structural dynamic analysis and performance-based earthquake engineering \cite{chopra2007dynamics, applied2009quantification, smerzini_engineering_2024}.

Predicting ground motion is challenging, and existing methods for ground motion synthesis have specific limitations. GMMs are empirical regression models that best fit the observed data as functions of first- and second-order predictor variables, such as magnitudes, recording distances, faulting mechanisms, and site conditions \cite{douglas2003earthquake, boore2014nga} and are often developed for specific regions or tectonic settings. They reduce the full wavefield to scalar properties like peak amplitudes or spectral accelerations, and are data-driven rather than physics-based, although the design of GMMs can be guided by physical considerations, e.g. via the functional form of distance-attenuation terms \cite{baker2021seismic}. While the different predictor variables in seismic waveforms are inherently correlated and physically linked, traditional GMMs have often considered them independently during the optimization of regression model parameters. Some more recent models account for these correlations by employing multivariate statistical techniques \cite{baker2017intensity, baker2021seismic}.

State-of-the-art methods for nonlinear structural dynamic analysis and performance-based earthquake engineering require full waveforms rather than static scalar ground motion features \cite{bommer2004use, luco2007does}. Owing to the scarcity of available short-distance records of large-magnitude events, a common practice is to scale the amplitude spectra of existing strong-motion databases until they meet an expected target spectrum \cite{baker2006spectral, katsanos2010selection, tariq2024BB}. These records are often sampled from relatively limited amounts of data. Notably, it is unclear whether the scaling process leads to realistic waveforms and whether the true ground motion variability is accurately represented in such scaled datasets.

Wavefields can also be modeled deterministically with physics-based numerical solutions of the wave equation. However, the accuracy of the resulting ground motions is limited by the accuracy and resolution with which seismic wave speeds and other relevant parameters are known. In addition, numerical simulators necessarily employ simplified physical models that can only approximate actual subsurface conditions. Together with high computational cost, this puts modeling the full range of hazard-relevant frequencies up to 10 Hz currently out of reach, except for applications with exceptionally well-studied regions on high-performance computing systems, e.g., \citeA{rodgers2020regional, paolucci_earthquake_2021, touhami2022sem3d, palgunadi2024rupture}. As a consequence, important hazard-relevant wave propagation phenomena such as scattering, site amplification, and sedimentary basin edge effects cannot be fully accounted for, peak accelerations may be outside of the resolved frequency range, and dominant modes of structures, including the high-frequency fundamental modes of low-rise buildings, may not be captured accurately. Computational cost and lack of detailed subsurface models also make it challenging to meaningfully account for the uncertainties in path and site effects in physics-based simulations \cite{mai2002spatial, savran2019ground}.

Alternatively, broadband waveforms can be modeled stochastically \cite{boore2003simulation}. The limitations of such stochastic simulations include: (i) they are based on simplified statistical representations of source, path, and site effects, incorporating some physical concepts but lacking detailed physics of wave propagation and source mechanisms which may result in less accurate synthesis of ground motions, especially for large-magnitude earthquakes \cite{boore2003simulation}; (ii) they heavily depend on empirical parameters derived from historical earthquake data \cite{boore1997site}; (iii) they assume the phase of the seismic waves to be random \cite{boore2003simulation, graves2010broadband}; (iv) and they may underestimate the correlation between amplitudes at different frequencies \cite{bayless2019empirical}.

Hybrid methods combine physics-based deterministic simulation at low frequencies ($<1$ Hz) with stochastic simulations at high frequencies ($\geq 1$ Hz) (e.g., \citeA{mai2010hybrid, graves2010broadband, olsen2015sdsu, van2020hybrid, jayalakshmi2021hybrid}). Such hybrid models provide the opportunity to generate earthquake scenarios for a limited number of ruptures in relatively well-studied regions, typically targeting large-magnitude events and near-fault distances (e.g. \citeA{paolucci_bbspeedset_2021}). However, they face similar challenges to those inherent to deterministic simulation, namely high computational cost and uncertainties in source characterization and seismic velocity models \cite{hartzell1999calculation, douglas2008survey}.
Furthermore, classic hybrid methods can suffer from parameterization issues when extrapolating between models where the two signals are easily matched but the phase spectrum is not \cite{mai2003hybrid, graves2010broadband}. 

Despite significant advances, a challenge remains in seismic hazard assessment: generating realistic, physics-consistent broadband seismic waveforms efficiently across the entire hazard-relevant range of earthquake scenarios. This challenge presents opportunities to improve (1) probabilistic seismic hazard assessments that incorporate waveform characteristics beyond scalar intensity measures and (2) ground motion models that better capture the complex relationships between earthquake parameters and the resulting ground motions at all frequencies. Current methods often involve tradeoffs between physical realism and computational efficiency, particularly when modeling high-frequency content.

Some of the limitations of established methods can potentially be overcome with machine learning techniques. In earthquake engineering, machine learning models have been used for the prediction of peak ground acceleration and response or Fourier spectra (e.g. \citeA{derras2012adapting, esfahani2021exploring, jozinovic2022transfer, lilienkamp2022ground}). Alternatively, neural networks can be used to enhance simulated waveforms. For instance, \citeA{paolucci2018broadband} generated band-limited waveforms with physics-based earthquake simulations and then trained an artificial neural network to enhance the spectral acceleration at lower periods. \citeA{gatti2020towards} used generative adversarial networks (GAN) to extract high-frequency features from ground-motion records and utilize them to enhance low-frequency physics-based simulated time series.  \citeA{tariq2024BB} used a combination of GAN and Fourier Neural Operator (FNO) \cite{li2020neural} for a similar purpose. GANs have also been used for seismic data augmentation in earthquake detection problems \cite{li2020seismic, wang2021seismogen}, and to train signal/noise discriminators for Earthquake Early Warning algorithms \cite{li2018machine}.

Conditional deep learning-based generative models can be designed to synthesize three-component seismic waveforms from scratch, conditional on ground motion predictor variables. \citeA{florez2022data} used a Wasserstein GAN to generate realistic broadband seismograms conditional on magnitude ($M$ 4.5–7.5), hypocentral distance ($R$ = 0–180 km), and $V_{S30}$ (0 - 1100 m/s). Among other things, their study showed that the model can reproduce a wide range of features observed in real waveforms. Furthermore, it accurately interpolates for conditional parameter ranges where no data is available, but it is not quite accurate for very large magnitude events ($M > 8$). Inspired by \citeA{florez2022data}, \citeA{esfahani2023tfcgan} trained a GAN to synthesize time-frequency representations of seismograms (TFCGAN) considering similar conditioning parameters ($M$ 3.8–7.5, $R<120$ km, $V_{S30} <1200$ m/s). A study by \citeA{shi2024broadband} developed a combination of GAN and neural operators, presenting the conditional ground-motion synthesis algorithm (cGM-GANO). Their algorithm is conditioned on moment magnitude ($M$ 4.5–8.0), closest-point-on-the-rupture to site distance ($R$ = 0–300 km), style of faulting, $V_{S30}$ (100–1100 m/s), and tectonic environment type (subduction or shallow crustal event). The most recent study by \citeA{matsumoto2024GAN} introduced additional input parameters focused on site-specific conditions consisting of the shear-wave velocities in the top 5, 10, and 20 m and the depth to the layer with shear-wave velocities of 1000 and 1400 m/s. They also used a modified GAN method based on the pre-existing StyleGAN by \citeA{karras2020analyzing}. The algorithm was trained with $M > 5$, rupture distance of $\leq 100$ km, and hypocentral depth of less than 30 km. While these advancements show the potential of generative waveform modeling in seismology, GAN-based models can suffer from mode collapse and other training instabilities \cite{goodfellow2014generative}. For a detailed overview of the use of GAN for earthquake-related engineering fields, the reader is referred to \citeA{marano2024generative}. 

In this study, we present the High Frequency Earthquake Model (\emph{HighFEM}), a novel latent denoising diffusion model for the conditional synthesis of seismic waveforms that addresses some of the discussed limitations. Compared to GANs, diffusion-based models may offer improved training stability and efficiency. Our approach uses an autoencoder to map spectrogram representations of observed ground motions into a lower-dimensional submanifold. This latent space formulation helps balance physical consistency across frequency bands while enabling conditional generation based on earthquake parameters. Our approach is straightforward to train, efficient during inference, and capable of generating diverse yet physically plausible ground motions, potentially supporting applications in seismic hazard assessment. We release the code and pre-trained models as part of 'This-Quake-Does-Not-Exist' (\texttt{tqdne}), an open-source Python library, with the hope of instigating a community effort on open generative seismic waveform modeling, similar to established efforts in the GMM community.

The manuscript is structured as follows: in Section~\ref{sec:methods}, we describe the model and training process. Data descriptions and preprocessing steps are presented in Section \ref{sec:data_description}. In Section~\ref{sec:results}, we evaluate to what extent the model can generate realistic seismic waveforms, using metrics from both seismology and machine learning. In Section~\ref{sec:discussion}, we discuss the potential applications of our approach and similar GWMs for both scientific and engineering applications.

\section{Methods}
\label{sec:methods}
\begin{sidewaysfigure}
    \centering
    \includegraphics[width=0.9\textwidth]{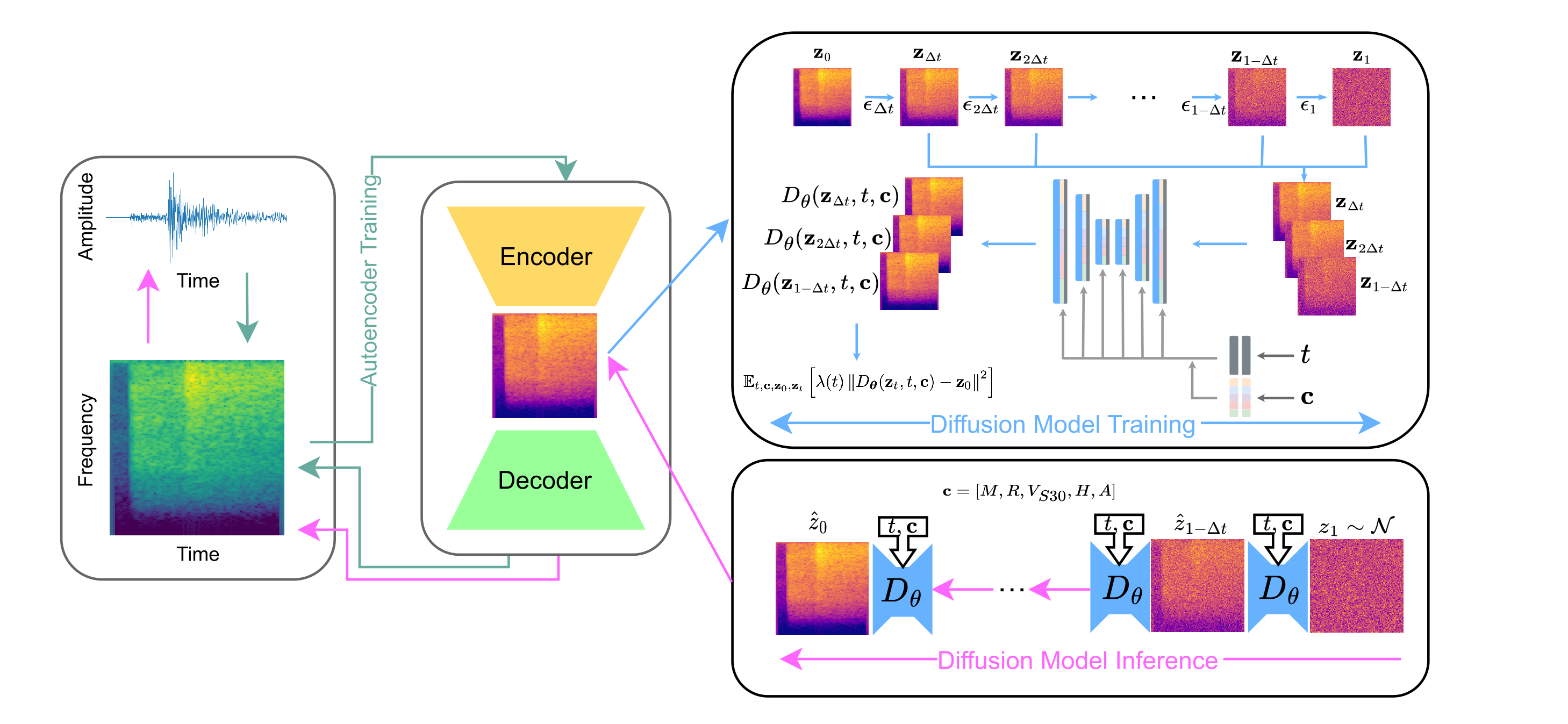}
    \caption{\textbf{HighFEM pipeline}. Waveforms are converted to spectrograms (left) and an autoencoder is trained to compress them (middle). A conditional denoising diffusion model is then trained to generate the autoencoder's latent representations (purple to yellow image in middle block), conditional on a low-dimensional parameter vector, $\bm{c}$ (right top). Training proceeds by generating a batch of encoded seismograms that are corrupted by noise $\epsilon$ of different magnitudes and attempting to predict the original noiseless encoded seismograms. The target is a simple weighted MSE loss. During inference (right bottom), samples are drawn from the denoising diffusion model for given conditioning parameters by sequentially denoising a white noise sample into a latent representation, mapped back to spectrograms using the decoder of the autoencoder, and finally inversely converted to waveforms.}
    \label{fig:pipeline}
\end{sidewaysfigure}

The HighFEM approach to generating high-resolution seismic waveforms with a latent diffusion model comprises three primary components. First, we transform the seismic waveforms into spectrogram representations (Section \ref{subsection:data_representation}), which are more amenable to generative modeling than time-domain signals. Second, we employ a convolutional variational autoencoder to compress these high-dimensional spectrograms into a lower-dimensional latent space (Section \ref{subsection:dimensional_reduction}). Third, we train a denoising diffusion model to generate samples in this latent space (Section \ref{sec:denoising_diffusion}). To arrive at the raw waveforms again, we map back the synthesized samples from the latent space to the data space, and convert them from their spectrogram representation to their raw waveform representation. In this section, we provide a comprehensive description of each component of the HighFEM GWM pipeline (Figure \ref{fig:pipeline}). Detailed explanations of the neural network architectures, which we omit here for brevity, are given in Appendix \ref{app:architectures}.

\subsection{Data representation}\label{subsection:data_representation}
Processing seismograms directly with neural networks is challenging due to their high-frequency content and amplitude variance, both within and across samples. This issue is particularly problematic in generative modeling, where high-amplitude areas dominate the loss function, compromising reconstruction accuracy in low-amplitude regions. To address these challenges, we transform the 40-second-long waveforms into spectrogram representations (Appendix \ref{sec:representations}), a technique commonly used in seismology and in audio signal generation, where typically waveforms of much higher frequency content are modeled \cite{kong2021diffwave,defossez2023high}. 
We apply a Short-Time Fourier Transform (STFT) on the signals which yields a $128 \times 128$-dimensional complex matrix. We project the matrix through a logarithmic transformation of its element-wise absolute values, yielding a spectrogram representation that resolves the frequency content ($\Delta f = 0.39$ Hz and $\Delta t = 0.32$ s) of the signal over specific time intervals and which we denote with $\vx \in \mathbb{R}^n$, $n = 128 \times 128$ (see Figure~\ref{fig:waveform_spectrogram}).

\subsection{Dimensionality reduction}\label{subsection:dimensional_reduction}
Generative modeling of high-dimensional data is computationally intensive during both training and inference. This limits its applicability to researchers and practitioners with access to high-performance computing environments. To make our GWM available to a broad scientific audience and mitigate the computational difficulties, we first train a variational autoencoder (VAE) to compress the data into a more manageable, lower-dimensional latent space before we train a denoising diffusion model on this learned space \cite{rombach2022highresolution}. Formally, let $\vx \sim p_{\text{data}}$ denote the (empirical) probability distribution of the data (which in our case is the distribution of the spectrograms) and where $\vx$ denotes a sample from the data set. We employ a variational autoencoder to model the distribution of variables $\vz \in \mathbb{R}^m$, $m=32 \times 32$. Here, we represent the latent variables $\vz$ as a spectrogram with smaller dimensions than the original spectrogram (Figure~\ref{fig:pipeline}). Formally, we employ as an encoder a convolutional neural network to model the distribution of $\vz$ as a conditional Gaussian distribution:
\begin{equation}
    q_{\vphi}(\vz | \vx) = \mathcal{N}(\vz \mid \vmu_{\vphi}(\vx), \diag(\vsigma^2_{\vphi}(\vx))),
\end{equation}
where the variational distribution $q_{\vphi}(\vz | \vx)$ is the encoding component of the VAE. The distribution is defined by its mean and standard deviation functions $\vmu_{\vphi}, \vsigma_{\vphi}: \R^n \to \R^m$, which are parameterized by a CNN with parameters $\vphi$. We employ a deterministic decoder, $G_{\vpsi}: \R^m \to \R^n$, which 
is another CNN parameterized by weights $\vpsi$ that maps latent variables back to the original data space. The encoder and decoder are trained jointly to minimize
\begin{equation}
    \E_{q_{\vphi}(\vz | \vx)} \left\| \vx - G_{\vpsi}(\vz) \right\|^2 + \beta \mathbb{K}\mathbb{L}[q_{\vphi}(\vz | \vx) \| \mathcal{N}_m(0, 1)],
\end{equation}
where the Kullback-Leibler divergence, $\mathbb{K}\mathbb{L}$, between the encoder distribution and a multivariate standard normal distribution is used to regularize the latent space. Since we employ the variational autoencoder solely for data compression rather than as a generative model, we set the regularization strength to a tiny value ($\beta=1\mathrm{e}{-6}$) to ensure high reconstruction quality \cite{higgins2017beta}.

\subsection{Generative modeling of the latent space} 
\label{sec:denoising_diffusion}
After compressing the data into a compact latent representation, we employ denoising diffusion models (DDMs; \citeA{song2019generative, ho2020denoising, song2020scorebased} to fit a generative model over the latent space. DDMs have gained popularity due to their outstanding performance in tasks such as image, audio, or video generation. Unlike previous techniques, DDMs do not rely on approximate variational inference which can produce blurry samples as in variational autoencoders \cite{kingma2013auto}, but they utilize a simple mean-squared error loss. Unlike adversarial training which is typically unstable, difficult to diagnose and suffers from mode collapse \cite{goodfellow2014generative}, DDMs have been shown to be robust during training and their convergence is simple to diagnose.

Specifically, we aim to fit a conditional denoising diffusion model to approximate the distribution of the lower-dimensional representation $\vz$ given a set of physical parameters $\vc$
\begin{equation}
    p_{\text{latent}}(\vz | \vc)
\end{equation}
where $\vc \in \R^5$ is a set of conditional parameters, which we expect to determine the shape and statistical properties of the seismic waves: earthquake magnitude, hypocentral distance, site conditions ($V_{S30}$ as a proxy), hypocenter depth, and azimuthal gap (Section \ref{sec:data_description}).

This formulation allows generating seismic waveforms with these parameters as properties. Within the framework of denoising diffusion models, this requires the definition of a pair of stochastic processes: a fixed forward(-time) noising process and a complementary (reverse-time) backward process which we describe in detail below.

\subsubsection{Forward process}
The forward process is a conventional Ito-process of the form
\begin{equation}
    \dd \vz_t = \vmu(\vz_t, t) \, \dd t + \sigma(y) \, \dd \vw,
    \label{eq:general_forward_process}
\end{equation}
where, $0\leq t \leq T$, $\vmu: \R^m \times \mathbb{R}^+ \to \R^m$ and $\sigma(t): \mathbb{R}^+ \to \mathbb{R}^+$ are fixed, pre-specified drift and diffusion terms, respectively. The stochastic differential equation (SDE) in \eqref{eq:general_forward_process} takes a sample $\vz_0 = \vz$ from the data set, in our case the latent spectrogram representation of a seismic waveform, and adds Gaussian noise in the form of a Wiener process $\vw$ to it until it resembles an isotropic Gaussian. A Wiener process $\vw(t)$ is a continuous stochastic process with initial condition $\vw(0) = 0$ which is distributed according to $\vw(t) \sim \mathcal{N}(0, t)$ with independent increments $\vw(t) - \vw(s) \sim \mathcal{N}(0, t - s)$. For the generation of spectrograms, which are images, we adopt the parameterizations proposed by \citeA{karras2022elucidating} which excel in image generation tasks. We utilize their \textit{variance exploding} version by setting: 
\begin{equation*}
\begin{split}
\vmu(\cdot, t) &= \vzero\\
\sigma(t) &= \sqrt{2 t}
\end{split}
\end{equation*}
leading to the forward process
\begin{align}
    \dd \vz_t = \sqrt{2 t} \, \dd \vw.
    \label{eq:variance_exploding}
\end{align}

Conveniently, integrating the forward process from time $0$ to $t$ can be done analytically and yields the transition kernel $p_t(\vz_t | \vz_0) = \mathcal{N}(\vz_t | \vz_0, t^2 \mathbf{I})$, i.e., one can sample from this distribution by adding Gaussian noise with zero mean and variance $t^2$ to the original sample $\vz_0$.

\subsubsection{Backward process}
The backward process is also an SDE and has the following form:
\begin{equation}
    \dd \vz_t = [\vmu(\vz_t, t) - \sigma(t)^2 \nabla_{\vz_t} \log p_t(\vz_t | \vc)] \, \dd t + \sigma(t) \, \dd \bar{\vw},
    \label{eq:general_reverse_process}
\end{equation}
where $\bar{\vw}$ is a Wiener process in reverse-time. As can be seen, the backward process requires access to the gradient of the logarithm of distribution $p_t$ (generally referred to as \textit{score} of the distribution and defined below)
which is not directly accessible (i.e., computable) and requires approximation using a neural network. Conversely to the forward process, the backward process commences from the aforementioned isotropic Gaussian and aims to invert the effect of the forward process, recovering samples that approximately follow the data distribution. By applying the variance-exploding noise schedule (Equation~\ref{eq:variance_exploding}), the backward process simplifies to
\begin{equation}
    \dd \vz_t = [-\sigma(t)^2 \nabla_{\vz_t} \log p_t(\vz_t | \vc)] \, \dd t + \sigma(t) \, \dd \bar{\vw},
    \label{eq:general_reverse_process}
\end{equation}
the reverse process during inference time, samples from the data distribution, i.e., in this case seismic waveforms, can be generated.

\subsubsection{Training the diffusion model}
We aim to fit a model $s: \R^m \times [0, T] \times \R^c \to \R^m$ to approximate the score $ \nabla_{\vz_t}\log p_t(\vz_t | \vc)$ using a neural network $ D_{\vtheta}: \R^m \times [0, T] \times \R^c \to \R^m$ and the parameterization
\begin{equation}
    s(\vz_t, t, \vc)  = (D_{\vtheta}(\vz_t, t, \vc) - \vz_t) / t^2
\end{equation}
Here, $D_{\vtheta}$ is trained to estimate the original sample $\vz_0$ of the data from its noisy version $\vz_t$ after having been perturbed using the forward-process given the conditioning $\vc$.  

The training objective with a time-dependent weighting function $\lambda: [0, T] \to \R_+$ is then given by
\begin{equation}
    \mathcal{L}(\vtheta) = \E_{t, \vc, \vz_0, \vz_t} \left[ \lambda(t) \left\| D_{\vtheta}(\vz_t, t, \vc) - \vz_0 \right\|^2\right],
    \label{eq:denoising_loss}
\end{equation}
where the expectation is taken over the time $t$, the conditional distribution $p_{\text{cond}}(\vc)$, the distribution over 
the encoded latent sample $\vz_0 \sim q_{\phi}(\vz | \vx)$ and its perturbed version $\vz_t \sim p_t(\vz_t | \vz_0)$.
We follow \citeA{karras2022elucidating} in parameterizing the loss weighting functions.

\subsubsection{Generating synthetic seismic waveforms}
Intringuingly, the optimal denoising model $D_{\vtheta}^*$ satisfies $\nabla_{\vz_t} \log p_t(\vz_t | \vc) = (D_{\vtheta}^*(\vz_t, t, \vc) - \vz_t) / t^2$ almost surely for all $t \in [0, T]$ \cite{vincent2011score}. The trained network $D_{\vtheta}$ can then be used to solve the backward process for sampling synthetic seismic waveforms.

\citeA{song2020scorebased} identify a deterministic process with the same marginal distributions as the reverse process which using the variance-exploding noise schedule (\eqref{eq:variance_exploding}) is characterized by the ordinary differential equation:
\begin{equation}
    \dd \vz_t = [ - \frac{1}{2} \sigma(t)^2 \nabla_{\vz_t} \log p_t(\vz_t | \vc)] \, \dd t.
    \label{eq:general_probability_flow}
\end{equation}
In practice, integrating this deterministic process enables efficient sampling as simulating a stochastic process typically requires more time discretization steps. Here, we utilize a second-order Heun method with $25$ steps corresponding to a total of $50$ model evaluations to sample from the model \cite{karras2022elucidating}.

\section{Data and Model Training}
\label{sec:data_description}
We use three-component strong-motion seismic waveforms recorded between 1996 and 2022 by the K-NET and KiK-net stations in Japan, provided by the National Research Institute for Earth Science and Disaster Resilience \cite{knet2019}. Before training our model, we preprocess the raw traces by removing the scalar gain factor. We then apply a causal second-order Butterworth high-pass filter with a corner frequency of 0.1 Hz and resample, via interpolation, to a common time vector, with 100 samples per second. We use the PhaseNet picker \cite{zhu2019phasenet} to estimate P-wave onset times, and save 40 second long waveform segments, with the P-wave onset at 5 s after the start of the trimmed record. Finally, we rotate the horizontal components from the original east–west and north–south axes to radial and transverse orientations. Additional details on the preprocessing steps and data distributions are provided in the supplementary material (Text S1 and S2).

We consider events with hypocentral depths between 0 and 100 km and magnitudes $M \geq 4.0$. Station–hypocenter distances range from 1 to 200 km. Because $V_{S30}$ is available for most stations, we discard those lacking a $V_{S30}$ estimate for the sake of simplicity. The remaining $V_{S30}$ values span 76–2100 m/s. We also compute the azimuthal gap per earthquake—the largest angle spanned at the epicenter between two adjacent stations, which in our dataset ranges from $7.9^\circ$ to $360^\circ$. 
We include the azimuthal gap as a conditioning parameter as a simple proxy for how well recorded a particular quake is. Well-recorded events beneath the seismic network have small azimuthal gaps. Offshore events, on the other hand - for which source parameter uncertainties tend to be larger - have larger gaps.
In total, we analyze 266,399 three-component records. We randomly divide the available data into $85\%$ for training, $5\%$ for validation, and $10\%$ for testing, where we use the test set to evaluate our model. Training of our model was conducted on four NVIDIA Grace-Hopper GPUs, requiring approximately 38 hours for the first-stage autoencoder and an additional 15 hours for the second-stage diffusion model. However, using a larger number of samples (e.g., longer waveforms or smaller sampling rates) can result in longer training times. In contrast, increasing the number of conditioning parameters does not affect training time.

The K-NET and KiK-net waveform data set used here is openly available but cannot be redistributed—for example, in a machine-learning-ready, pre-processed format. Therefore, readers must download and pre-process the raw KiK-net and K-NET data. To provide a reproducibility test, we also supply an equivalent model and perform the same evaluations on a subset of the Stanford Earthquake Dataset (STEAD) \cite{mousavi2019stanford}; see Supplementary Materials (Text S3).

\section{Results}
\label{sec:results}
The design goal of the HighFEM GWM is to synthesize ground motion records that are statistically indistinguishable from real records, across a wide range of frequencies and conditioning parameters, namely source magnitudes, hypocentral distances, $V_{S30}$, hypocenter depth, and azimuthal gap. In the following section, we discuss the extent to which the HighFEM model can achieve that goal and compare its outputs (i) to real data,  (ii) to commonly used Ground Motion Models (GMMs), and (iii) to the Generative Adversarial Neural Operator (GANO) model of \cite{shi2024broadband}. 
First, we compare the distributions of time-domain signal envelopes (Section \ref{sec:time_domain_signal}) and of Fourier Amplitude Spectra (Section \ref{sec:fourier_amplitude_spectra}) between the real data and the GWM synthetics. Next, we evaluate how well the GWM predicts scalar ground motion intensity measures, in terms of prediction accuracy (Section \ref{sec:model_bias}), and variability (Section \ref{sec:variability_amplitude}). Then we compare shaking durations of real data and GWM synthetics (Section \ref{sec:signal_durations}), and the scaling between peak amplitude statistics and the conditioning predictor variables (Sections \ref{sec:amps_vs_distance} - \ref{sec:amps_vs_vs30}). Finally, we evaluate relative and absolute model performances for different magnitude and hypocentral distance ranges, by computing average model probabilities (Section \ref{sec:probability}), and performance metrics from the image generation domain (Section \ref{sec:quantitative-evaluation}). 

\begin{figure}[htp!]
    \centering
    \includegraphics[width=0.9\textwidth]{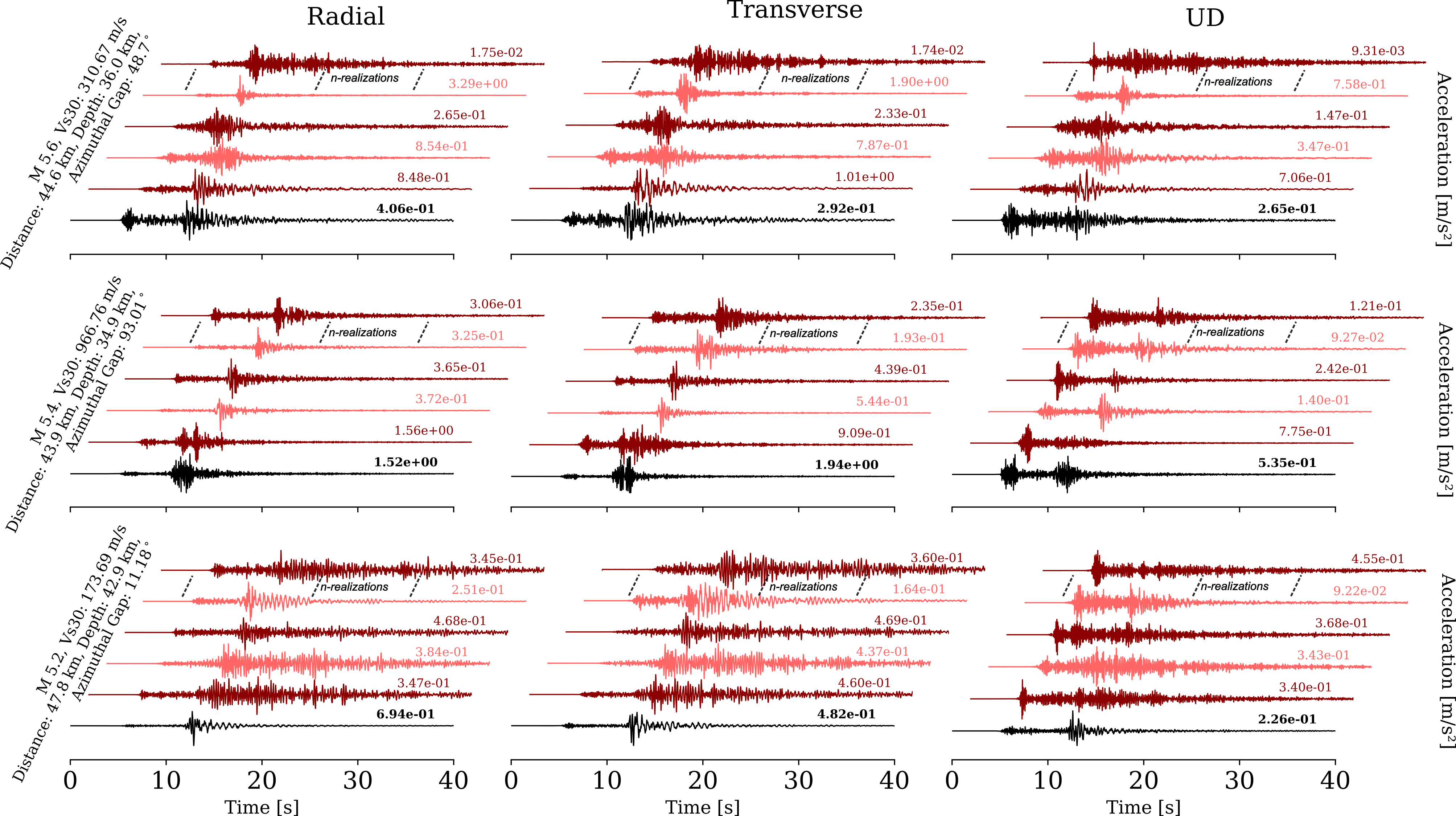}
    \caption{Real three-component acceleration seismograms (grey) and 5 randomly selected examples of GWM synthetics (alternating shades of red for improved visibility), for three sets of conditioning parameters: magnitude, hypocentral distance, $V_{S30}$, hypocenter depth, and azimuthal gap. Peak absolute amplitudes are given above each seismogram.}
    \label{fig:waveform-comparison}
\end{figure}

For each real seismogram, we use the trained HighFEM to produce a number of corresponding GWM synthetic seismograms with the conditioning parameters of the real record, i.e., their magnitude, hypocentral distance, $V_{S30}$ value, hypocenter depth, and azimuthal gap. We can then directly compare the real seismograms to their corresponding stochastic synthetic realizations (Figure \ref{fig:waveform-comparison}). The GWM synthetics appear to capture the first-order characteristics of the real seismograms: they have clear P- and S-phases with realistic phase arrival time differences, as well as realistic coda wave decays. For the same conditioning parameter choices, there is considerable variability between individual realizations, for example, in terms of peak amplitudes. In Section \ref{sec:variability_amplitude} we show that this variability closely matches the variability observed in the real data.

\subsection{Time domain signal envelopes}\label{sec:time_domain_signal}
\begin{figure}[htp!]
    \centering
    \includegraphics[width=0.9\textwidth]{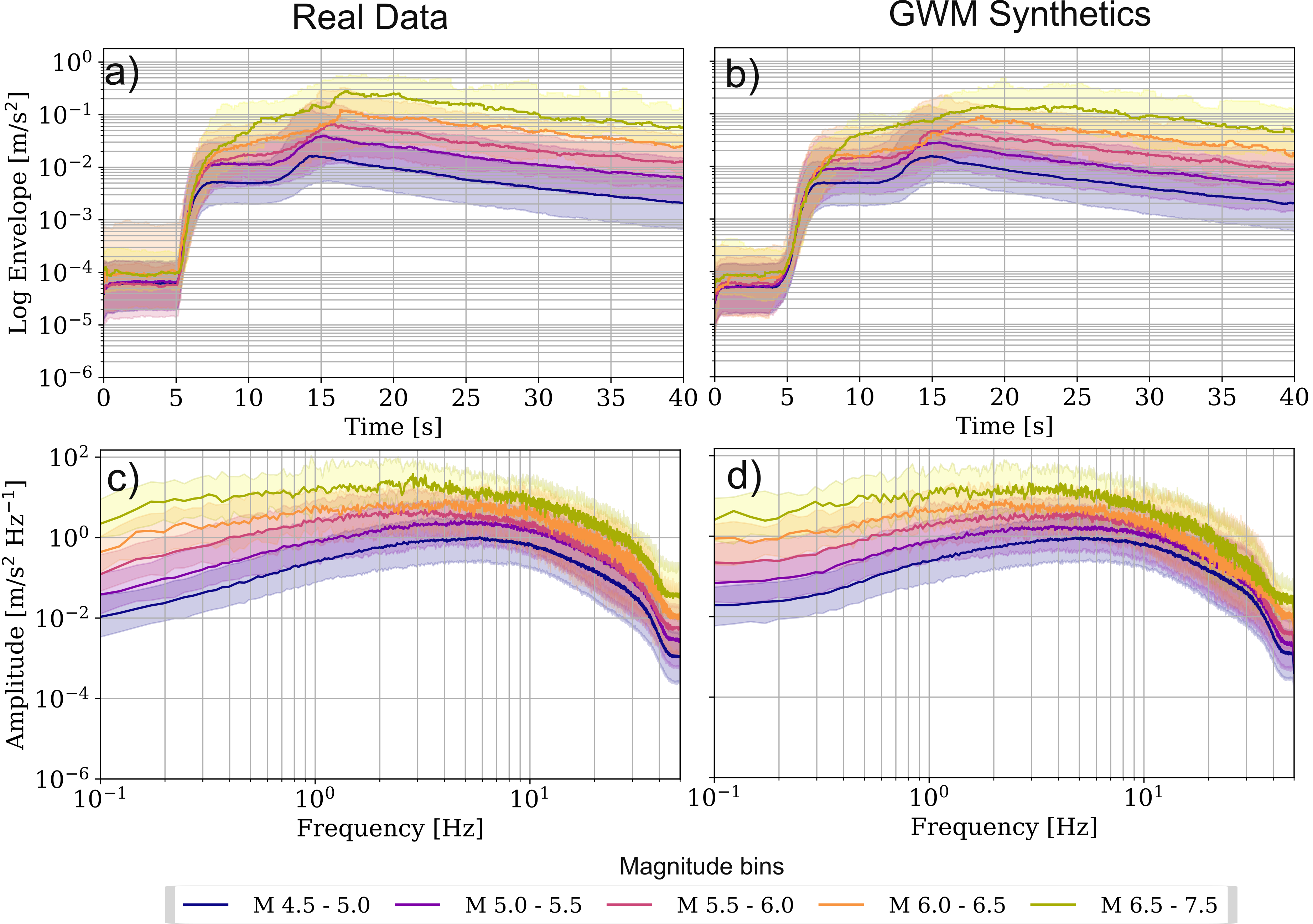}
    \caption{First order seismogram characteristics of real data (left) and GWM synthetics (right) for hypocentral distances of 50 - 70 km in 5 different magnitude bins. (a) and (b) Distribution of time-domain envelopes for radial component of the acceleration seismograms in terms of the mean (solid line) and the standard deviation (shaded areas). (c) and (d) Distribution of Fourier spectra log-amplitudes for radial component of the acceleration seismograms. The sample counts for each bin, in ascending order of magnitude, are $882, 398, 140, 48, 24$.}
    \label{fig:envelope}
\end{figure}

To compare the real and synthetic waveforms quantitatively, we compute signal envelope time series for both sets. The signal envelopes are obtained by taking the moving average of the absolute waveform signals with a kernel size of 128, followed by a logarithmic transformation. This comparison shows that the GWM synthetics have very similar first-order shapes in the time domain compared to the real seismograms, across the entire range of magnitudes and recording distances for which the model was trained (Figures \ref{fig:envelope}a and \ref{fig:envelope}b). The low-noise amplitudes before the P-wave onsets are followed by an impulsive P-wave amplitude increase. This amplitude growth is similar for both small and large magnitudes until the smaller magnitude records reach their maximum P-wave amplitude, whereas the large magnitude records continue to grow. The later-arriving S- and surface waves cause additional amplitude growth in the real waveforms, which is accurately mimicked by the GWM synthetics. The variability of the envelopes in each bin is symmetric around the median in log-space and is of the same order for both real and synthetic data. Additional figures in the supplementary material show different bins, and separate evaluations of tranverse and vertical components (supplementary Figures \ref{fig:supp_grid_plot} to \ref{fig:supp_asd_2}). 

We also compare the time-domain signal envelopes with the GANO model of \citeA{shi2024broadband} (see Supplementary Meterials Text S4, and Figures \ref{fig:supp_envelope_gano_0}–\ref{fig:supp_envelope_gano_2}). Similar to our GWM, the GANO model successfully reproduces P- and S-phase arrivals as well as increasing ground-motion amplitudes with magnitude. However, it produces lower peak amplitudes than those of either our GWM or the real data, including at large distances, where the different distance parameterisations (hypocentral distance for HighFEM, rupture distance for GANO) should not matter.

\subsection{Fourier amplitude spectra}\label{sec:fourier_amplitude_spectra}
Next, we compare the logarithmic Fourier amplitude spectra between the real data and the HighFEM synthetics. These spectra are obtained by performing a Fourier transform of the time-domain signals, calculating the magnitudes of the resulting complex values, and then applying a logarithmic transformation. The resulting distributions are similar in terms of mean log-amplitudes and variability (Figures \ref{fig:envelope}c and \ref{fig:envelope}d). Equivalent evaluations for other parameter bins are shown in supplementary Figures \ref{fig:supp_envelope_0} to \ref{fig:supp_asd_2}. Additionally, Section~\ref{sec:quantitative-evaluation} discusses the use of Fréchet distance to compare distributions of log-amplitude spectra for real data and GWM synthetics. The lower overall amplitudes of the GANO model observed in the time domain envelopes are visible particularly at low frequencies ($\leq 1$ Hz) (Figures \ref{fig:supp_gano_0}–\ref{fig:supp_gano_2}).

\subsection{Scalar ground motion statistics}
\label{sec:IM_evaluation}

For earthquake engineering and seismic hazard applications, peak ground motion amplitude statistics are of particular importance. Here, we compare how various peak amplitude statistics of the GWM synthetics compare with the real data, how they correlate with the conditioning predictor variables, and how they compare with predictions from widely used Ground Motion Models (GMMs). Specifically, we compute peak ground acceleration (PGA), peak ground velocity (PGV), and pseudo-spectral acceleration (SA) for both real data and GWM synthetics. We use the orientation-independent GMRotD50 statistic \cite{boore2006orientation}, which represents the median of the horizontal components, rotated over all possible rotation angles. We utilize GMMs from \citeA{boore2014nga}, optimized for a global database of shallow crustal earthquakes in active tectonic regions, and from \citeA{kanno2006new}, which used a database of strong ground motion records from shallow and deep crustal earthquakes in Japan between 1963 and 2003. For this comparison with GMMs we only use the subset of the test dataset with M $\geq$ 5, because the GMMs were constructed with data from that range.

\subsubsection{Accuracy of predicted peak amplitudes}\label{sec:model_bias}

\begin{figure}[htp!]
    \centering
    \includegraphics[width=\textwidth]{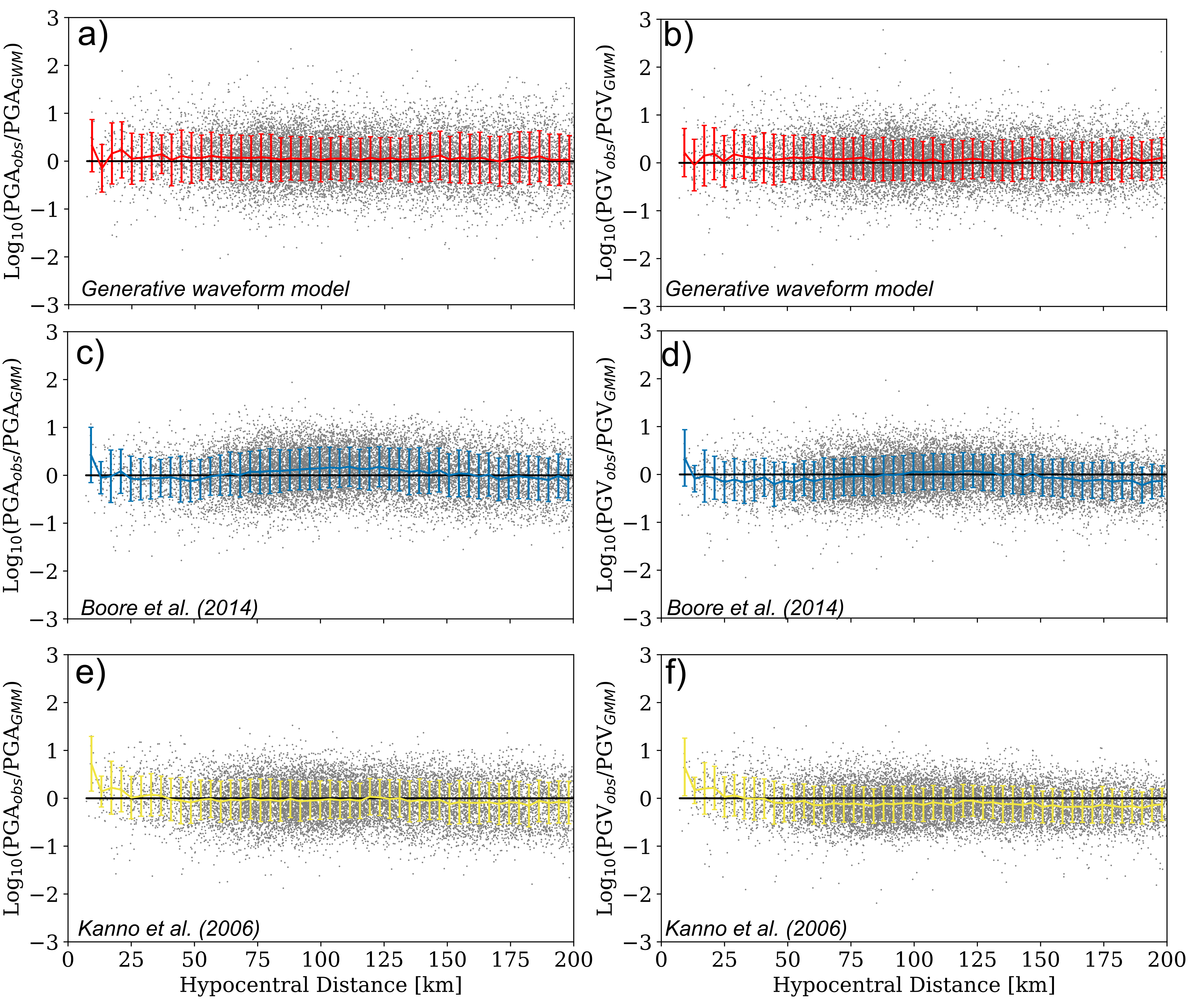}
    \caption{Model bias as a function of hypocentral distance for the generative waveform model (red), GMMs by \protect\citeA{boore2014nga} (blue), and \protect\citeA{kanno2006new} (yellow) for PGA (a, c, and e) and PGV (b, d, and f), with respect to real data. Colored lines represent the mean of the ratio in 50 distance bins of 3.67 km width. The bars represent the standard deviation in each bin.}
    \label{fig:bias_IM}
\end{figure}

For each record in the real waveform test dataset, we compute a single GWM synthetic waveform, using the conditioning parameter of the real data (magnitude, hypocentral distance, hypocentral depth, $V_{S30}$, and azimuthal gap). We then compare the PGA and PGV values between the real data and the GWM synthetics, and find that the GWM predictions are at least as accurate as those from the GMMs, and that they have similar prediction variability (Figure \ref{fig:bias_IM}).

Specifically, we compute the model bias as the logarithm of the ratio between observed and predicted peak amplitudes. For the GWM synthetics, the mean of the distribution of this log-ratio is close to zero for both PGA and PGV: the mean model bias across all distances is 0.061 log-units for PGA and 0.066 log-units for PGV. This corresponds to an underprediction by 15\% and 16\%, respectively. At very short hypocentral distances ($< 30$ km), the GWM underestimates observations by 22\% (0.088 logarithmic units) for PGA and 24\% (0.094 logarithmic units) for PGV (Figures \ref{fig:bias_IM}a and \ref{fig:bias_IM}b). 

In comparison, the GMM by \citeA{boore2014nga} overestimates the real PGA for distances $< 75$ km by 4.3 $\%$, underestimates for distances 75 - 130 km by 35$\%$, and predicts with minimum bias for $> 130$ km (Figure \ref{fig:bias_IM}c). For PGV, this GMM overestimates by 7.5 $\%$ on average (Figure \ref{fig:bias_IM}d).
The GMM by \citeA{kanno2006new} overestimates PGA by an average of 11$\%$ (Figure \ref{fig:bias_IM}e) across all distances and underestimates by 16$\%$ for distances $< 25$ km. It accurately predicts PGV (23$\%$ over-prediction across all distances) (Figure \ref{fig:bias_IM}f), except at short distances ($< 25$ km), where the under-prediction amounts to 31\%. 
These comparisons indicate that the bias in GWM peak amplitude predictions is similar to, and sometimes smaller than, that of the GMMs.
The average GANO underpredictions amount to $38\%$ for PGA and $65\%$ for PGV for distances $>50$ km, where differences in rupture distance parameterizations should not matter (Figure \ref{fig:bias_gano}, Text S4).

\subsubsection{Variability of predicted peak amplitudes}\label{sec:variability_amplitude} 
Another important criterion for ground motion prediction methods is that the variability in predicted peak amplitudes is accurately characterized. To evaluate the variability of the GWM predictions, we compare their total standard deviation to the variability in the real data, and to the predictions of the two GMMs. 

To measure prediction residuals, we fit a simple, custom GMM to the PGA and PGV of the real data, as a function of magnitude $M$, hypocentral distance $R$, and $V_{S30}$. We use ordinary least squares and find:

\begin{equation}\label{eq:GMPE_PGA}
\begin{split}
    \log_{10}{\left(\text{PGA} \right)} = & 0.4750 + 0.3715 \times M - 0.3612 \times \log_{10}{\left(V_{S30}\right)} \\
    &\ - 1.2479 \times \log_{10}{\left(\text{R}\right)}
\end{split}
\end{equation}
\begin{equation}\label{eq:GMPE_PGV}
\begin{split}
    \log_{10}{\left(\text{PGV} \right)} = &-1.4531 + 0.5281 \times M - 0.5383 \times \log_{10}{\left(V_{S30}\right)} \\
    &\ - 1.1600 \times \log_{10}{\left(\text{R}\right).}
\end{split}
\end{equation}

\begin{figure}
    \centering
    \includegraphics[width=\linewidth]{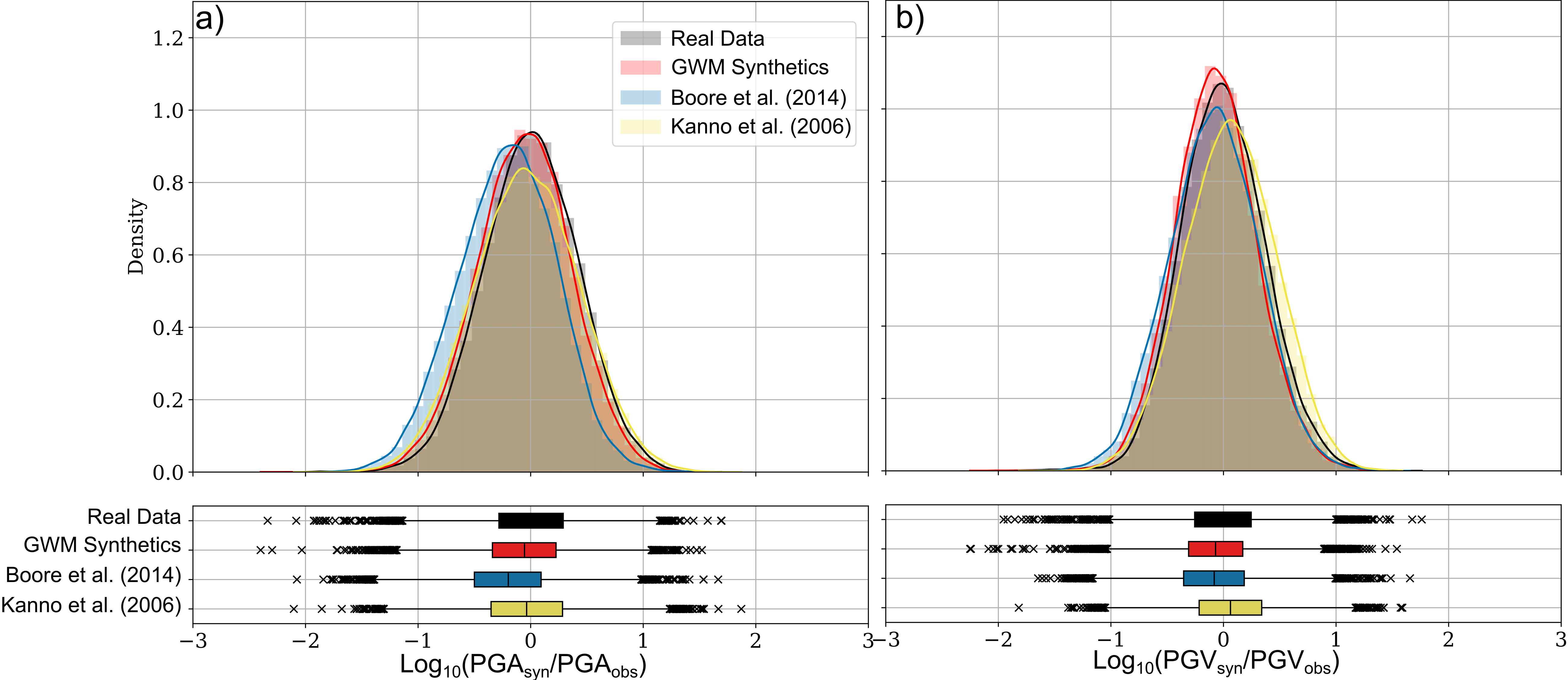}
    \caption{Histogram of (a) PGA and (b) PGV residuals showing the spread of the real data (black), of the GWM synthetics (red), of the \protect\citeA{boore2003simulation}  GMM (blue), and of the \protect\citeA{kanno2006new} GMM (yellow), with respect to the simple fitted ground motion models (equations \ref{eq:GMPE_PGA} and \ref{eq:GMPE_PGV}) on a $log_{10}$ scale. The box plot shows the median values, quantiles, and extreme values.}
    \label{fig:histogram_stats_PGA_PGV}
\end{figure}

We then calculate logarithmic residuals by subtracting the predictions of equations \ref{eq:GMPE_PGA} and \ref{eq:GMPE_PGV} (i) from the real data, (ii) from the GWM, (iii) from the GMM of \citeA{boore2014nga}, and (iv) from the GMM of \citeA{kanno2006new}. The distribution of these residuals (Figure \ref{fig:histogram_stats_PGA_PGV}) is similar between the real data and the GWM synthetics for both PGA and PGV. This indicates that the GWM synthetics have a similar spread around the median prediction as the real data set. 

The standard deviations of the distribution of residuals are very similar for all four cases: for PGA they are 0.422 (real-data), 0.420 (GWM synthetics), 0.433 (\citeA{boore2014nga} GMM) and 0.465 (\citeA{kanno2006new} GMM) in logarithmic units. For PGV they are 0.376, 0.366, 0.398, and 0.404, respectively. That is, for all four cases, about 68$\%$ of amplitudes fall between $40\%$ ($10^{-0.4}$) and $251\%$ ($10^{0.4}$) of the median predicted amplitude value.

\subsubsection{Shaking durations}
\label{sec:signal_durations}
\begin{figure}
    \centering
    \includegraphics[width=\linewidth]{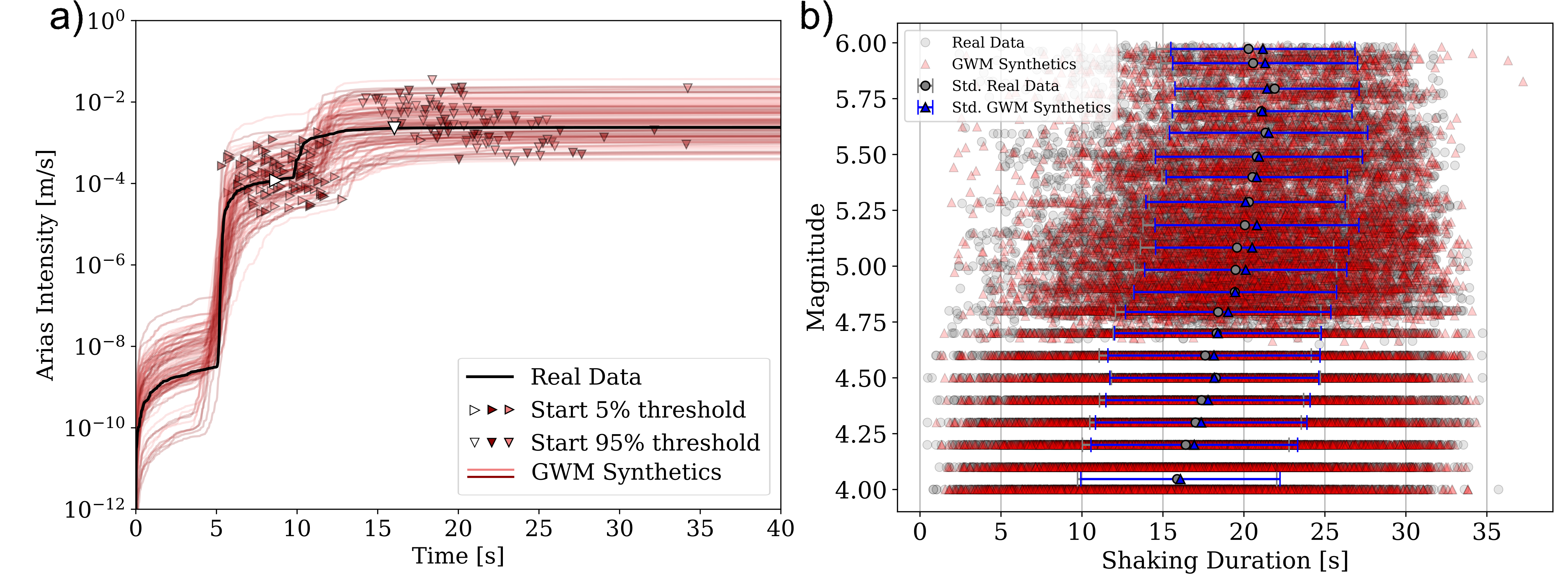}
    \caption{Shaking duration estimated using cumulative Arias Intensity (cAI). (a) cAI for a real example waveform (black line) and 100 GWM synthetics (red lines). Triangle-right and triangle-down symbols represent 5\% and 95\% of the maximum cAI for the real data (white) and GWM synthetics (red), respectively. (b) Shaking duration for real data (grey circles) and one GWM synthetic per real record (red triangles), with corresponding conditioning parameters. For each magnitude bin (every 0.08) from $M$ 4.0 - 6.0, grey dots and lines show the mean and standard deviation of the real data, while blue triangles and lines show the mean and standard deviation of the GWM synthetics.}
    \label{fig:arias_intensity}
\end{figure}

Shaking duration is another signal characteristic that is important for earthquake engineering applications. We use the cumulative Arias Intensity (cAI) metric \cite{arias1970measure} to compare the significant shaking duration of the GWM synthetics and of the real data. The significant shaking duration is defined by the times corresponding to 5\% and 95\% of the maximum cAI. Figure \ref{fig:arias_intensity}a shows the cAI curve for an example record from the real data set with $M$ 5.4, recorded at $R = 44$ km, on a site with $V_{S30} = 966$ m/s, hypocenter depth = 34.9 km, and azimuthal gap of $93^\circ$, along with the cAI curves from 100 GWM realizations with the same conditioning parameters. When we compute 1 GWM realization for each real record of the filtered test data (i.e., only showing for M $\leq 6$) (Figure \ref{fig:arias_intensity}b), we find very similar duration distributions between real data and GWM synthetics. The growth of shaking duration with earthquake magnitude is limited to the smaller events of this data set, because the signal duration of 40 s is too short to represent the full seismic waveforms of the larger events. 

\subsubsection{Predicting distributions of peak amplitudes}\label{sec:spectral_acceleration}
\begin{figure}
    \centering
    \includegraphics[width=0.5\linewidth]{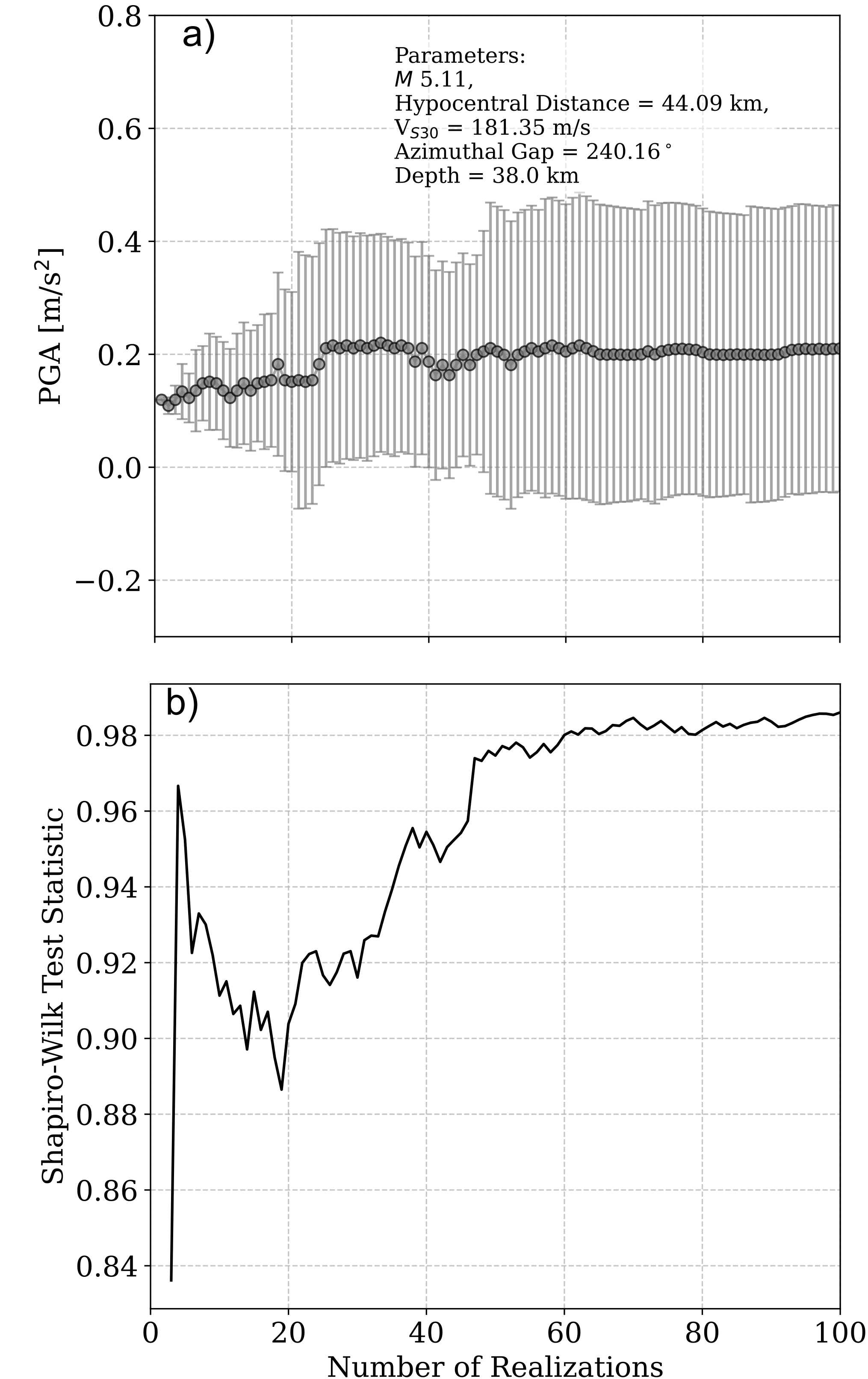}
    \caption{Statistics of the GWM realizations. a) Median and standard deviation of PGA values with different numbers of GWM realizations. b) Shapiro-Wilk test statistic for different numbers of realizations.}
    \label{fig:realization}
\end{figure}

Because we can generate any number of synthetics with the GWM, we can use the model to predict distributions of ground motion statistics, much like we commonly would with GMMs. For instance, we can generate $n$ synthetic waveforms for a set of conditioning parameters, and then compute the median and standard deviation of, e.g., PGA. 
It takes on the order of 60 GWM realizations for the median and the standard deviation estimates to stabilize (Figure \ref{fig:realization}a). To establish this we generate between 1 and 100 realizations using $M = 5.11$, $R = 44.09$ km, $V_{S30} = 181.35$ m/s, hypocenter depth = $38$ km, and azimuthal gap = $240.16^\circ$ as conditioning parameters, and analyze the median and standard deviation of the corresponding PGA values. Furthermore, we compute the Shapiro-Wilk test statistic (Figure \ref{fig:realization}b), to confirm that the peak amplitude predictions from the GWM are indeed log-normally distributed, as is the case for real data. This test statistic compares a data distribution to a normal distribution by evaluating $(\Sigma_{i=1}^n a_i x_{(i)})^2/\Sigma_{i=1}^n (x_i - \Bar{x})^2$, i.e. the ratio between the square of the sorted weighted sample values to the sum of the squared sample deviations from the mean, where $a_i$ is the weight defined by the test statistics and $x_i$ is the sample. A value close to 1 indicates normally distributed data and values close to 0 imply non-normal distribution. The test values for this particular set of conditioning parameters exceed 95\% at $n~>20$, suggesting that the model has correctly learned to generate waveforms with log-normally distributed peak amplitudes associated with standard deviation of 0.175 in log scale. This is a first-order characteristic of real data, and it also implies that we can accurately represent the amplitude distributions with only a mean and a standard deviation.

\subsubsection{Pseudo-spectral acceleration versus hypocentral distance}
\label{sec:amps_vs_distance}
\begin{figure}[htp!]
    \centering
    \includegraphics[width=.98\textwidth]{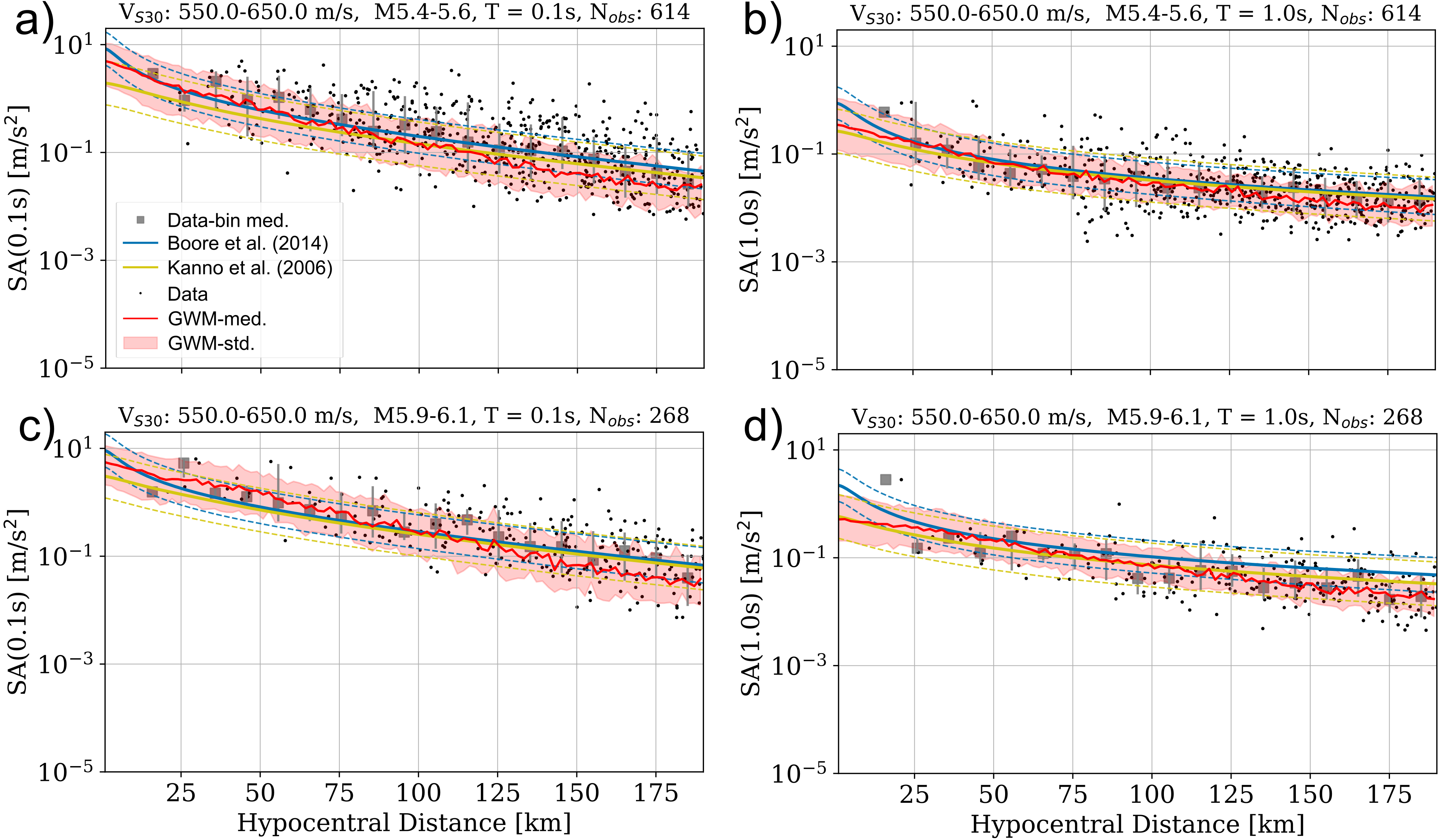}
    \caption{RotD50 pseudo-spectral acceleration (SA) with a damping factor 5\% versus hypocentral distance for periods ($T$) of 0.1 s and 1.0 s. Median prediction of the GWM (red line) and standard deviation (red shaded area), along with median prediction (solid lines) and standard deviation (dashed lines) of the \protect\citeA{boore2014nga} GMM (blue), and the \protect\citeA{kanno2006new} GMM (yellow), using $M=5.5$ and $V_{S30}=600$ m/s (a and b), and $M=6.0$ and $V_{S30}=600$ m/s (c and d). The data are sampled from narrow magnitude and $V_{S30}$ bins, as written in the figure titles, and shown by their median (grey squares) and standard deviations (grey lines).}
    \label{fig:peak_amps_vs_distance}
\end{figure}

We can use the GWM in this sense to directly compare the distributions it predicts with predicted distributions from the GMMs. To study, for instance, how peak amplitudes decay with distance, we compute 100 GWM synthetics for a vector of evenly spaced distances, every 1.9 km between 0.1 km to 200 km, and calculate the median and standard deviation of the peak amplitudes in each distance bin, for fixed magnitude, $V_{S30}$ value, hypocentral depth and azimuthal gap (Figure \ref{fig:peak_amps_vs_distance}).

For pseudo-spectral acceleration at 0.1 and 1.0 second periods with 5\% damping, the resulting GWM predictions decay smoothly with distance, similar to the GMM predictions. This includes very short distances, where both GWM and GMMs are under-constrained by available data. 
Depending on the selected period and parameter bin, the GWM matches the test data either similarly well (Figure \ref{fig:peak_amps_vs_distance}a and \ref{fig:peak_amps_vs_distance}c) as the GMMs or better (Figure \ref{fig:peak_amps_vs_distance}b and \ref{fig:peak_amps_vs_distance}d). The GWM also matches the real data rather well in cases when the data diverge from the typical GMM decay, which is sometimes observed for distances greater than 100 km (supplementary Figures \ref{fig:SA_GMM_5.0_150}-\ref{fig:SA_GMM_7.0_600}).

\subsubsection{Pseudo-spectral acceleration versus magnitude}\label{sec:amps_vs_magnitude}
\begin{figure}[htp!]
    \centering
    \includegraphics[width=0.99\textwidth]{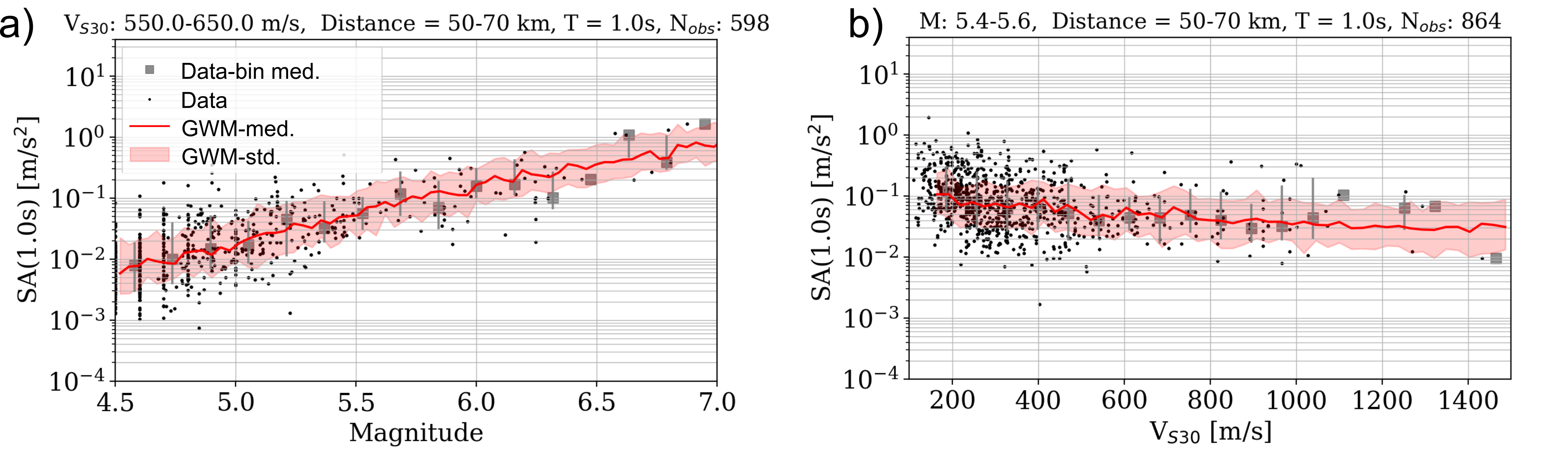}
    \caption{RotD50 pseudo-spectral acceleration (SA) with a damping factor of 5\% versus magnitude (a) and versus $V_{S30}$ (b) at $T = 1.0$ s. GWM prediction in terms of their median (red lines) and standard deviation (red shaded areas), real data in terms of their median (gray squares) and standard deviation (vertical grey lines), and using $R = 60$ km, for $V_{S30}=600$ m/s in (a) and $M=5.5$ in (b). All realizations are with conditioning parameters of hypocenter depth of 20 km and azimuthal gap of $130^\circ$. The data (black dots) are sampled from narrow magnitude, $R$, and $V_{S30}$ bins, as written in the figure titles.}
    \label{fig:peak_amps_vs_magnitude}
\end{figure}

In a similar sense, we can investigate how the GWM predictions grow with magnitude. We produce 100 GWM synthetics for a vector of magnitudes, evenly spaced every 0.035 from $M$ 4.5 to 7, for fixed distance and $V_{S30}$ values. The predictions for SA at $T=1.0$ s show relatively smooth, monotonous growth, up to a saturation at $M ~ 7.0$ (Figure \ref{fig:peak_amps_vs_magnitude}a), consistent with the (very sparse) real data. The same trend is observed for other conditioning parameter combinations (supplementary Figures \ref{fig:SA_magnitude_0-30} - \ref{fig:SA_magnitude_110-150}). 
It is interesting, and encouraging, that the GWM predictions are well-behaved in conditioning parameter ranges where the training data are very sparse, such as at $M > 6.5$, or for $R<20$ km.

\subsubsection{Pseudo-spectral acceleration versus $V_{S30}$}\label{sec:amps_vs_vs30}
To assess the scaling of pseudo-spectral acceleration with $V_{S30}$,
we generate another 100 GWM realizations for a vector of $V_{S30}$ values, evenly spaced every 13.63 m/s from 150 m/s to 1500 m/s (Figure \ref{fig:peak_amps_vs_magnitude}b). Generally, the GWM predictions follow the real data distribution, with SA values slightly decreasing with increasing $V_{S30}$. Towards very low $V_{S30}$, the GWM synthetics somewhat underpredict the strong growth of SA. 
Interestingly, for SA at $T \geq 1.0$ s, the SA values decrease up to $V_{S30} = 800$ m/s, then remain stable up to approximately $V_{S30} = 1200$ m/s. This is observed for a wide range of magnitude and distance bins (supplementary Figures \ref{fig:SA_vs30_0-30}-\ref{fig:SA_vs30_110-150}).
Given the limitations of $V_{S30}$ as a site response proxy  \cite{bergamo_2019_proxies}, a strong correlation of $V_{S30}$ and SA is not necessarily expected.

\subsection{Model probabilities given the data}\label{sec:probability}
In addition to studying model performance in selected parameter bins, we can assess the GWM and the GMMs across a wide range of magnitude and distance combinations by computing cumulative probabilities of the models, given the observed data. For each observed logarithmic spectral acceleration value $SA_i$, the GWM and GMMs predict a Gaussian normal distribution of expected logarithmic SA amplitudes, with a predicted mean $\mu$, and a standard deviation $\sigma$. We can then evaluate the probability of each data point under the predicted distribution. Assuming a uniform prior distribution, this probability is proportional to the probability of the model, given the data. By summing up these probabilities across all data points in a magnitude and distance bin, and then normalizing it with the number of data in the bin $N$, we can compute an average, relative model probability: 
\begin{equation}\label{eq:probability_given_data}
    P^{k}= \frac{1}{N}\sum_i^N \frac{1}{\sqrt{2\pi}\sigma_k} \exp\left( -\frac{(\text{SA}_i - \mu_k)^2}{2\sigma_k^2} \right),
\end{equation}
where $\mu_k$ and $\sigma_k$ are the predicted mean and standard deviation for the $k^{th}$ bin. This probability is high only if the model accurately predicts both the mean and standard deviation of the data. Therefore, with these probabilities, we can readily assess the agreement between the data and model predictions for a large number of conditioning parameter combinations. 

\begin{figure}[htp!]
    \centering
    \includegraphics[width=.99\textwidth]{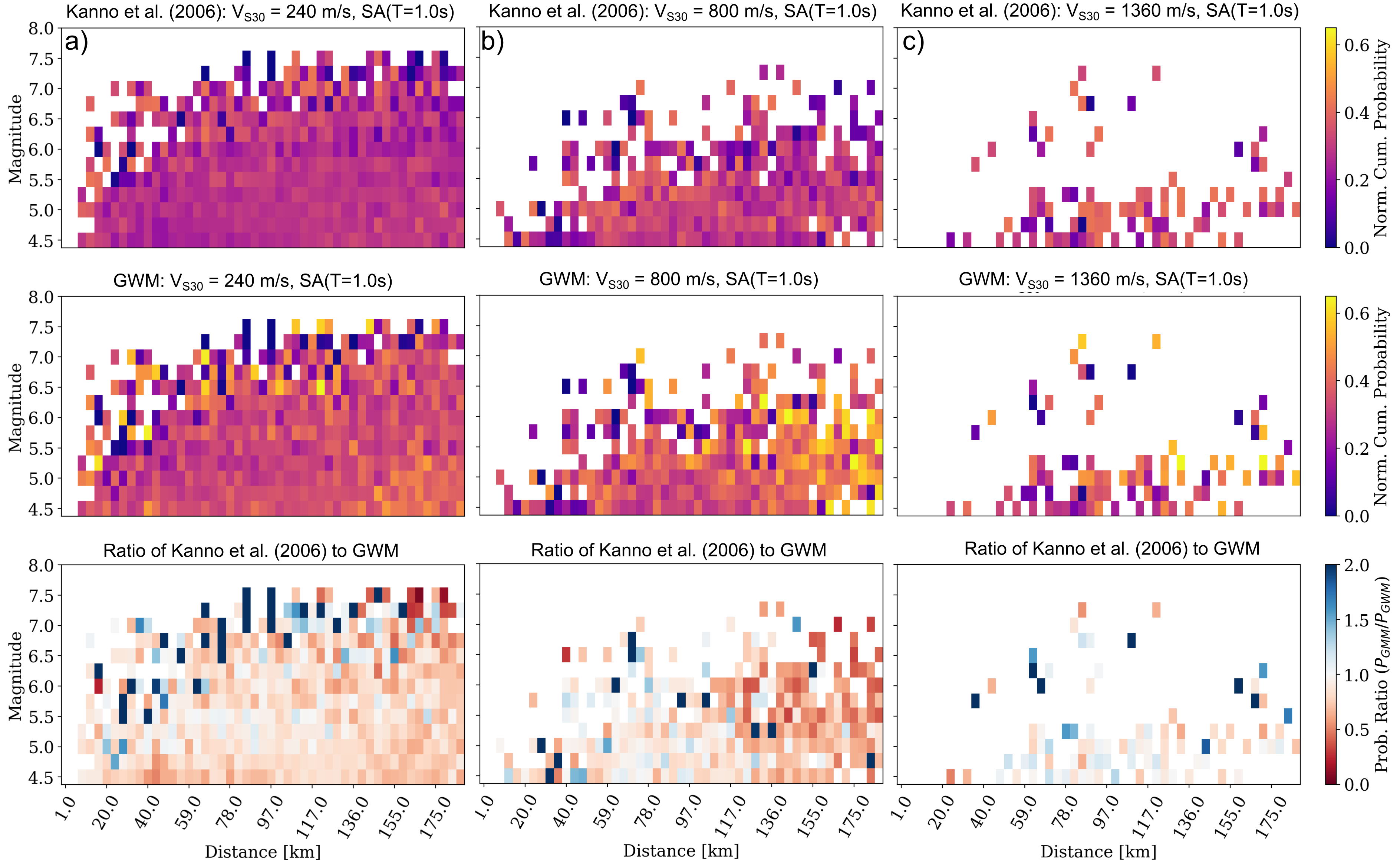}
    \caption{Average model probabilities given the SA data at $T$ = 1.0 s, in bins of magnitude and $R$, for $V_{S30} = 240$ m/s (a), 800 m/s (b), and 1360 m/s (c), for the \protect\citeA{kanno2006new} GMM (top row) and the GWM (middle row). The ratio between the two probabilities (bottom row) shows which model explains the observed SA data better.}
    \label{fig:GM_likelihood}
\end{figure}

We compute these model probabilities for the GWM ($P_{GWM}$) and for the two GMMs ($P_{GMM}$). We use the GMMs to compute the mean SA for the center of each bin, and use their reported standard deviations. For the GWM, we use 100 GWM realizations to compute the mean and standard deviation for each bin, likewise using the magnitude and distance of each bin center. We use hypocenter depth = 20 km and azimuthal gap = $130^\circ$, and repeat the computations for a number of $V_{S30}$ values, to compute the average probabilities for each bin after Eq. \ref{eq:probability_given_data}.
Figure \ref{fig:GM_likelihood} shows the model average probabilities for the GMM by \citeA{kanno2006new} (top row) and the GWM (middle row). To compare the two models we compute the probability ratio $r = P_{GMM}/P_{GWM}$ (bottom row). Equivalent figures for the \citeA{boore2014nga} GMM are shown in supplementary Figures \ref{fig:prob_GMM_boore_240}, \ref{fig:prob_GMM_boore_800}, and \ref{fig:prob_GMM_boore_1360}. 

In general, the GWM and GMMs have similar probabilities for most magnitude-distance combinations ($r \sim1$). For the largest magnitudes, the probability ratios are more variable, with GMM probabilities often higher than those of the GWM. 
The GMM by \citeA{boore2014nga} is comparable to the GWM for small magnitudes and moderate hypocentral distances (supplementary Figure \ref{fig:prob_GMM_boore_240}a). At lower periods $T < 1.0$ s and larger magnitudes ($M > 5.5$), the GWM performs better than either of the GMMs considered in this study (supplementary Figures \ref{fig:prob_GMM_boore_240} and \ref{fig:prob_GMM_kanno_240}). These patterns are also observed for $V_{S30} \leq 800$ m/s (Figures \ref{fig:GM_likelihood}b,\ref{fig:GM_likelihood}c and supplementary Figures \ref{fig:prob_GMM_boore_520} - \ref{fig:prob_GMM_kanno_800}). Unfortunately, for $V_{S30} > 800$ m/s, the test dataset is limited. Therefore, the model probabilities are more scattered (Supplementary Figures \ref{fig:prob_GMM_boore_1080} - \ref{fig:prob_GMM_kanno_1360}). 

\subsection{Fréchet Distances and classifier accuracy}
\label{sec:quantitative-evaluation}

In addition to the commonly used seismological performance metrics, we introduce additional, novel metrics that quantify the similarity between real data and GWM synthetics. The metrics are inspired by well-established performance metrics from the
image generation community and could play a crucial role in a systematic and quantitative comparison of various proposed GWMs, e.g., within the framework of future community efforts for GWM evaluations.

\subsubsection{Fréchet Distance of Fourier amplitude spectra}
\label{sec:asp_fd}
We use the Fréchet Distance (FD) to measure the distance between the Fourier amplitude spectra of the real and GWM synthetics. Ideally, realistic GWM waveforms should have a Fourier amplitude spectrum that is statistically indistinguishable from that of the real data. Larger FD values would indicate differences between the GWM synthetics and real data. The FD is equivalent to the Wasserstein-2 distance, and measures the minimum effort required, in the L2 sense, to transform one distribution into another. 
As in section \ref{sec:model_bias}, we compute one GWM synthetic waveform for each real waveform in the data set, using the conditioning parameter corresponding to the real data. The amplitude spectrum of each waveform is represented by $2033$ amplitude values ($1/2$ signal length + 1). We treat each vector element as an independent Gaussian, and compute the element-wise mean $\vmu$ and standard deviation $\vsigma$ of the log-amplitudes for each vector element, across all waveforms.  
The Fréchet Distance $d$ between observed real and GWM synthetic waveforms is then given by:
\begin{equation} \label{eq:fd}
    d = \| \vmu_{\text{obs}} - \vmu_{\text{syn}} \|^2 + \|\vsigma_{\text{obs}} - \vsigma_{\text{syn}} \|^2.
\end{equation}

Computing the FD for the entire data set, we find that it ranges from 25 to 40 for the three spatial components of the seismograms (Table \ref{tab:combined_results}). To provide a baseline for these FD values, we also compute the FD between the training and test sets (Section \ref{sec:data_description}), i.e. between sub-sets of the real data. This provides a baseline for ideal model performance, and suggests that, despite the good performance shown in sections \ref{sec:time_domain_signal} - \ref{sec:probability}, there are significant differences between real and synthetic spectra and that there is, therefore, room for model improvement. 
Furthermore, we also use this FD metric to show that the model performance decreases substantially if we leave out the auto-encoder, or if we choose a signal representation other than spectrograms (see ablation studies in \ref{app:ablations}).

\begin{figure}[h]
    \centering
    \includegraphics[width=.99\textwidth]{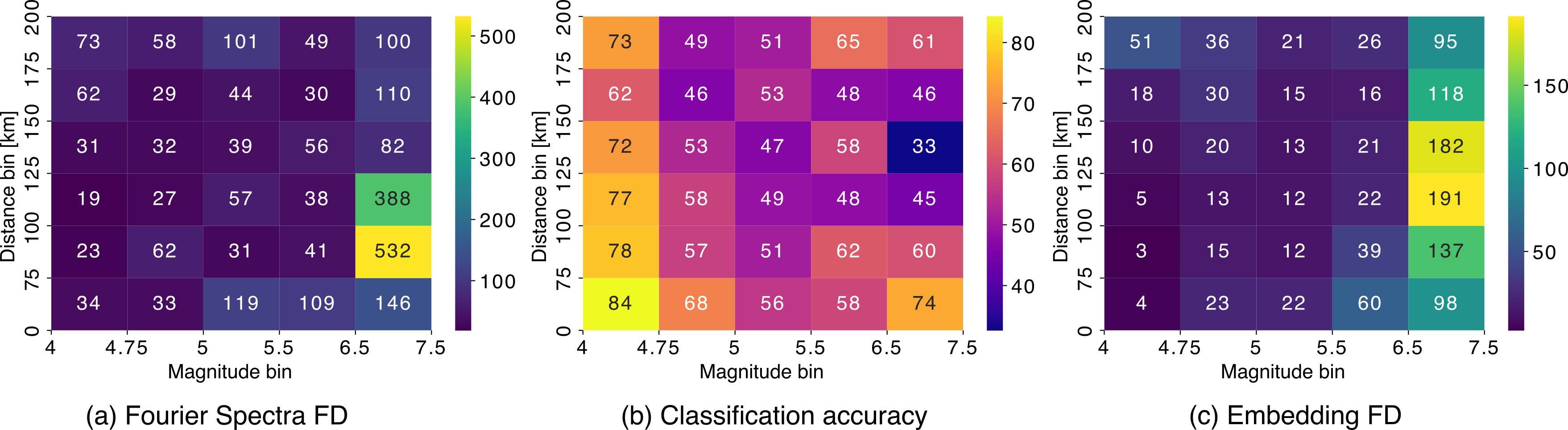}
    \caption{Evaluations on test data set. Fréchet Distances between real data and GWM synthetics in bins of magnitude and hypocentral distance, for the Fourier amplitude spectra of the radial component of the acceleration seismograms (a), classifier accuracy on the GWM synthetics (b), and FD of the classifier embeddings between training data and GWM synthetics, in bins of magnitude and hypocentral distance (c). We do not show the 7.5 - 9.1 magnitude bin, due to low-sample size in the test set.}
    \label{fig:FD}
\end{figure}

Furthermore, we can use the FD to compare the waveform generation for different magnitude and distance bins (Figure \ref{fig:FD}a and supplementary Figure \ref{fig:supp_asd}). The FDs are systematically higher for larger magnitude and shorter distance recordings, indicating a poorer fit between real and GWM synthetic data. This may be due to the relative scarcity of training data, and/or due to the inherently higher complexity of these records, making them more challenging for a model to mimic.

\begin{table}[h]
    \centering
    \resizebox{\linewidth}{!}{
    \begin{tabular}{lSSSSS}
        \toprule
        & \multicolumn{3}{c}{Fourier spectra FD $\downarrow$} & \multicolumn{2}{c}{Classifier} \\
        \cmidrule(lr){2-4} \cmidrule(lr){5-6}
        Distribution & {Radial} & {Transverse} & {Vertical} & {Accuracy (\%) $\uparrow$} & {Embedding FD $\downarrow$} \\
        \midrule
        GWM synthetics &  22.46 &  20.56 &  20.65 & 67.74 & 0.26 \\        
        Test set        &  0.29 &  0.38 &  0.42 & 69.31 & 0.10 \\
        \bottomrule
    \end{tabular}}
    \caption{Fréchet Distances of the Fourier amplitude spectra classifier accuracies and FDs of classifier embeddings, between real data and GWM synthetics. The Fourier spectra and classifier FDs are computed between the training data and corresponding GWM synthetics (first row), and between the training and test sets (second row).}
    \label{tab:combined_results}
\end{table}

\subsubsection{Classification accuracy}
\label{sec:classifier_accuracy}
Inspired by the common practice in image synthesis to evaluate the quality of generated images with a pre-trained classifier \cite{heusel2017gans}, we adopt a similar approach: we train a classifier to categorize seismic data into bins of magnitude and distance. We divide our dataset into five magnitude and five distance bins, resulting in 25 classes, with each class containing a similar number of samples (Appendix \ref{app:classifier_binning}). 
We then train a classifier to predict the magnitude-distance bin for each record. For the classifier, we use a convolutional neural network (CNN) architecture that, like the GWM, operates on spectrogram representations of the waveforms (Appendix \ref{app:architectures}). 
The classifier achieves a test accuracy of $69.31\%$, correctly predicting the magnitude-distance bin for this fraction of real waveforms. When applied to synthetic waveforms, the classifier's accuracy is $67.74\%$ (Table~\ref{tab:combined_results}). Ideally, if synthetic waveforms were indistinguishable from real ones, the classifier would maintain the same accuracy for both. The 36 magnitude-distance classes are not perfectly balanced (Figure~\ref{fig:binning}b); a naive random guessing approach would yield $1/36$ accuracy, while always predicting the largest magnitude-distance bin would achieve approximately $20.00\%$ accuracy (largest class in the training data ($38946/193983$)). Although not perfect, the classifier's performance on synthetic waveforms ($67.74\%$) strongly exceeds both these baseline accuracies, indicating that the synthetic waveforms encapsulate substantial information about first-order statistics, enabling the classifier to make informed predictions. Figure~\ref{fig:FD}b illustrates the accuracy achieved on the generated dataset across different magnitude-distance bins. Notably, some of the most difficult parameter ranges have high classification accuracy.

\subsubsection{Fréchet Distance of classifier embeddings}
\label{sec:classifier_embedding_fd}
For similar input data, not only should the classifiers' performance be comparable, but its internal representations should also align. Based on this intuition, we use the FD to measure the similarity of the classifier's hidden representations between the real and synthetic waveforms. We collect the $256$-dimensional hidden representations from the penultimate layer of the classifier for both real and synthetic waveforms. Unlike the FD of Fourier amplitude spectra (Section \ref{sec:asp_fd}) where each dimension was treated independently, the reduced dimensionality of this representation ($256$ instead of $2033$) allows us to compute correlations between dimensions. Thus, we calculate the entire covariance matrix $\mSigma$ of the hidden representations for both the real and generated waveforms. The Fréchet Distance $d$ between the two sets of hidden representations is then given by:
\begin{equation} \label{eq:general_fd}
    d = \|\vmu_{\text{obs}} - \vmu_{\text{gen}}\|^2 + \text{Tr}(\mSigma_{\text{obs}} + \mSigma_{\text{gen}} - 2(\mSigma_{\text{obs}}\mSigma_{\text{gen}})^{1/2}).
\end{equation}
This is a generalization of the Fréchet Distance in Eq. \ref{eq:fd} to general multivariate (non-isotropic) Gaussians. In image generation literature, this metric is known as the Fréchet Inception Distance (FID) \cite{heusel2017gans}, with "Inception" referring to the classifier architecture employed. 

This FD of the classifier embeddings between the training data and the GWM synthetics is close to the FD between training and test data (Table \ref{tab:combined_results}) implying that the synthetics share characteristics with real waveforms. When computed separately for the magnitude-distance bins, we see again how the model performs worse for larger magnitude records (Figure \ref{fig:FD}c), and how the presented model performs significantly better than the ablated models evaluated in \ref{app:ablations}.

In summary, all three metrics provide an objective, relative measure of synthesis quality. As such they can readily be used to, for example, compare different proposed GWMs, model versions, or to evaluate the model performance on a particular subset of the data. Importantly, ideal lower bounds for the FD, and ideal upper bounds for the classifier, can be estimated by computing the metrics using just real data (second row in Table \ref{tab:combined_results}).

\section{Discussion} 
\label{sec:discussion}
Generative Waveform Models (GWMs) are rapidly advancing and have the potential to significantly improve earthquake hazard assessment and earthquake engineering studies \cite{florez2022data, esfahani2023tfcgan, shi2024broadband, matsumoto2024GAN}. Unlike GMMs, which predict scalar ground motion metrics,
GWMs can synthesize fully realistic waveforms, complete with realistic frequency- and time-domain properties.

The ability to predict waveforms enables studies that rely on waveform content, such as building response simulations \cite{bommer2004use}. Any scalar ground motion metric can be derived from the predicted waveforms. A key advantage of this approach is that the waveforms and their derivatives are equally realistic across the entire frequency range (0.1 - 50 Hz). This may contrast with hybrid methods, which add high-frequency spectra in the second stage, either using stochastic methods \cite{saikia1997simulated, mai2010hybrid, graves2010broadband} or neural networks \cite{paolucci2018broadband, gatti2020towards, okazaki2021simulation}. The former can introduce artifacts at the transition between the deterministic and stochastic parts of the predictions \cite{mai2003hybrid, graves2010broadband, tang2023stochastic}.

Another advantage of GWMs is their ability to accurately learn the correlation between scalar statistics of the same waveform (e.g., spectral accelerations at different frequency ranges). This factor has been considered in only a few GMMs to date \cite{baker2017intensity, baker2021seismic}.

This study introduces HighFEM, a new GWM that builds on the recent successes of Denoising Diffusion Models in image, audio, and video generation \cite{song2019generative, ho2020denoising, song2020scorebased, dhariwal2021diffusion, nichol2021improved, kong2021diffwave, rombach2022highresolution, ho2022video}. Diffusion models are capable of generating high-resolution and diverse samples while being simple to implement and train. Unlike GANs used for synthesizing waveforms \cite{florez2022data, esfahani2023tfcgan, matsumoto2024GAN}, denoising diffusion models do not suffer from mode collapse or related training challenges. Additionally, no dedicated neural network architecture is required, as with neural operators \cite{shi2024broadband}.

The proposed GWM operates on the spectrogram representation of waveform data to address scaling issues associated with the wide range of amplitudes contained in seismic waveforms. It incorporates an autoencoder to compress these spectrograms into a more compact form, thereby enhancing the efficiency of both the training and generation processes while simultaneously improving the resolved frequency of the generated waveforms. This efficiency may also lead to better model performance in scenarios where data scarcity is a concern, a common issue in many earthquake seismology problems. Moreover, it is important for applications where large numbers of forward computations are required, such as in probabilistic seismic hazard assessment.

The proposed generative latent denoising diffusion model generates highly realistic waveforms across the entire hazard-relevant frequency range. The GWM synthetics have realistic time domain shapes (Figure \ref{fig:envelope}a), Fourier amplitude spectra (Figure \ref{fig:envelope}c), and shaking durations (Figure \ref{fig:arias_intensity}). The predicted peak amplitude statistics are largely unbiased (Figure \ref{fig:bias_IM}) and have log-normally distributed amplitude variation (Figure \ref{fig:realization}), with the same amount of variability as the real data (Figure \ref{fig:histogram_stats_PGA_PGV}), 

With repeated inference for the same conditioning parameter sets, GWMs can be used to predict distributions of scalar amplitude statistics, like is commonly done with GMMs, albeit at a higher computational cost. We show that our model predicts peak amplitudes of strong motion seismic data in Japan at least as accurately and as precisely as two prominent and reliable GMMs.

While GWM-based predictions are computationally more expensive than GMM computations, they offer new possibilities for both practical and scientific applications. For instance, next-generation seismic hazard models could include model branches containing a catalog of representative waveforms - rather than just amplitude statistics - which are expected over the hazard target period. This would provide a much more detailed description of the anticipated ground motion in a target region. It would also expand the applicability of hazard calculations to use cases where full waveforms are required or desirable, such as non-linear structural dynamic analyses. In this sense, GWMs have the potential to unify earthquake hazard and engineering studies.

GWMs could eventually complement the need to re-scale limited data pools of observed seismic waveforms to match expected target spectra, or to use hybrid methods for enriching simulated waveforms with high-frequency seismic energy. Instead, a set of fully realistic broadband waveforms can be generated from a single, self-consistent synthesis process.

One important question in this context will be which among existing and future GWMs generates the most realistic waveforms for a particular application. To facilitate quantitative, representative and fair model comparisons, we propose that the emerging GWM community embrace open model and code standards and participate in existing community efforts for the comparison and evaluation of ground motion models (e.g., \citeA{maechling2015scec}), with yet-to-be-established benchmark data sets. Performance metrics like the ones introduced in Section \ref{sec:quantitative-evaluation} could facilitate meaningful comparison between models and model versions. Such model comparison efforts could also include blind signal classification exercises, where trained classifier models would attempt to distinguish between real and synthetic waveforms.

\subsection{This-Quake-Does-Not-Exist ('tqdne') Python Library}
For the HighFEM model presented in this study, we introduce an openly available and user-friendly Python library that can be used to generate waveforms using the pre-trained GWM from this study or to train custom GWMs. Generating a three-component waveform with the pre-trained GWMs requires approximately 2 seconds per realization on a typical MacBook M2 laptop and approximately 0.3 seconds on a single GPU NVIDIA A100. The library facilitates saving of the waveforms in SeisBench format \cite{woollam2022seisbench}. The name of the library is inspired by the popular \citeA{thispersondoesnotexist} application, which uses the StyleGAN algorithm \cite{karras2019style, karras2020analyzing} to generate human portrait images.

\subsection{Application of the Generative Waveform Model}
Our HighFEM can produce GMM-type models similar to those shown in Figure \ref{fig:peak_amps_vs_distance} and Figures \ref{fig:SA_GMM_5.0_150}–\ref{fig:SA_GMM_7.0_150}. The data-driven approach allows the generation of an arbitrarily large ensemble of GWM, exhibiting a log-scale standard deviation of $\approx$0.175. This scatter represents the lower and upper bounds (i.e., standard deviation) of the empirical distribution when conditioned on hypocentral distance, moment magnitude, $V_{S30}$, hypocenter depth, and azimuthal gap. Because the method is waveform-based, the GWM can be readily regionalized, e.g., by retraining with local earthquake records if a large dataset is available or by fine-tuning, which is sufficient to adapt the model to any target area worldwide.

\subsection{Limitations}
While the presented GWM arguably achieves high seismic waveform synthesis performance, there are several limitations that future models can aspire to overcome. We list the limitations as follows:

\begin{itemize}
    \item \textbf{Stochastic nature of the generated waveforms}. Fundamentally, the generated waveforms are stochastic representations of real seismograms. There is no underlying physical model for wave excitation and propagation. Although the GWM synthetics exhibit clear energy packets that closely resemble P-, S-, and coda waves, they do not represent any wavefield phases in a deterministic sense.

    \item \textbf{Limited training data}. As is the case for all models of strong ground motion, the limited number of short-distance recordings of large magnitude earthquakes is a major restriction. This limitation affects the synthesis performance of this crucial data regime. GWMs can, in principle, be used to augment such data sets, but it is currently an open question how well the models extrapolate beyond the parameter ranges for which they have been trained, and how well they perform at the data-scarce edges of the parameter ranges. 

    \item \textbf{Point source assumption}. Our model assumes that the earthquake source is a point source and neglects finite fault source characteristics such as fault geometry and distance, source roughness, directivity, and unilateral or bilateral rupture modes.

    \item \textbf{Uncorrelated stations}. The current model does not explicitly take into account the correlation of observations across different records of the same quake. Each generated waveform is an independent realization of the denoising forward process. This may lead to an underestimation of the correlation of observed ground motions from the same quake, and might limit the ability of the model to generalize to new stations. 

    \item \textbf{P-wave onset times}. The current model was trained on a waveform dataset whose traces were aligned with the PhaseNet onset detector \cite{zhu2019phasenet}, re-trained on the Japan Meteorological Agency unified dataset \cite{naoi2024neural}. Because this detector is not perfectly accurate, the resulting GWM synthetics show slight variability in their P-wave onset times.

    \item \textbf{Signal length}. The 40-second long seismograms are sufficient to describe the ground motion from quakes with magnitudes of up to M$\sim 7.5$. For even larger quakes, the source duration alone may exceed this signal length. Producing longer sequences without compromising temporal resolution presents some challenges, even for our efficient model. Addressing this issue may necessitate an approach with more favorable asymptotic behavior, which is a subject for future research.

    \item \textbf{Lower spectral amplitude}. Our model slightly underestimates the spectral amplitude of the ground motion compared to the real data (Figures \ref{fig:envelope}b and \ref{fig:bias_IM}a). This discrepancy is observed exclusively in the model operating on the spectrogram representation. We hypothesize that this is due to the model generating a slightly blurred version of the encoded spectrograms, similar to the effect of a Gaussian filter. While this blurring is inconsequential for image generation tasks, as it is imperceptible to the human eye, it may result in a lower spectral amplitude (e.g., averaging 0.04 m/s$^2$Hz$^{-1}$ at frequency $<$ 30 Hz, (supplementary Figure \ref{fig:amp_residual})). Partially, the underestimation may also stem from the spectrogram autoencoder (Figure \ref{fig:spectrogram_autoencoder_asd}). A potential solution could involve incorporating additional loss terms, such as adversarial loss, to encourage the model to generate sharper spectrograms. These could for instance be included in the autoencoder stage. Alternatively, exploring different, potentially smoother spectral representations that are less sensitive to blurring may also be beneficial.
\end{itemize}

\section{Conclusion}
\label{sec:conclusion}
We present High Frequency Earthquake Model (HighFEM), a data-driven, conditional generative model for synthesizing three-component strong motion seismograms. This generative waveform model (GWM) combines a convolutional auto-encoder with a state-of-the-art latent denoising diffusion model, which generates encoded - rather than raw - spectrogram representations of the seismic signals.

We trained the openly available model on Japanese strong motion data with hypocentral distances of 1–200 km, moment magnitudes $\geq$ 4.0, and $V_{S30}$ values of 76–2100 m/s.
Using a variety of commonly used and novel evaluation metrics, we demonstrate that the GWM synthetics accurately capture the statistical properties of the observed data in both the time and frequency domains, across a wide range of conditioning parameters, and up to the highest hazard-relevant frequencies.

Furthermore, we systematically compare the peak ground motion statistics of the GWM synthetics to predictions from commonly used GMMs. The GWM predictions are largely unbiased and exhibit the same level of amplitude variability as the real data. As a result, they may be useful for practical applications, such as probabilistic seismic hazard assessment and structural dynamic analyses.

With GWMs, seismic hazard models can potentially expand their scope to include applications that require waveform representations, rather than just scalar amplitude statistics. Future community efforts to benchmark and compare GWMs would provide guidance for which models to best use in practical and scientific applications, and may accelerate GWM innovation.



\appendix
\section{Generative model details}
This section provides additional details on the model architectures and representations used in the experiments. 

\subsection{Advantages and disadvantages of diffusion models}
Denoising diffusion models (DDMs) have rapidly become the state of the art in generative modeling \cite{dhariwal2021diffusion,yang2023diffusion}, offering a multitude of advantages. Importantly, they have repeatedly been shown to convincingly outperform GANs, VAEs, normalizing flows, and other state-of-the-art generative models across modalities such as images, video, audio, and text. In general, DDMs provide exceptional mode coverage and sample diversity. Compared to GANs, they are more robust during training and exhibit interpretable loss curves, making convergence easier to monitor. GANs, by contrast, often suffer from mode collapse or vanishing gradients, and selecting an appropriate training optimizer can be brittle. Additionally, because DDMs have access to the score function, a key statistical quantity, they can be applied to, for instance, inverse problems \cite{daras2024survey}, approximate inference \cite{blessing2024beyond}, and reinforcement learning \cite{zhu2023diffusion}.

DDMs do have some disadvantages. Theoretically, generating samples requires solving the entire reverse SDE. However, by using a Heun sampler (see Section~\ref{sec:denoising_diffusion}), we reduce the number of sampling steps to only $50$. This still involves more neural network evaluations than GANs or VAEs. As a potential remedy, distillation techniques, such as consistency distillation \cite{song23consistency}, can be applied to reduce the sampling process to a single step, though this requires additional training. Given the acceptable inference speed of our model, we did not find such an approach necessary. Furthermore, DDMs can converge more slowly during training than GANs or VAEs due to their theoretical formulation.

\subsection{Neural network architectures}
\label{app:architectures}
All our models are based on the widely used U-Net architecture presented in \citeA{song2020scorebased}. The U-Net consists of
three components which we denote \textit{left} (encoder), \textit{middle}, and \textit{right} (decoder). All components use several residual blocks that use two convolutional layers (2D for the spectrogram and 1D for the moving average envelope), with pre-layer group normalization and SiLU activation functions. In addition, the downsampling component uses a convolutional layer to reduce the dimensionality of an input between each pair of residual blocks, while the upsampling component uses upsampling operations to double the dimensionality of an input between each pair of residual blocks. An overview of the architectures is given in Table~\ref{tab:hyperparameters}.

The most critical hyperparameters for the DDMs are the number of channels they are trained on (provided by the autoencoder) and the number of hidden channels. We non-exhaustively evaluated different combinations of these two hyperparameters and found that the DDMs are highly robust to them. For instance, we found that DDMs trained on 8 or 4 channels perform very similarly, indicating that already 4 latent channels would have sufficed. Similarly, reducing the number of hidden channels by a factor of two only minimally hurts performance. In the end, we chose both autoencoders and DDMs based on the lowest validation losses during training. Other hyperparameters, such as the 'Attention Levels' or 'Dropout Rate' have been adopted from previous literature and not been tuned. Due to data and model sizes, for this study exhaustive hyperparameter-searches are not computationally feasible.

\begin{figure}[h]
    \centering
    \includegraphics[width=\linewidth]{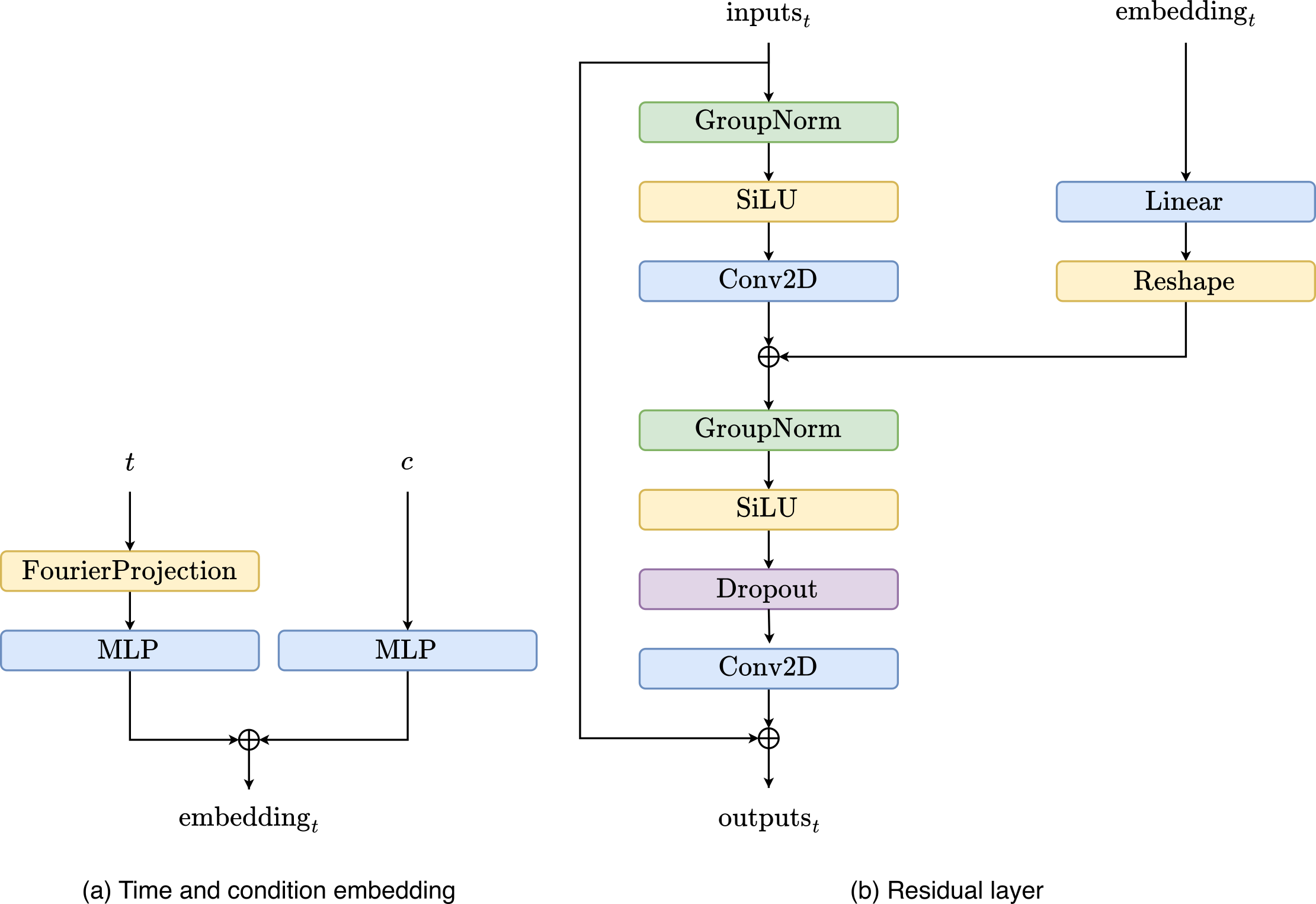}
    \caption{Using low-dimensional features for conditioning the diffusion model. (a) For the scalar value $t$, we first compute $256$-dimensional Fourier features and embed both via separate MLP neural networks. The two embeddings are combined by simply adding them elementwise. (b) For each residual block, we condition the synthesized spectrogram ($\text{inputs}_t$) using the combined time-feature embedding by adding them to the hidden representation of the residual layer. For that, we take the $256$-dimensional embedding vector, transform it through a linear layer with $K$ output neurons, and reshape it to match the size of the hidden representation. Concretely, if the hidden representation has dimensionality
    $N \times H \times W \times K$ where $N$ is the batch size and $H \times W$ is the spectrogram size, we repeat the embedding $N \times H \times W$ times, reshape the resulting tensor to match the dimensionality of the hidden representation, and add the hidden representation and conditioning information elementwise (in deep learning libraries like \texttt{PyTorch} this can be efficiently done).}
    \label{fig:conditioning-residual-block}
\end{figure}

\textbf{Denoising Diffusion}
The neural network for the diffusion model uses four residual blocks on encoder and decoder components, with an additional residual block in the middle. After the first three levels, we include a downsampling operation on the encoder an upsampling operation on the decoder side. In addition, the central blocks incorporate a self-attention module. As per convention, conditioning information is injected within each residual block by concatenating projections of the conditioning vector $\vc$ and the time embedding $t$ to the intermediate representations of an input $\vz_t$ (see Figure~\ref{fig:conditioning-residual-block} and \citeA{song2020scorebased,karras2022elucidating}). 

\paragraph*{Autoencoder} The autoencoder architecture comprises the same blocks as the denoising model but lacks self-attention modules and only uses $3$ residual blocks. It uses downsampling and upsampling operations between each residual block. As a consequence, the autoencoder compresses the input by a factor of four in each spatial dimension.

\paragraph*{Classifier} The classifier is a convolutional neural network consisting of four residual blocks, each followed by a downsampling operation. It includes a self-attention layer at the end, followed by a global average pooling operation, an output multi-layer perceptron (MLP), a linear layer, and a softmax activation function. When extracting embeddings from the classifier, we utilize the output of the MLP prior to the linear layer. The classifier is trained on the spectrogram representation of the data.

\begin{table}[h]
    \centering
    \resizebox{\linewidth}{!}{
    \begin{tabular}{lccccccc}
        \toprule
        Hyperparameters & {Moving Average Diffusion} & \multicolumn{2}{c}{Moving Average Latent Diffusion} & {Spectrogram Diffusion} & \multicolumn{2}{c}{Spectrogram Latent Diffusion} & {Classifier} \\
        & & {Autoencoder} & {Diffusion Model} & & {Autoencoder} & {Diffusion Model} & \\
        \midrule
        Convolution Kernel Size & 5 & 5 & 5 & $3 \times 3$ & $3 \times 3$ & $3 \times 3$ & $3 \times 3$ \\
        Hidden Channels & [64, 128, 256, 256] & [64, 128, 256] & [64, 128, 256, 256] & [128, 256, 256, 512] & [64, 128, 256] & [128, 256, 256, 512] & [64, 128, 256, 256] \\
        Latent dimensionality & - & 16 & - & - & 8 & - & - \\
        Attention Levels & [4] & - & [4] & [4] & - & [4] & [4] \\
        Dropout Rate & 0.1 & 0.1 & 0.1 & 0.1 & 0.1 & 0.1 & 0.1 \\
        KL Weight & - & $10^{-6}$ & - & - & $10^{-6}$ & - & - \\
        Optimizer & Adam & Adam & Adam & Adam & Adam & Adam & Adam \\
        Learning Rate & $10^{-4}$ & $10^{-4}$ & $10^{-4}$ & $10^{-4}$ & $10^{-4}$ & $10^{-4}$ & $10^{-4}$ \\
        EMA Decay & 0.999 & 0.999 & 0.999 & 0.999 & 0.999 & 0.999 & 0.999 \\
        Batch Size & 256 & 256 & 256 & 64 & 128 & 256 & 128 \\
        Epochs & 300 & 200 & 300 & 300 & 300 & 300 & 100 \\
        \bottomrule
    \end{tabular}}
    \caption{Hyperparameters for the various models used in the experiments.} 
    \label{tab:hyperparameters}
\end{table}

\subsection{Training}

For all architectures, we use a conventional Adam optimizer \cite{kingma2015adam} with a cosine annealing schedule which decays the learning rate over the number of training epochs from the initial learning rate to zero (\citeA{loshchilov2017sgdr}; see Table~\ref{tab:hyperparameters} for hyperparameters). We found both autoencoder as well as DDM highly robust to parameterization of these optimizers and did not observe any instabilities during training.

\subsection{Representations} 
\label{sec:representations}
We choose two different data representations for generative modelling of the seismic waveforms which we describe in detail below.

\subsubsection*{Spectrogram Representation} 
To transform each of the three channels in the original waveform into a spectrogram, we utilize a Short-Time Fourier Transform (STFT) with 256 frequency bins and a hop length of 32 samples. Due to the symmetry of the spectrogram, only half of the frequency bins are used. To prevent padding issues, the original waveform is truncated to $4064$ samples, resulting in a complex-valued matrix of size $128 \times 128$. We then take the magnitude of this matrix and apply a logarithmic transformation to obtain the spectrogram, discarding the phase information due to its high-frequency nature, which is challenging to model accurately. To reconstruct the original waveform, we employ the Griffin-Lim algorithm \cite{griffin1984signal, perraudin2013fast}, which reliably estimates the phase from the magnitude spectrogram.

\begin{figure}[h]
    \centering
    \includegraphics[width=\linewidth]{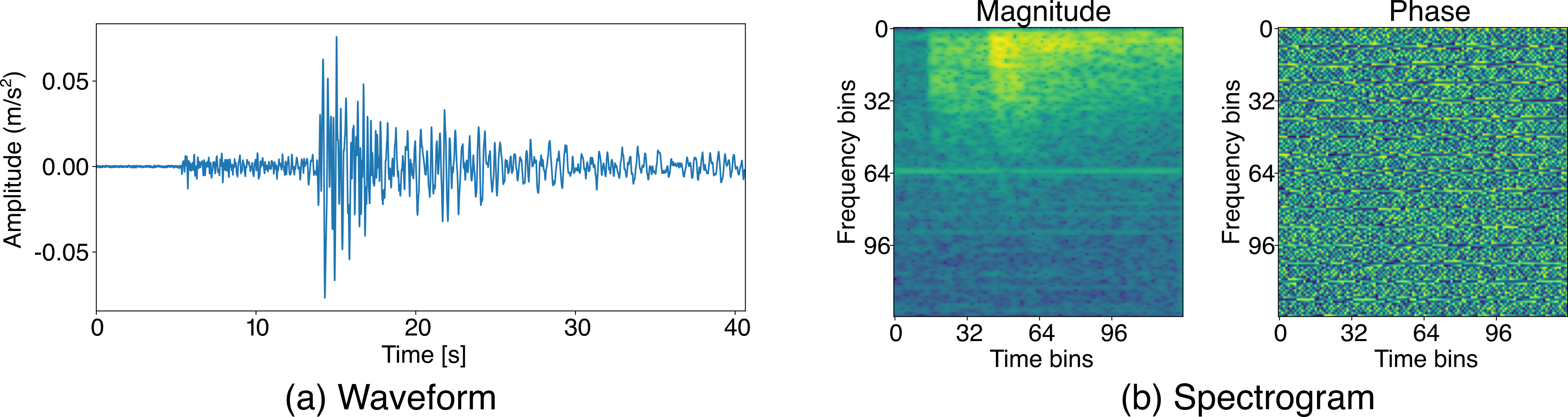}
    \caption{Seismic waveform and its corresponding spectrogram representation.}
    \label{fig:waveform_spectrogram}
\end{figure}

\subsubsection*{Moving Average Envelope Representation}
The moving average envelope is computed by convolving the absolute waveform signal with an averaging boxcar filter of length 128. The final representation is the concatenation of the original waveform divided by the envelope and the logarithm of the envelope.

\subsection{Classifier training data binning}
\label{app:classifier_binning}
\begin{figure}[h]
    \centering
    \includegraphics[width=\linewidth]{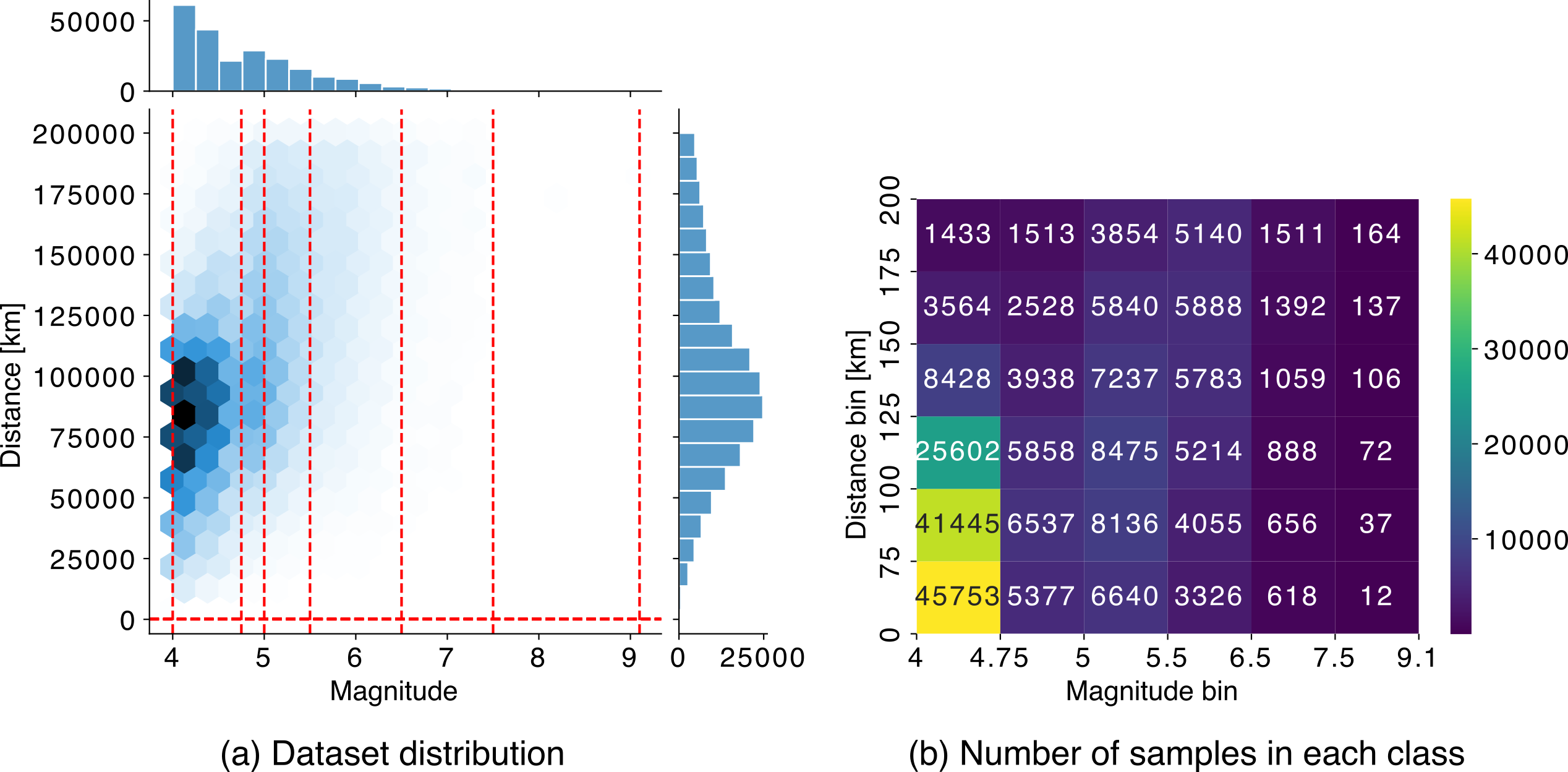}
    \caption{Binning of the data into classes for the classifier, based on magnitude and distance.}
    \label{fig:binning}
\end{figure}

When training the classifier to categorize data based on earthquake magnitude and distance, we divide the data into five magnitude and five distance bins. Figure~\ref{fig:binning} displays the sample count in each bin, ensuring a balanced sample distribution across classes.

\section{Ablation Studies}
\label{app:ablations}
This section evaluates the significance of the three components in our proposed model. The time-domain representation of seismic data shows significant amplitude variation, making direct processing of raw waveforms ineffective. Figure~\ref{fig:raw_autoencoder_asd} illustrates the logarithm of the Fourier amplitude spectrum between an autoencoder's input and output when trained on raw waveforms, revealing poor reconstruction, especially in the high-frequency range. It is important to note that this is not a generative model but an autoencoder tasked to reconstruct the input. Despite this, the waveforms are poorly reconstructed, particularly in the high-frequency range.

To address this, we explore alternative time-domain representations. We decompose the signal into its positive envelope and residual signal, with the envelope being a smoothed version of the absolute signal, as detailed in Section~\ref{sec:representations}. Figure~\ref{fig:envelope_autoencoder_asd} shows that an autoencoder trained on this representation performs better than one trained on raw waveforms but still struggles with high-frequency components.

Thirdly, we experiment with a spectrogram representation, specifically the log-transformed magnitude of the short-time Fourier transform. This representation is smooth and robust to amplitude variations. Figure~\ref{fig:spectrogram_autoencoder_asd} demonstrates that an autoencoder trained on this representation nearly perfectly reconstructs the original signal, with the exception of a minor underestimation of the spectral amplitude, as discussed in Section~\ref{sec:discussion}. This highlights the effectiveness of the spectrogram representation for the task of ground motion synthesis.

Finally, we assess the performance of the diffusion model trained on different representations and the impact of incorporating the autoencoder stage. Table~\ref{tab:ablation} summarizes our findings, showing that the spectrogram representation significantly outperforms the envelope representation across all metrics. Additionally, the autoencoder stage improves the spectral fit. Overall, the latent diffusion approach with the spectrogram representation is the most effective configuration.

\begin{figure}[h]
    \centering
    \includegraphics[width=\linewidth]{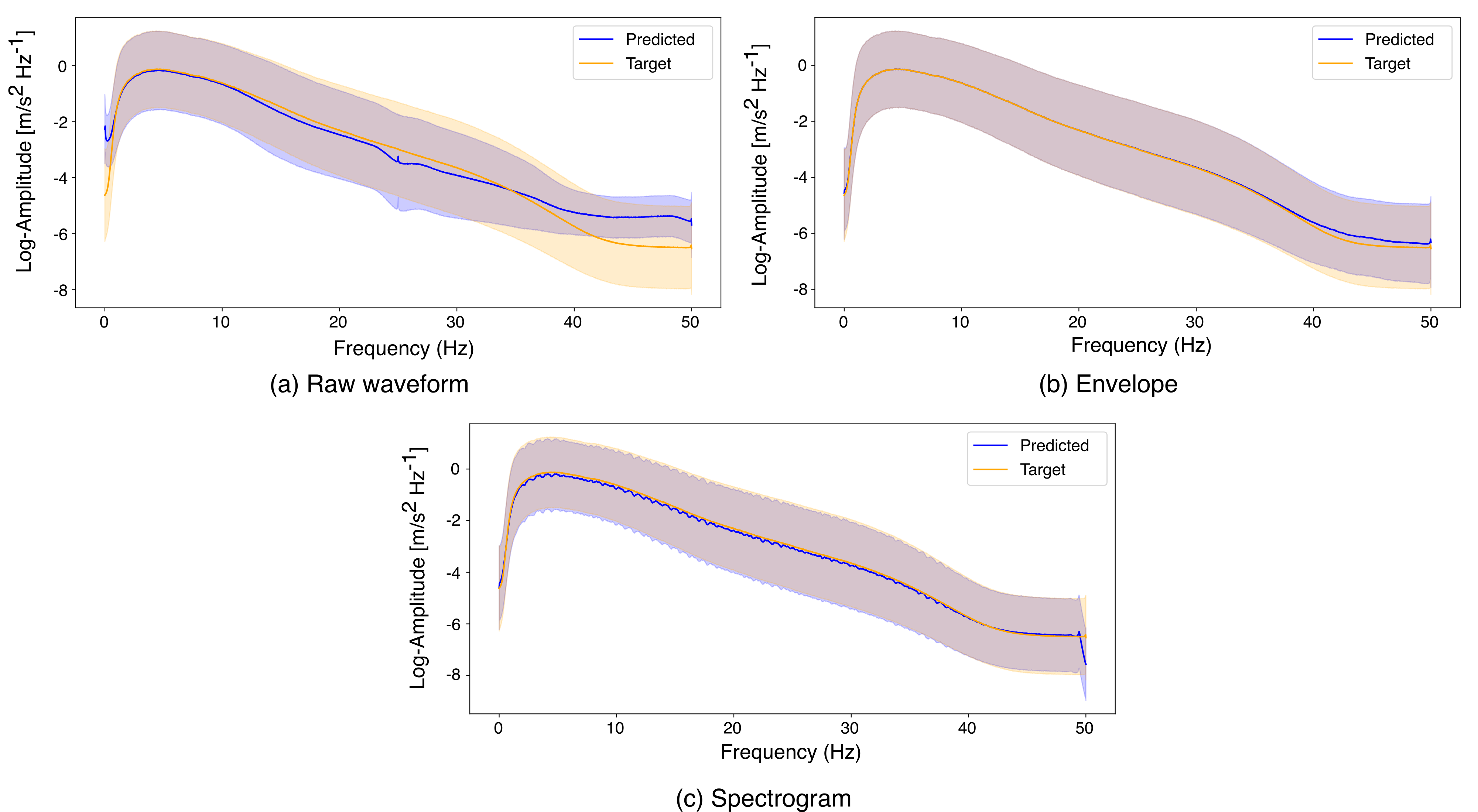}
    \caption{Fourier spectra log-amplitude comparison between the input and output of an autoencoder trained on different representations. The East-West component of the three-channel signal is used for visualization.}
    \label{fig:asd_autoencoder}
\end{figure}

\begin{table}[h]
    \centering
    \resizebox{\linewidth}{!}{
    \begin{tabular}{lcccccc}
        \toprule
        & & \multicolumn{3}{c}{Fourier spectra FD $\downarrow$} & \multicolumn{2}{c}{Classifier} \\
        \cmidrule(lr){3-5} \cmidrule(lr){6-7}
        Representation & Latent space & {Radial} & {Transverse} & {Vertical} & {Accuracy (\%) $\uparrow$} & {Embedding FD $\downarrow$} \\
        \midrule
        Envelope & \xmark & 2071.57 & 2131.29 & 2056.52 & 10.58 & 705.07\\
        Envelope & \cmark & 204.26 & 202.50 & 207.72 & 34.08 &  244.04\\
        Spectrogram & \xmark & 908.00 & 935.87 & 873.15 & 62.40 & 43.22 \\
        Spectrogram & \cmark & 24.08 & 21.10 & 26.45 & 67.74 &  1.48\\
        \bottomrule
    \end{tabular}}
    \caption{Ablation study comparing the performance of the diffusion model when trained on the moving average envelope and the spectrogram representation. Results are shown for both direct training on representations and training on the latent space of an autoencoder. The Fréchet Distance (FD) for the log-amplitude Fourier spectra and classifier embeddings is reported between the test set and the generated samples. Classifier accuracy is reported for the generated samples. In both cases conditioning variables from the test set have been used.}
    \label{tab:ablation}
\end{table}

%
%
\clearpage
\section*{Open Research Section}\label{sec:data_resources}
The three-component raw strong-motion data from the Kyosin network (K-NET) and the Kiban-Kyoshin network (KiK-net) time series waveforms are provided by the National Research Institute for Earth Science and Disaster Prevention of Japan and can be freely downloaded at {\footnotesize \mbox{\url{https://www.kyoshin.bosai.go.jp/}}} \cite{knet2019}.

The This-Quake-Does-Not-Exist (’tqdne’) Python Library is available in a Zenodo online repository at {\footnotesize \mbox{\url{https://zenodo.org/records/16405538}}} \cite{bergmeister_2024_13952381} and on GitHub at {\footnotesize \mbox{\url{https://github.com/highfem/tqdne/tags}}}. The repositories contain the trained model, including weights, and the code to train the model. The source code can be used to reproduce all experimental results, to train neural networks, and to generate synthetic waveforms. All online pages were last accessed on November 12\textsuperscript{th}, 2024. The supplementary material provides additional information and figures to complement the main content of the primary text, offering a deeper understanding and further validation of the presented results.

\section*{Declaration of Competing Interest}
The authors declare that there are no conflicts of interest.

\acknowledgments
This work was supported by grant number C22-10 (HighFEM) of the Swiss Data Science Center (SDSC), Ecole Polytechnique Fédérale de Lausanne and ETH Zürich awarded to M-A. Meier, L. Ermert and M. Koroni. L. Ermert was supported by Swiss National Science Foundation grant 209941. M. Koroni is supported by the Swiss Federal Nuclear Safety Inspectorate (ENSI) under contract number CTR00830. We thank Donat Fäh and Paolo Bergamo for useful discussions on ground motion models and earthquake engineering. We also thank Editor John Rundle, Randy Harsuko, and three other reviewers for their constructive comments, which significantly improved our original manuscript.  We thank CSCS Swiss National Computing Center (Piz Daint under projects sd28 and s1165) and Swiss Seismological Service ``Bigstar" Cluster for providing computational resources for this research.

%
%
\clearpage
\bibliography{references}

\begin{thebibliography}{}

\bibitem [\protect \citeauthoryear {%
{Applied Technology Council}%
}{%
{Applied Technology Council}%
}{%
{\protect \APACyear {2009}}%
}]{%
applied2009quantification}
\APACinsertmetastar {%
applied2009quantification}%
\begin{APACrefauthors}%
{Applied Technology Council}.%
\end{APACrefauthors}%
\unskip\
\newblock
\APACrefYear{2009}.
\newblock
\APACrefbtitle {{Quantification of building seismic performance factors}}
  {{Quantification of building seismic performance factors}}.
\newblock
\APACaddressPublisher{}{US Department of Homeland Security, FEMA}.
\PrintBackRefs{\CurrentBib}

\bibitem [\protect \citeauthoryear {%
Aquib%
\ \BBA {} Mai%
}{%
Aquib%
\ \BBA {} Mai%
}{%
{\protect \APACyear {2024}}%
}]{%
tariq2024BB}
\APACinsertmetastar {%
tariq2024BB}%
\begin{APACrefauthors}%
Aquib, A\BPBI T.%
\BCBT {}\ \BBA {} Mai, P\BPBI M.%
\end{APACrefauthors}%
\unskip\
\newblock
\APACrefYearMonthDay{2024}{}{}.
\newblock
{\BBOQ}\APACrefatitle {{{Broadband Ground‐Motion Simulations with
  Machine‐Learning‐Based High‐Frequency Waves from Fourier Neural
  Operators}}} {{{Broadband Ground‐Motion Simulations with
  Machine‐Learning‐Based High‐Frequency Waves from Fourier Neural
  Operators}}}.{\BBCQ}
\newblock
\APACjournalVolNumPages{Bulletin of the Seismological Society of
  America}{}{}{}.
\newblock
\begin{APACrefDOI} \doi{10.1785/0120240027} \end{APACrefDOI}
\PrintBackRefs{\CurrentBib}

\bibitem [\protect \citeauthoryear {%
Arias%
}{%
Arias%
}{%
{\protect \APACyear {1970}}%
}]{%
arias1970measure}
\APACinsertmetastar {%
arias1970measure}%
\begin{APACrefauthors}%
Arias, A.%
\end{APACrefauthors}%
\unskip\
\newblock
\APACrefYearMonthDay{1970}{}{}.
\newblock
{\BBOQ}\APACrefatitle {{A measure of earthquake intensity}} {{A measure of
  earthquake intensity}}.{\BBCQ}
\newblock
\APACjournalVolNumPages{Seismic design for nuclear plants}{}{}{438--483}.
\PrintBackRefs{\CurrentBib}

\bibitem [\protect \citeauthoryear {%
Baker%
\ \BBA {} Allin~Cornell%
}{%
Baker%
\ \BBA {} Allin~Cornell%
}{%
{\protect \APACyear {2006}}%
}]{%
baker2006spectral}
\APACinsertmetastar {%
baker2006spectral}%
\begin{APACrefauthors}%
Baker, J.%
\BCBT {}\ \BBA {} Allin~Cornell, C.%
\end{APACrefauthors}%
\unskip\
\newblock
\APACrefYearMonthDay{2006}{}{}.
\newblock
{\BBOQ}\APACrefatitle {{Spectral shape, epsilon and record selection}}
  {{Spectral shape, epsilon and record selection}}.{\BBCQ}
\newblock
\APACjournalVolNumPages{Earthquake Engineering \& Structural
  Dynamics}{35}{9}{1077--1095}.
\newblock
\begin{APACrefDOI} \doi{10.1002/eqe.571} \end{APACrefDOI}
\PrintBackRefs{\CurrentBib}

\bibitem [\protect \citeauthoryear {%
Baker%
\ \BBA {} Bradley%
}{%
Baker%
\ \BBA {} Bradley%
}{%
{\protect \APACyear {2017}}%
}]{%
baker2017intensity}
\APACinsertmetastar {%
baker2017intensity}%
\begin{APACrefauthors}%
Baker, J.%
\BCBT {}\ \BBA {} Bradley, B.%
\end{APACrefauthors}%
\unskip\
\newblock
\APACrefYearMonthDay{2017}{}{}.
\newblock
{\BBOQ}\APACrefatitle {{Intensity measure correlations observed in the
  NGA-West2 database, and dependence of correlations on rupture and site
  parameters}} {{Intensity measure correlations observed in the NGA-West2
  database, and dependence of correlations on rupture and site
  parameters}}.{\BBCQ}
\newblock
\APACjournalVolNumPages{Earthquake Spectra}{33}{1}{145--156}.
\newblock
\begin{APACrefDOI} \doi{10.1193/060716eqs095m} \end{APACrefDOI}
\PrintBackRefs{\CurrentBib}

\bibitem [\protect \citeauthoryear {%
Baker%
, Bradley%
\BCBL {}\ \BBA {} Stafford%
}{%
Baker%
\ \protect \BOthers {.}}{%
{\protect \APACyear {2021}}%
}]{%
baker2021seismic}
\APACinsertmetastar {%
baker2021seismic}%
\begin{APACrefauthors}%
Baker, J.%
, Bradley, B.%
\BCBL {}\ \BBA {} Stafford, P.%
\end{APACrefauthors}%
\unskip\
\newblock
\APACrefYear{2021}.
\newblock
\APACrefbtitle {{Seismic hazard and risk analysis}} {{Seismic hazard and risk
  analysis}}.
\newblock
\APACaddressPublisher{}{Cambridge University Press}.
\newblock
\begin{APACrefDOI} \doi{10.1017/9781108425056} \end{APACrefDOI}
\PrintBackRefs{\CurrentBib}

\bibitem [\protect \citeauthoryear {%
Bayless%
\ \BBA {} Abrahamson%
}{%
Bayless%
\ \BBA {} Abrahamson%
}{%
{\protect \APACyear {2019}}%
}]{%
bayless2019empirical}
\APACinsertmetastar {%
bayless2019empirical}%
\begin{APACrefauthors}%
Bayless, J.%
\BCBT {}\ \BBA {} Abrahamson, N\BPBI A.%
\end{APACrefauthors}%
\unskip\
\newblock
\APACrefYearMonthDay{2019}{}{}.
\newblock
{\BBOQ}\APACrefatitle {An empirical model for the interfrequency correlation of
  epsilon for {F}ourier amplitude spectra} {An empirical model for the
  interfrequency correlation of epsilon for {F}ourier amplitude
  spectra}.{\BBCQ}
\newblock
\APACjournalVolNumPages{Bulletin of the Seismological Society of
  America}{109}{3}{1058--1070}.
\newblock
\begin{APACrefDOI} \doi{10.1785/0120180238} \end{APACrefDOI}
\PrintBackRefs{\CurrentBib}

\bibitem [\protect \citeauthoryear {%
Bergamo%
, Hammer%
\BCBL {}\ \BBA {} Fäh%
}{%
Bergamo%
\ \protect \BOthers {.}}{%
{\protect \APACyear {2019}}%
}]{%
bergamo_2019_proxies}
\APACinsertmetastar {%
bergamo_2019_proxies}%
\begin{APACrefauthors}%
Bergamo, P.%
, Hammer, C.%
\BCBL {}\ \BBA {} Fäh, D.%
\end{APACrefauthors}%
\unskip\
\newblock
\APACrefYearMonthDay{2019}{}{}.
\newblock
\APACrefbtitle {{SERA WP7/NA5 - Deliverable 7.4: Towards improvement of site
  condition indicators}} {{SERA WP7/NA5 - Deliverable 7.4: Towards improvement
  of site condition indicators}}\ \APACbVolEdTR {}{Report}.
\newblock
\APACaddressInstitution{Zurich}{ETH Zurich}.
\newblock
\begin{APACrefDOI} \doi{10.3929/ethz-b-000467564} \end{APACrefDOI}
\PrintBackRefs{\CurrentBib}

\bibitem [\protect \citeauthoryear {%
Bergmeister%
\ \protect \BOthers {.}}{%
Bergmeister%
\ \protect \BOthers {.}}{%
{\protect \APACyear {2024}}%
}]{%
bergmeister_2024_13952381}
\APACinsertmetastar {%
bergmeister_2024_13952381}%
\begin{APACrefauthors}%
Bergmeister, A.%
, Palgunadi, K\BPBI H.%
, Bosisio, A.%
, Ermert, L.%
, Koroni, M.%
, Perraudin, N.%
, Dirmeier, S.%
\BCBL {}\ \BBA {} Meier, M\BHBI A.%
\end{APACrefauthors}%
\unskip\
\newblock
\APACrefYearMonthDay{2024}{}{}.
\newblock
\APACrefbtitle {{Software package "tqdne" for paper titled "High Resolution
  Seismic Waveform Generation using Denoising Diffusion"}.} {{Software package
  "tqdne" for paper titled "High Resolution Seismic Waveform Generation using
  Denoising Diffusion"}.}
\newblock
\APACaddressPublisher{}{Zenodo}.
\newblock
\begin{APACrefDOI} \doi{10.5281/zenodo.14017182} \end{APACrefDOI}
\PrintBackRefs{\CurrentBib}

\bibitem [\protect \citeauthoryear {%
Blessing%
, Jia%
, Esslinger%
, Vargas%
\BCBL {}\ \BBA {} Neumann%
}{%
Blessing%
\ \protect \BOthers {.}}{%
{\protect \APACyear {2024}}%
}]{%
blessing2024beyond}
\APACinsertmetastar {%
blessing2024beyond}%
\begin{APACrefauthors}%
Blessing, D.%
, Jia, X.%
, Esslinger, J.%
, Vargas, F.%
\BCBL {}\ \BBA {} Neumann, G.%
\end{APACrefauthors}%
\unskip\
\newblock
\APACrefYearMonthDay{2024}{}{}.
\newblock
{\BBOQ}\APACrefatitle {{Beyond ELBOs: A Large-Scale Evaluation of Variational
  Methods for Sampling}} {{Beyond ELBOs: A Large-Scale Evaluation of
  Variational Methods for Sampling}}.{\BBCQ}
\newblock
\BIn{} \APACrefbtitle {Forty-first International Conference on Machine
  Learning.} {Forty-first international conference on machine learning.}
\newblock
\begin{APACrefDOI} \doi{10.5555/3692070.3692239} \end{APACrefDOI}
\PrintBackRefs{\CurrentBib}

\bibitem [\protect \citeauthoryear {%
Bommer%
\ \BBA {} Acevedo%
}{%
Bommer%
\ \BBA {} Acevedo%
}{%
{\protect \APACyear {2004}}%
}]{%
bommer2004use}
\APACinsertmetastar {%
bommer2004use}%
\begin{APACrefauthors}%
Bommer, J.%
\BCBT {}\ \BBA {} Acevedo, A.%
\end{APACrefauthors}%
\unskip\
\newblock
\APACrefYearMonthDay{2004}{}{}.
\newblock
{\BBOQ}\APACrefatitle {{The use of real earthquake accelerograms as input to
  dynamic analysis}} {{The use of real earthquake accelerograms as input to
  dynamic analysis}}.{\BBCQ}
\newblock
\APACjournalVolNumPages{Journal of Earthquake Engineering}{8}{spec01}{43--91}.
\newblock
\begin{APACrefDOI} \doi{10.1080/13632460409350521} \end{APACrefDOI}
\PrintBackRefs{\CurrentBib}

\bibitem [\protect \citeauthoryear {%
Boore%
}{%
Boore%
}{%
{\protect \APACyear {2003}}%
}]{%
boore2003simulation}
\APACinsertmetastar {%
boore2003simulation}%
\begin{APACrefauthors}%
Boore, D\BPBI M.%
\end{APACrefauthors}%
\unskip\
\newblock
\APACrefYearMonthDay{2003}{}{}.
\newblock
{\BBOQ}\APACrefatitle {{Simulation of ground motion using the stochastic
  method}} {{Simulation of ground motion using the stochastic method}}.{\BBCQ}
\newblock
\APACjournalVolNumPages{Pure and applied geophysics}{160}{}{635--676}.
\newblock
\begin{APACrefDOI} \doi{10.1007/PL00012553} \end{APACrefDOI}
\PrintBackRefs{\CurrentBib}

\bibitem [\protect \citeauthoryear {%
Boore%
\ \BBA {} Joyner%
}{%
Boore%
\ \BBA {} Joyner%
}{%
{\protect \APACyear {1997}}%
}]{%
boore1997site}
\APACinsertmetastar {%
boore1997site}%
\begin{APACrefauthors}%
Boore, D\BPBI M.%
\BCBT {}\ \BBA {} Joyner, W\BPBI B.%
\end{APACrefauthors}%
\unskip\
\newblock
\APACrefYearMonthDay{1997}{}{}.
\newblock
{\BBOQ}\APACrefatitle {{Site amplifications for generic rock sites}} {{Site
  amplifications for generic rock sites}}.{\BBCQ}
\newblock
\APACjournalVolNumPages{Bulletin of the seismological society of
  America}{87}{2}{327--341}.
\newblock
\begin{APACrefDOI} \doi{10.1785/BSSA0870020327} \end{APACrefDOI}
\PrintBackRefs{\CurrentBib}

\bibitem [\protect \citeauthoryear {%
Boore%
, Stewart%
, Seyhan%
\BCBL {}\ \BBA {} Atkinson%
}{%
Boore%
\ \protect \BOthers {.}}{%
{\protect \APACyear {2014}}%
}]{%
boore2014nga}
\APACinsertmetastar {%
boore2014nga}%
\begin{APACrefauthors}%
Boore, D\BPBI M.%
, Stewart, J\BPBI P.%
, Seyhan, E.%
\BCBL {}\ \BBA {} Atkinson, G\BPBI M.%
\end{APACrefauthors}%
\unskip\
\newblock
\APACrefYearMonthDay{2014}{}{}.
\newblock
{\BBOQ}\APACrefatitle {{NGA-West2 equations for predicting PGA, PGV, and 5\%
  damped PSA for shallow crustal earthquakes}} {{NGA-West2 equations for
  predicting PGA, PGV, and 5\% damped PSA for shallow crustal
  earthquakes}}.{\BBCQ}
\newblock
\APACjournalVolNumPages{Earthquake Spectra}{30}{3}{1057--1085}.
\newblock
\begin{APACrefDOI} \doi{10.1193/070113EQS184} \end{APACrefDOI}
\PrintBackRefs{\CurrentBib}

\bibitem [\protect \citeauthoryear {%
Boore%
, Watson-Lamprey%
\BCBL {}\ \BBA {} Abrahamson%
}{%
Boore%
\ \protect \BOthers {.}}{%
{\protect \APACyear {2006}}%
}]{%
boore2006orientation}
\APACinsertmetastar {%
boore2006orientation}%
\begin{APACrefauthors}%
Boore, D\BPBI M.%
, Watson-Lamprey, J.%
\BCBL {}\ \BBA {} Abrahamson, N\BPBI A.%
\end{APACrefauthors}%
\unskip\
\newblock
\APACrefYearMonthDay{2006}{}{}.
\newblock
{\BBOQ}\APACrefatitle {{Orientation-independent measures of ground motion}}
  {{Orientation-independent measures of ground motion}}.{\BBCQ}
\newblock
\APACjournalVolNumPages{Bulletin of the Seismological Society of
  America}{96}{4A}{1502--1511}.
\newblock
\begin{APACrefDOI} \doi{10.1785/0120050209} \end{APACrefDOI}
\PrintBackRefs{\CurrentBib}

\bibitem [\protect \citeauthoryear {%
Chopra%
}{%
Chopra%
}{%
{\protect \APACyear {2007}}%
}]{%
chopra2007dynamics}
\APACinsertmetastar {%
chopra2007dynamics}%
\begin{APACrefauthors}%
Chopra, A\BPBI K.%
\end{APACrefauthors}%
\unskip\
\newblock
\APACrefYear{2007}.
\newblock
\APACrefbtitle {{Dynamics of structures}} {{Dynamics of structures}}.
\newblock
\APACaddressPublisher{}{Pearson Education India}.
\PrintBackRefs{\CurrentBib}

\bibitem [\protect \citeauthoryear {%
Daras%
\ \protect \BOthers {.}}{%
Daras%
\ \protect \BOthers {.}}{%
{\protect \APACyear {2024}}%
}]{%
daras2024survey}
\APACinsertmetastar {%
daras2024survey}%
\begin{APACrefauthors}%
Daras, G.%
, Chung, H.%
, Lai, C\BHBI H.%
, Mitsufuji, Y.%
, Ye, J\BPBI C.%
, Milanfar, P.%
, Dimakis, A\BPBI G.%
\BCBL {}\ \BBA {} Delbracio, M.%
\end{APACrefauthors}%
\unskip\
\newblock
\APACrefYearMonthDay{2024}{}{}.
\newblock
{\BBOQ}\APACrefatitle {{A survey on diffusion models for inverse problems}} {{A
  survey on diffusion models for inverse problems}}.{\BBCQ}
\newblock
\APACjournalVolNumPages{arXiv preprint arXiv:2410.00083}{}{}{}.
\newblock
\begin{APACrefDOI} \doi{10.48550/arXiv.2410.00083} \end{APACrefDOI}
\PrintBackRefs{\CurrentBib}

\bibitem [\protect \citeauthoryear {%
D{\'e}fossez%
, Copet%
, Synnaeve%
\BCBL {}\ \BBA {} Adi%
}{%
D{\'e}fossez%
\ \protect \BOthers {.}}{%
{\protect \APACyear {2023}}%
}]{%
defossez2023high}
\APACinsertmetastar {%
defossez2023high}%
\begin{APACrefauthors}%
D{\'e}fossez, A.%
, Copet, J.%
, Synnaeve, G.%
\BCBL {}\ \BBA {} Adi, Y.%
\end{APACrefauthors}%
\unskip\
\newblock
\APACrefYearMonthDay{2023}{}{}.
\newblock
{\BBOQ}\APACrefatitle {{High Fidelity Neural Audio Compression}} {{High
  Fidelity Neural Audio Compression}}.{\BBCQ}
\newblock
\APACjournalVolNumPages{Transactions on Machine Learning Research}{}{}{}.
\newblock
\begin{APACrefDOI} \doi{10.48550/arXiv.2210.13438} \end{APACrefDOI}
\PrintBackRefs{\CurrentBib}

\bibitem [\protect \citeauthoryear {%
Derras%
, Bard%
, Cotton%
\BCBL {}\ \BBA {} Bekkouche%
}{%
Derras%
\ \protect \BOthers {.}}{%
{\protect \APACyear {2012}}%
}]{%
derras2012adapting}
\APACinsertmetastar {%
derras2012adapting}%
\begin{APACrefauthors}%
Derras, B.%
, Bard, P\BHBI Y.%
, Cotton, F.%
\BCBL {}\ \BBA {} Bekkouche, A.%
\end{APACrefauthors}%
\unskip\
\newblock
\APACrefYearMonthDay{2012}{}{}.
\newblock
{\BBOQ}\APACrefatitle {{Adapting the neural network approach to PGA prediction:
  An example based on the KiK-net data}} {{Adapting the neural network approach
  to PGA prediction: An example based on the KiK-net data}}.{\BBCQ}
\newblock
\APACjournalVolNumPages{Bulletin of the Seismological Society of
  America}{102}{4}{1446--1461}.
\newblock
\begin{APACrefDOI} \doi{10.1785/0120110088} \end{APACrefDOI}
\PrintBackRefs{\CurrentBib}

\bibitem [\protect \citeauthoryear {%
Dhariwal%
\ \BBA {} Nichol%
}{%
Dhariwal%
\ \BBA {} Nichol%
}{%
{\protect \APACyear {2021}}%
}]{%
dhariwal2021diffusion}
\APACinsertmetastar {%
dhariwal2021diffusion}%
\begin{APACrefauthors}%
Dhariwal, P.%
\BCBT {}\ \BBA {} Nichol, A.%
\end{APACrefauthors}%
\unskip\
\newblock
\APACrefYearMonthDay{2021}{}{}.
\newblock
{\BBOQ}\APACrefatitle {{Diffusion models beat GANs on image synthesis}}
  {{Diffusion models beat GANs on image synthesis}}.{\BBCQ}
\newblock
\APACjournalVolNumPages{Advances in neural information processing
  systems}{34}{}{8780--8794}.
\newblock
\begin{APACrefDOI} \doi{10.48550/arXiv.2105.05233} \end{APACrefDOI}
\PrintBackRefs{\CurrentBib}

\bibitem [\protect \citeauthoryear {%
Douglas%
}{%
Douglas%
}{%
{\protect \APACyear {2003}}%
}]{%
douglas2003earthquake}
\APACinsertmetastar {%
douglas2003earthquake}%
\begin{APACrefauthors}%
Douglas, J.%
\end{APACrefauthors}%
\unskip\
\newblock
\APACrefYearMonthDay{2003}{}{}.
\newblock
{\BBOQ}\APACrefatitle {{Earthquake ground motion estimation using strong-motion
  records: a review of equations for the estimation of peak ground acceleration
  and response spectral ordinates}} {{Earthquake ground motion estimation using
  strong-motion records: a review of equations for the estimation of peak
  ground acceleration and response spectral ordinates}}.{\BBCQ}
\newblock
\APACjournalVolNumPages{Earth-Science Reviews}{61}{1-2}{43--104}.
\newblock
\begin{APACrefDOI} \doi{10.1016/S0012-8252(02)00112-5} \end{APACrefDOI}
\PrintBackRefs{\CurrentBib}

\bibitem [\protect \citeauthoryear {%
Douglas%
\ \BBA {} Aochi%
}{%
Douglas%
\ \BBA {} Aochi%
}{%
{\protect \APACyear {2008}}%
}]{%
douglas2008survey}
\APACinsertmetastar {%
douglas2008survey}%
\begin{APACrefauthors}%
Douglas, J.%
\BCBT {}\ \BBA {} Aochi, H.%
\end{APACrefauthors}%
\unskip\
\newblock
\APACrefYearMonthDay{2008}{}{}.
\newblock
{\BBOQ}\APACrefatitle {{A survey of techniques for predicting earthquake ground
  motions for engineering purposes}} {{A survey of techniques for predicting
  earthquake ground motions for engineering purposes}}.{\BBCQ}
\newblock
\APACjournalVolNumPages{Surveys in geophysics}{29}{}{187--220}.
\newblock
\begin{APACrefDOI} \doi{10.1007/s10712-008-9046-y} \end{APACrefDOI}
\PrintBackRefs{\CurrentBib}

\bibitem [\protect \citeauthoryear {%
Esfahani%
, Cotton%
, Ohrnberger%
\BCBL {}\ \BBA {} Scherbaum%
}{%
Esfahani%
\ \protect \BOthers {.}}{%
{\protect \APACyear {2023}}%
}]{%
esfahani2023tfcgan}
\APACinsertmetastar {%
esfahani2023tfcgan}%
\begin{APACrefauthors}%
Esfahani, R\BPBI D.%
, Cotton, F.%
, Ohrnberger, M.%
\BCBL {}\ \BBA {} Scherbaum, F.%
\end{APACrefauthors}%
\unskip\
\newblock
\APACrefYearMonthDay{2023}{}{}.
\newblock
{\BBOQ}\APACrefatitle {{TFCGAN}: {N}onstationary Ground-Motion Simulation in
  the Time-Frequency Domain Using Conditional Generative Adversarial Network
  (CGAN) and Phase Retrieval Methods} {{TFCGAN}: {N}onstationary ground-motion
  simulation in the time-frequency domain using conditional generative
  adversarial network (cgan) and phase retrieval methods}.{\BBCQ}
\newblock
\APACjournalVolNumPages{Bulletin of the Seismological Society of
  America}{113}{1}{453--467}.
\newblock
\begin{APACrefDOI} \doi{10.1785/0120220068} \end{APACrefDOI}
\PrintBackRefs{\CurrentBib}

\bibitem [\protect \citeauthoryear {%
Esfahani%
\ \protect \BOthers {.}}{%
Esfahani%
\ \protect \BOthers {.}}{%
{\protect \APACyear {2021}}%
}]{%
esfahani2021exploring}
\APACinsertmetastar {%
esfahani2021exploring}%
\begin{APACrefauthors}%
Esfahani, R\BPBI D.%
, Vogel, K.%
, Cotton, F.%
, Ohrnberger, M.%
, Scherbaum, F.%
\BCBL {}\ \BBA {} Kriegerowski, M.%
\end{APACrefauthors}%
\unskip\
\newblock
\APACrefYearMonthDay{2021}{}{}.
\newblock
{\BBOQ}\APACrefatitle {{Exploring the dimensionality of ground-motion data by
  applying autoencoder techniques}} {{Exploring the dimensionality of
  ground-motion data by applying autoencoder techniques}}.{\BBCQ}
\newblock
\APACjournalVolNumPages{Bulletin of the Seismological Society of
  America}{111}{3}{1563--1576}.
\newblock
\begin{APACrefDOI} \doi{10.1785/0120200285} \end{APACrefDOI}
\PrintBackRefs{\CurrentBib}

\bibitem [\protect \citeauthoryear {%
Florez%
\ \protect \BOthers {.}}{%
Florez%
\ \protect \BOthers {.}}{%
{\protect \APACyear {2022}}%
}]{%
florez2022data}
\APACinsertmetastar {%
florez2022data}%
\begin{APACrefauthors}%
Florez, M\BPBI A.%
, Caporale, M.%
, Buabthong, P.%
, Ross, Z\BPBI E.%
, Asimaki, D.%
\BCBL {}\ \BBA {} Meier, M\BHBI A.%
\end{APACrefauthors}%
\unskip\
\newblock
\APACrefYearMonthDay{2022}{}{}.
\newblock
{\BBOQ}\APACrefatitle {{Data-driven synthesis of broadband earthquake ground
  motions using artificial intelligence}} {{Data-driven synthesis of broadband
  earthquake ground motions using artificial intelligence}}.{\BBCQ}
\newblock
\APACjournalVolNumPages{Bulletin of the Seismological Society of
  America}{112}{4}{1979--1996}.
\newblock
\begin{APACrefDOI} \doi{10.1785/0120210264} \end{APACrefDOI}
\PrintBackRefs{\CurrentBib}

\bibitem [\protect \citeauthoryear {%
Gatti%
\ \BBA {} Clouteau%
}{%
Gatti%
\ \BBA {} Clouteau%
}{%
{\protect \APACyear {2020}}%
}]{%
gatti2020towards}
\APACinsertmetastar {%
gatti2020towards}%
\begin{APACrefauthors}%
Gatti, F.%
\BCBT {}\ \BBA {} Clouteau, D.%
\end{APACrefauthors}%
\unskip\
\newblock
\APACrefYearMonthDay{2020}{}{}.
\newblock
{\BBOQ}\APACrefatitle {{Towards blending physics-based numerical simulations
  and seismic databases using generative adversarial network}} {{Towards
  blending physics-based numerical simulations and seismic databases using
  generative adversarial network}}.{\BBCQ}
\newblock
\APACjournalVolNumPages{Computer Methods in Applied Mechanics and
  Engineering}{372}{}{113421}.
\newblock
\begin{APACrefDOI} \doi{10.1016/j.cma.2020.113421} \end{APACrefDOI}
\PrintBackRefs{\CurrentBib}

\bibitem [\protect \citeauthoryear {%
Goodfellow%
\ \protect \BOthers {.}}{%
Goodfellow%
\ \protect \BOthers {.}}{%
{\protect \APACyear {2014}}%
}]{%
goodfellow2014generative}
\APACinsertmetastar {%
goodfellow2014generative}%
\begin{APACrefauthors}%
Goodfellow, I.%
, Pouget-Abadie, J.%
, Mirza, M.%
, Xu, B.%
, Warde-Farley, D.%
, Ozair, S.%
, Courville, A.%
\BCBL {}\ \BBA {} Bengio, Y.%
\end{APACrefauthors}%
\unskip\
\newblock
\APACrefYearMonthDay{2014}{}{}.
\newblock
{\BBOQ}\APACrefatitle {{Generative Adversarial Nets}} {{Generative Adversarial
  Nets}}.{\BBCQ}
\newblock
\BIn{} \APACrefbtitle {Advances in Neural Information Processing Systems.}
  {Advances in neural information processing systems.}
\newblock
\begin{APACrefDOI} \doi{10.48550/arXiv.1406.2661} \end{APACrefDOI}
\PrintBackRefs{\CurrentBib}

\bibitem [\protect \citeauthoryear {%
Graves%
\ \BBA {} Pitarka%
}{%
Graves%
\ \BBA {} Pitarka%
}{%
{\protect \APACyear {2010}}%
}]{%
graves2010broadband}
\APACinsertmetastar {%
graves2010broadband}%
\begin{APACrefauthors}%
Graves, R.%
\BCBT {}\ \BBA {} Pitarka, A.%
\end{APACrefauthors}%
\unskip\
\newblock
\APACrefYearMonthDay{2010}{}{}.
\newblock
{\BBOQ}\APACrefatitle {{Broadband ground-motion simulation using a hybrid
  approach}} {{Broadband ground-motion simulation using a hybrid
  approach}}.{\BBCQ}
\newblock
\APACjournalVolNumPages{Bulletin of the Seismological Society of
  America}{100}{5A}{2095--2123}.
\newblock
\begin{APACrefDOI} \doi{10.1785/0120100057} \end{APACrefDOI}
\PrintBackRefs{\CurrentBib}

\bibitem [\protect \citeauthoryear {%
Griffin%
\ \BBA {} Lim%
}{%
Griffin%
\ \BBA {} Lim%
}{%
{\protect \APACyear {1984}}%
}]{%
griffin1984signal}
\APACinsertmetastar {%
griffin1984signal}%
\begin{APACrefauthors}%
Griffin, D.%
\BCBT {}\ \BBA {} Lim, J.%
\end{APACrefauthors}%
\unskip\
\newblock
\APACrefYearMonthDay{1984}{}{}.
\newblock
{\BBOQ}\APACrefatitle {{Signal estimation from modified short-time Fourier
  transform}} {{Signal estimation from modified short-time Fourier
  transform}}.{\BBCQ}
\newblock
\APACjournalVolNumPages{IEEE Transactions on Acoustics, Speech, and Signal
  Processing}{32}{2}{236--243}.
\newblock
\begin{APACrefDOI} \doi{10.1109/TASSP.1984.1164317} \end{APACrefDOI}
\PrintBackRefs{\CurrentBib}

\bibitem [\protect \citeauthoryear {%
Hartzell%
, Harmsen%
, Frankel%
\BCBL {}\ \BBA {} Larsen%
}{%
Hartzell%
\ \protect \BOthers {.}}{%
{\protect \APACyear {1999}}%
}]{%
hartzell1999calculation}
\APACinsertmetastar {%
hartzell1999calculation}%
\begin{APACrefauthors}%
Hartzell, S.%
, Harmsen, S.%
, Frankel, A.%
\BCBL {}\ \BBA {} Larsen, S.%
\end{APACrefauthors}%
\unskip\
\newblock
\APACrefYearMonthDay{1999}{}{}.
\newblock
{\BBOQ}\APACrefatitle {{Calculation of broadband time histories of ground
  motion: Comparison of methods and validation using strong-ground motion from
  the 1994 Northridge earthquake}} {{Calculation of broadband time histories of
  ground motion: Comparison of methods and validation using strong-ground
  motion from the 1994 Northridge earthquake}}.{\BBCQ}
\newblock
\APACjournalVolNumPages{Bulletin of the Seismological Society of
  America}{89}{6}{1484--1504}.
\newblock
\begin{APACrefDOI} \doi{10.1785/BSSA0890061484} \end{APACrefDOI}
\PrintBackRefs{\CurrentBib}

\bibitem [\protect \citeauthoryear {%
Heusel%
, Ramsauer%
, Unterthiner%
, Nessler%
\BCBL {}\ \BBA {} Hochreiter%
}{%
Heusel%
\ \protect \BOthers {.}}{%
{\protect \APACyear {2017}}%
}]{%
heusel2017gans}
\APACinsertmetastar {%
heusel2017gans}%
\begin{APACrefauthors}%
Heusel, M.%
, Ramsauer, H.%
, Unterthiner, T.%
, Nessler, B.%
\BCBL {}\ \BBA {} Hochreiter, S.%
\end{APACrefauthors}%
\unskip\
\newblock
\APACrefYearMonthDay{2017}{}{}.
\newblock
{\BBOQ}\APACrefatitle {{GANs Trained by a Two Time-Scale Update Rule Converge
  to a Local Nash Equilibrium}} {{GANs Trained by a Two Time-Scale Update Rule
  Converge to a Local Nash Equilibrium}}.{\BBCQ}
\newblock
\BIn{} \APACrefbtitle {Advances in Neural Information Processing Systems.}
  {Advances in neural information processing systems.}
\newblock
\begin{APACrefDOI} \doi{10.48550/arXiv.1706.08500} \end{APACrefDOI}
\PrintBackRefs{\CurrentBib}

\bibitem [\protect \citeauthoryear {%
Higgins%
\ \protect \BOthers {.}}{%
Higgins%
\ \protect \BOthers {.}}{%
{\protect \APACyear {2017}}%
}]{%
higgins2017beta}
\APACinsertmetastar {%
higgins2017beta}%
\begin{APACrefauthors}%
Higgins, I.%
, Matthey, L.%
, Pal, A.%
, Burgess, C.%
, Glorot, X.%
, Botvinick, M.%
, Mohamed, S.%
\BCBL {}\ \BBA {} Lerchner, A.%
\end{APACrefauthors}%
\unskip\
\newblock
\APACrefYearMonthDay{2017}{}{}.
\newblock
{\BBOQ}\APACrefatitle {beta-{VAE}: Learning Basic Visual Concepts with a
  Constrained Variational Framework} {beta-{VAE}: Learning basic visual
  concepts with a constrained variational framework}.{\BBCQ}
\newblock
\BIn{} \APACrefbtitle {International Conference on Learning Representations.}
  {International conference on learning representations.}
\newblock
\begin{APACrefURL} \url{https://openreview.net/forum?id=Sy2fzU9gl}
  \end{APACrefURL}
\PrintBackRefs{\CurrentBib}

\bibitem [\protect \citeauthoryear {%
Ho%
, Jain%
\BCBL {}\ \BBA {} Abbeel%
}{%
Ho%
\ \protect \BOthers {.}}{%
{\protect \APACyear {2020}}%
}]{%
ho2020denoising}
\APACinsertmetastar {%
ho2020denoising}%
\begin{APACrefauthors}%
Ho, J.%
, Jain, A.%
\BCBL {}\ \BBA {} Abbeel, P.%
\end{APACrefauthors}%
\unskip\
\newblock
\APACrefYearMonthDay{2020}{}{}.
\newblock
{\BBOQ}\APACrefatitle {Denoising Diffusion Probabilistic Models} {Denoising
  diffusion probabilistic models}.{\BBCQ}
\newblock
\BIn{} \APACrefbtitle {{Advances in Neural Information Processing Systems}.}
  {{Advances in Neural Information Processing Systems}.}
\newblock
\begin{APACrefDOI} \doi{10.48550/arXiv.2006.11239} \end{APACrefDOI}
\PrintBackRefs{\CurrentBib}

\bibitem [\protect \citeauthoryear {%
Ho%
\ \protect \BOthers {.}}{%
Ho%
\ \protect \BOthers {.}}{%
{\protect \APACyear {2022}}%
}]{%
ho2022video}
\APACinsertmetastar {%
ho2022video}%
\begin{APACrefauthors}%
Ho, J.%
, Salimans, T.%
, Gritsenko, A.%
, Chan, W.%
, Norouzi, M.%
\BCBL {}\ \BBA {} Fleet, D\BPBI J.%
\end{APACrefauthors}%
\unskip\
\newblock
\APACrefYearMonthDay{2022}{}{}.
\newblock
{\BBOQ}\APACrefatitle {Video Diffusion Models} {Video diffusion models}.{\BBCQ}
\newblock
\BIn{} \APACrefbtitle {{Advances in Neural Information Processing Systems}.}
  {{Advances in Neural Information Processing Systems}.}
\PrintBackRefs{\CurrentBib}

\bibitem [\protect \citeauthoryear {%
Jayalakshmi%
, Dhanya%
, Raghukanth%
\BCBL {}\ \BBA {} Mai%
}{%
Jayalakshmi%
\ \protect \BOthers {.}}{%
{\protect \APACyear {2021}}%
}]{%
jayalakshmi2021hybrid}
\APACinsertmetastar {%
jayalakshmi2021hybrid}%
\begin{APACrefauthors}%
Jayalakshmi, S.%
, Dhanya, J.%
, Raghukanth, S.%
\BCBL {}\ \BBA {} Mai, P\BPBI M.%
\end{APACrefauthors}%
\unskip\
\newblock
\APACrefYearMonthDay{2021}{}{}.
\newblock
{\BBOQ}\APACrefatitle {{Hybrid broadband ground motion simulations in the
  Indo-Gangetic basin for great Himalayan earthquake scenarios}} {{Hybrid
  broadband ground motion simulations in the Indo-Gangetic basin for great
  Himalayan earthquake scenarios}}.{\BBCQ}
\newblock
\APACjournalVolNumPages{Bulletin of Earthquake Engineering}{19}{}{3319--3348}.
\newblock
\begin{APACrefDOI} \doi{10.1007/s10518-021-01094-0} \end{APACrefDOI}
\PrintBackRefs{\CurrentBib}

\bibitem [\protect \citeauthoryear {%
Jozinovi{\'c}%
, Lomax%
, {\v{S}}tajduhar%
\BCBL {}\ \BBA {} Michelini%
}{%
Jozinovi{\'c}%
\ \protect \BOthers {.}}{%
{\protect \APACyear {2022}}%
}]{%
jozinovic2022transfer}
\APACinsertmetastar {%
jozinovic2022transfer}%
\begin{APACrefauthors}%
Jozinovi{\'c}, D.%
, Lomax, A.%
, {\v{S}}tajduhar, I.%
\BCBL {}\ \BBA {} Michelini, A.%
\end{APACrefauthors}%
\unskip\
\newblock
\APACrefYearMonthDay{2022}{}{}.
\newblock
{\BBOQ}\APACrefatitle {{Transfer learning: Improving neural network based
  prediction of earthquake ground shaking for an area with insufficient
  training data}} {{Transfer learning: Improving neural network based
  prediction of earthquake ground shaking for an area with insufficient
  training data}}.{\BBCQ}
\newblock
\APACjournalVolNumPages{Geophysical Journal International}{229}{1}{704--718}.
\newblock
\begin{APACrefDOI} \doi{10.1093/gji/ggab488} \end{APACrefDOI}
\PrintBackRefs{\CurrentBib}

\bibitem [\protect \citeauthoryear {%
Kanno%
, Narita%
, Morikawa%
, Fujiwara%
\BCBL {}\ \BBA {} Fukushima%
}{%
Kanno%
\ \protect \BOthers {.}}{%
{\protect \APACyear {2006}}%
}]{%
kanno2006new}
\APACinsertmetastar {%
kanno2006new}%
\begin{APACrefauthors}%
Kanno, T.%
, Narita, A.%
, Morikawa, N.%
, Fujiwara, H.%
\BCBL {}\ \BBA {} Fukushima, Y.%
\end{APACrefauthors}%
\unskip\
\newblock
\APACrefYearMonthDay{2006}{}{}.
\newblock
{\BBOQ}\APACrefatitle {{A new attenuation relation for strong ground motion in
  Japan based on recorded data}} {{A new attenuation relation for strong ground
  motion in Japan based on recorded data}}.{\BBCQ}
\newblock
\APACjournalVolNumPages{Bulletin of the Seismological Society of
  America}{96}{3}{879--897}.
\newblock
\begin{APACrefDOI} \doi{10.1785/0120050138} \end{APACrefDOI}
\PrintBackRefs{\CurrentBib}

\bibitem [\protect \citeauthoryear {%
Karras%
, Aittala%
, Aila%
\BCBL {}\ \BBA {} Laine%
}{%
Karras%
\ \protect \BOthers {.}}{%
{\protect \APACyear {2022}}%
}]{%
karras2022elucidating}
\APACinsertmetastar {%
karras2022elucidating}%
\begin{APACrefauthors}%
Karras, T.%
, Aittala, M.%
, Aila, T.%
\BCBL {}\ \BBA {} Laine, S.%
\end{APACrefauthors}%
\unskip\
\newblock
\APACrefYearMonthDay{2022}{}{}.
\newblock
{\BBOQ}\APACrefatitle {{Elucidating the Design Space of Diffusion-Based
  Generative Models}} {{Elucidating the Design Space of Diffusion-Based
  Generative Models}}.{\BBCQ}
\newblock
\BIn{} \APACrefbtitle {{Advances in Neural Information Processing Systems}.}
  {{Advances in Neural Information Processing Systems}.}
\newblock
\begin{APACrefDOI} \doi{10.48550/arXiv.2206.00364} \end{APACrefDOI}
\PrintBackRefs{\CurrentBib}

\bibitem [\protect \citeauthoryear {%
Karras%
, Laine%
\BCBL {}\ \BBA {} Aila%
}{%
Karras%
\ \protect \BOthers {.}}{%
{\protect \APACyear {2019}}%
}]{%
karras2019style}
\APACinsertmetastar {%
karras2019style}%
\begin{APACrefauthors}%
Karras, T.%
, Laine, S.%
\BCBL {}\ \BBA {} Aila, T.%
\end{APACrefauthors}%
\unskip\
\newblock
\APACrefYearMonthDay{2019}{}{}.
\newblock
{\BBOQ}\APACrefatitle {{A style-based generator architecture for generative
  adversarial networks}} {{A style-based generator architecture for generative
  adversarial networks}}.{\BBCQ}
\newblock
\BIn{} \APACrefbtitle {Proceedings of the IEEE/CVF conference on computer
  vision and pattern recognition} {Proceedings of the ieee/cvf conference on
  computer vision and pattern recognition}\ (\BPGS\ 4401--4410).
\newblock
\begin{APACrefDOI} \doi{10.48550/arXiv.1812.04948} \end{APACrefDOI}
\PrintBackRefs{\CurrentBib}

\bibitem [\protect \citeauthoryear {%
Karras%
\ \protect \BOthers {.}}{%
Karras%
\ \protect \BOthers {.}}{%
{\protect \APACyear {2020}}%
}]{%
karras2020analyzing}
\APACinsertmetastar {%
karras2020analyzing}%
\begin{APACrefauthors}%
Karras, T.%
, Laine, S.%
, Aittala, M.%
, Hellsten, J.%
, Lehtinen, J.%
\BCBL {}\ \BBA {} Aila, T.%
\end{APACrefauthors}%
\unskip\
\newblock
\APACrefYearMonthDay{2020}{}{}.
\newblock
{\BBOQ}\APACrefatitle {{Analyzing and improving the image quality of StyleGAN}}
  {{Analyzing and improving the image quality of StyleGAN}}.{\BBCQ}
\newblock
\BIn{} \APACrefbtitle {Proceedings of the IEEE/CVF conference on computer
  vision and pattern recognition} {Proceedings of the ieee/cvf conference on
  computer vision and pattern recognition}\ (\BPGS\ 8110--8119).
\newblock
\begin{APACrefDOI} \doi{10.48550/arXiv.1912.04958} \end{APACrefDOI}
\PrintBackRefs{\CurrentBib}

\bibitem [\protect \citeauthoryear {%
Katsanos%
, Sextos%
\BCBL {}\ \BBA {} Manolis%
}{%
Katsanos%
\ \protect \BOthers {.}}{%
{\protect \APACyear {2010}}%
}]{%
katsanos2010selection}
\APACinsertmetastar {%
katsanos2010selection}%
\begin{APACrefauthors}%
Katsanos, E.%
, Sextos, A.%
\BCBL {}\ \BBA {} Manolis, G.%
\end{APACrefauthors}%
\unskip\
\newblock
\APACrefYearMonthDay{2010}{}{}.
\newblock
{\BBOQ}\APACrefatitle {{Selection of earthquake ground motion records: A
  state-of-the-art review from a structural engineering perspective}}
  {{Selection of earthquake ground motion records: A state-of-the-art review
  from a structural engineering perspective}}.{\BBCQ}
\newblock
\APACjournalVolNumPages{Soil dynamics and earthquake
  engineering}{30}{4}{157--169}.
\newblock
\begin{APACrefDOI} \doi{10.1016/j.soildyn.2009.10.005} \end{APACrefDOI}
\PrintBackRefs{\CurrentBib}

\bibitem [\protect \citeauthoryear {%
Kingma%
\ \BBA {} Ba%
}{%
Kingma%
\ \BBA {} Ba%
}{%
{\protect \APACyear {2015}}%
}]{%
kingma2015adam}
\APACinsertmetastar {%
kingma2015adam}%
\begin{APACrefauthors}%
Kingma, D\BPBI P.%
\BCBT {}\ \BBA {} Ba, J.%
\end{APACrefauthors}%
\unskip\
\newblock
\APACrefYearMonthDay{2015}{}{}.
\newblock
{\BBOQ}\APACrefatitle {{Adam: A Method for Stochastic Optimization}} {{Adam: A
  Method for Stochastic Optimization}}.{\BBCQ}
\newblock
\BIn{} \APACrefbtitle {International Conference on Learning Representations.}
  {International conference on learning representations.}
\newblock
\begin{APACrefDOI} \doi{10.48550/arXiv.1412.6980} \end{APACrefDOI}
\PrintBackRefs{\CurrentBib}

\bibitem [\protect \citeauthoryear {%
Kingma%
\ \BBA {} Welling%
}{%
Kingma%
\ \BBA {} Welling%
}{%
{\protect \APACyear {2014}}%
}]{%
kingma2013auto}
\APACinsertmetastar {%
kingma2013auto}%
\begin{APACrefauthors}%
Kingma, D\BPBI P.%
\BCBT {}\ \BBA {} Welling, M.%
\end{APACrefauthors}%
\unskip\
\newblock
\APACrefYearMonthDay{2014}{}{}.
\newblock
{\BBOQ}\APACrefatitle {{Auto-Encoding Variational Bayes}} {{Auto-Encoding
  Variational Bayes}}.{\BBCQ}
\newblock
\BIn{} \APACrefbtitle {{International Conference on Learning Representations}.}
  {{International Conference on Learning Representations}.}
\newblock
\begin{APACrefDOI} \doi{10.48550/arXiv.1312.6114} \end{APACrefDOI}
\PrintBackRefs{\CurrentBib}

\bibitem [\protect \citeauthoryear {%
Kong%
, Ping%
, Huang%
, Zhao%
\BCBL {}\ \BBA {} Catanzaro%
}{%
Kong%
\ \protect \BOthers {.}}{%
{\protect \APACyear {2021}}%
}]{%
kong2021diffwave}
\APACinsertmetastar {%
kong2021diffwave}%
\begin{APACrefauthors}%
Kong, Z.%
, Ping, W.%
, Huang, J.%
, Zhao, K.%
\BCBL {}\ \BBA {} Catanzaro, B.%
\end{APACrefauthors}%
\unskip\
\newblock
\APACrefYearMonthDay{2021}{}{}.
\newblock
{\BBOQ}\APACrefatitle {{DiffWave: A Versatile Diffusion Model for Audio
  Synthesis}} {{DiffWave: A Versatile Diffusion Model for Audio
  Synthesis}}.{\BBCQ}
\newblock
\BIn{} \APACrefbtitle {International Conference on Learning Representations.}
  {International conference on learning representations.}
\newblock
\begin{APACrefDOI} \doi{10.48550/arXiv.2009.09761} \end{APACrefDOI}
\PrintBackRefs{\CurrentBib}

\bibitem [\protect \citeauthoryear {%
Y.~Li%
, Ku%
, Zhang%
, Ahn%
\BCBL {}\ \BBA {} Ko%
}{%
Y.~Li%
\ \protect \BOthers {.}}{%
{\protect \APACyear {2020}}%
}]{%
li2020seismic}
\APACinsertmetastar {%
li2020seismic}%
\begin{APACrefauthors}%
Li, Y.%
, Ku, B.%
, Zhang, S.%
, Ahn, J\BHBI K.%
\BCBL {}\ \BBA {} Ko, H.%
\end{APACrefauthors}%
\unskip\
\newblock
\APACrefYearMonthDay{2020}{}{}.
\newblock
{\BBOQ}\APACrefatitle {{Seismic data augmentation based on conditional
  generative adversarial networks}} {{Seismic data augmentation based on
  conditional generative adversarial networks}}.{\BBCQ}
\newblock
\APACjournalVolNumPages{Sensors}{20}{23}{6850}.
\newblock
\begin{APACrefDOI} \doi{10.3390/s20236850} \end{APACrefDOI}
\PrintBackRefs{\CurrentBib}

\bibitem [\protect \citeauthoryear {%
Z.~Li%
\ \protect \BOthers {.}}{%
Z.~Li%
\ \protect \BOthers {.}}{%
{\protect \APACyear {2020}}%
}]{%
li2020neural}
\APACinsertmetastar {%
li2020neural}%
\begin{APACrefauthors}%
Li, Z.%
, Kovachki, N.%
, Azizzadenesheli, K.%
, Liu, B.%
, Bhattacharya, K.%
, Stuart, A.%
\BCBL {}\ \BBA {} Anandkumar, A.%
\end{APACrefauthors}%
\unskip\
\newblock
\APACrefYearMonthDay{2020}{}{}.
\newblock
{\BBOQ}\APACrefatitle {{Neural operator: Graph kernel network for partial
  differential equations}} {{Neural operator: Graph kernel network for partial
  differential equations}}.{\BBCQ}
\newblock
\APACjournalVolNumPages{arXiv}{}{}{}.
\newblock
\begin{APACrefDOI} \doi{10.48550/arXiv.2003.03485} \end{APACrefDOI}
\PrintBackRefs{\CurrentBib}

\bibitem [\protect \citeauthoryear {%
Z.~Li%
, Meier%
, Hauksson%
, Zhan%
\BCBL {}\ \BBA {} Andrews%
}{%
Z.~Li%
\ \protect \BOthers {.}}{%
{\protect \APACyear {2018}}%
}]{%
li2018machine}
\APACinsertmetastar {%
li2018machine}%
\begin{APACrefauthors}%
Li, Z.%
, Meier, M\BHBI A.%
, Hauksson, E.%
, Zhan, Z.%
\BCBL {}\ \BBA {} Andrews, J.%
\end{APACrefauthors}%
\unskip\
\newblock
\APACrefYearMonthDay{2018}{}{}.
\newblock
{\BBOQ}\APACrefatitle {{Machine learning seismic wave discrimination:
  Application to earthquake early warning}} {{Machine learning seismic wave
  discrimination: Application to earthquake early warning}}.{\BBCQ}
\newblock
\APACjournalVolNumPages{Geophysical Research Letters}{45}{10}{4773--4779}.
\newblock
\begin{APACrefDOI} \doi{10.1029/2018GL077870} \end{APACrefDOI}
\PrintBackRefs{\CurrentBib}

\bibitem [\protect \citeauthoryear {%
Lilienkamp%
, von Specht%
, Weatherill%
, Caire%
\BCBL {}\ \BBA {} Cotton%
}{%
Lilienkamp%
\ \protect \BOthers {.}}{%
{\protect \APACyear {2022}}%
}]{%
lilienkamp2022ground}
\APACinsertmetastar {%
lilienkamp2022ground}%
\begin{APACrefauthors}%
Lilienkamp, H.%
, von Specht, S.%
, Weatherill, G.%
, Caire, G.%
\BCBL {}\ \BBA {} Cotton, F.%
\end{APACrefauthors}%
\unskip\
\newblock
\APACrefYearMonthDay{2022}{}{}.
\newblock
{\BBOQ}\APACrefatitle {{Ground-motion modeling as an image processing task:
  Introducing a neural network based, fully data-driven, and nonergodic
  approach}} {{Ground-motion modeling as an image processing task: Introducing
  a neural network based, fully data-driven, and nonergodic approach}}.{\BBCQ}
\newblock
\APACjournalVolNumPages{Bulletin of the Seismological Society of
  America}{112}{3}{1565--1582}.
\newblock
\begin{APACrefDOI} \doi{10.1785/0120220008} \end{APACrefDOI}
\PrintBackRefs{\CurrentBib}

\bibitem [\protect \citeauthoryear {%
Loshchilov%
\ \BBA {} Hutter%
}{%
Loshchilov%
\ \BBA {} Hutter%
}{%
{\protect \APACyear {2017}}%
}]{%
loshchilov2017sgdr}
\APACinsertmetastar {%
loshchilov2017sgdr}%
\begin{APACrefauthors}%
Loshchilov, I.%
\BCBT {}\ \BBA {} Hutter, F.%
\end{APACrefauthors}%
\unskip\
\newblock
\APACrefYearMonthDay{2017}{}{}.
\newblock
{\BBOQ}\APACrefatitle {{SGDR}: Stochastic Gradient Descent with Warm Restarts}
  {{SGDR}: Stochastic gradient descent with warm restarts}.{\BBCQ}
\newblock
\BIn{} \APACrefbtitle {International Conference on Learning Representations.}
  {International conference on learning representations.}
\newblock
\begin{APACrefDOI} \doi{10.48550/arXiv.1608.03983} \end{APACrefDOI}
\PrintBackRefs{\CurrentBib}

\bibitem [\protect \citeauthoryear {%
Luco%
\ \BBA {} Bazzurro%
}{%
Luco%
\ \BBA {} Bazzurro%
}{%
{\protect \APACyear {2007}}%
}]{%
luco2007does}
\APACinsertmetastar {%
luco2007does}%
\begin{APACrefauthors}%
Luco, N.%
\BCBT {}\ \BBA {} Bazzurro, P.%
\end{APACrefauthors}%
\unskip\
\newblock
\APACrefYearMonthDay{2007}{}{}.
\newblock
{\BBOQ}\APACrefatitle {{Does amplitude scaling of ground motion records result
  in biased nonlinear structural drift responses?}} {{Does amplitude scaling of
  ground motion records result in biased nonlinear structural drift
  responses?}}{\BBCQ}
\newblock
\APACjournalVolNumPages{Earthquake Engineering \& Structural
  Dynamics}{36}{13}{1813--1835}.
\newblock
\begin{APACrefDOI} \doi{10.1002/eqe.695} \end{APACrefDOI}
\PrintBackRefs{\CurrentBib}

\bibitem [\protect \citeauthoryear {%
Maechling%
, Silva%
, Callaghan%
\BCBL {}\ \BBA {} Jordan%
}{%
Maechling%
\ \protect \BOthers {.}}{%
{\protect \APACyear {2015}}%
}]{%
maechling2015scec}
\APACinsertmetastar {%
maechling2015scec}%
\begin{APACrefauthors}%
Maechling, P\BPBI J.%
, Silva, F.%
, Callaghan, S.%
\BCBL {}\ \BBA {} Jordan, T\BPBI H.%
\end{APACrefauthors}%
\unskip\
\newblock
\APACrefYearMonthDay{2015}{}{}.
\newblock
{\BBOQ}\APACrefatitle {{SCEC broadband platform: System architecture and
  software implementation}} {{SCEC broadband platform: System architecture and
  software implementation}}.{\BBCQ}
\newblock
\APACjournalVolNumPages{Seismological Research Letters}{86}{1}{27--38}.
\newblock
\begin{APACrefDOI} \doi{10.1785/0220140125} \end{APACrefDOI}
\PrintBackRefs{\CurrentBib}

\bibitem [\protect \citeauthoryear {%
Mai%
\ \BBA {} Beroza%
}{%
Mai%
\ \BBA {} Beroza%
}{%
{\protect \APACyear {2002}}%
}]{%
mai2002spatial}
\APACinsertmetastar {%
mai2002spatial}%
\begin{APACrefauthors}%
Mai, P\BPBI M.%
\BCBT {}\ \BBA {} Beroza, G.%
\end{APACrefauthors}%
\unskip\
\newblock
\APACrefYearMonthDay{2002}{}{}.
\newblock
{\BBOQ}\APACrefatitle {{A spatial random field model to characterize complexity
  in earthquake slip}} {{A spatial random field model to characterize
  complexity in earthquake slip}}.{\BBCQ}
\newblock
\APACjournalVolNumPages{Journal of Geophysical Research: Solid
  Earth}{107}{B11}{ESE--10}.
\newblock
\begin{APACrefDOI} \doi{10.1029/2001JB000588} \end{APACrefDOI}
\PrintBackRefs{\CurrentBib}

\bibitem [\protect \citeauthoryear {%
Mai%
\ \BBA {} Beroza%
}{%
Mai%
\ \BBA {} Beroza%
}{%
{\protect \APACyear {2003}}%
}]{%
mai2003hybrid}
\APACinsertmetastar {%
mai2003hybrid}%
\begin{APACrefauthors}%
Mai, P\BPBI M.%
\BCBT {}\ \BBA {} Beroza, G.%
\end{APACrefauthors}%
\unskip\
\newblock
\APACrefYearMonthDay{2003}{}{}.
\newblock
{\BBOQ}\APACrefatitle {{A hybrid method for calculating near-source, broadband
  seismograms: Application to strong motion prediction}} {{A hybrid method for
  calculating near-source, broadband seismograms: Application to strong motion
  prediction}}.{\BBCQ}
\newblock
\APACjournalVolNumPages{Physics of the Earth and Planetary
  Interiors}{137}{1-4}{183--199}.
\newblock
\begin{APACrefDOI} \doi{10.1016/S0031-9201(03)00014-1} \end{APACrefDOI}
\PrintBackRefs{\CurrentBib}

\bibitem [\protect \citeauthoryear {%
Mai%
, Imperatori%
\BCBL {}\ \BBA {} Olsen%
}{%
Mai%
\ \protect \BOthers {.}}{%
{\protect \APACyear {2010}}%
}]{%
mai2010hybrid}
\APACinsertmetastar {%
mai2010hybrid}%
\begin{APACrefauthors}%
Mai, P\BPBI M.%
, Imperatori, W.%
\BCBL {}\ \BBA {} Olsen, K\BPBI B.%
\end{APACrefauthors}%
\unskip\
\newblock
\APACrefYearMonthDay{2010}{}{}.
\newblock
{\BBOQ}\APACrefatitle {{Hybrid broadband ground-motion simulations: Combining
  long-period deterministic synthetics with high-frequency multiple S-to-S
  backscattering}} {{Hybrid broadband ground-motion simulations: Combining
  long-period deterministic synthetics with high-frequency multiple S-to-S
  backscattering}}.{\BBCQ}
\newblock
\APACjournalVolNumPages{Bulletin of the Seismological Society of
  America}{100}{5A}{2124--2142}.
\newblock
\begin{APACrefDOI} \doi{10.1785/0120080194} \end{APACrefDOI}
\PrintBackRefs{\CurrentBib}

\bibitem [\protect \citeauthoryear {%
Marano%
, Rosso%
, Aloisio%
\BCBL {}\ \BBA {} Cirrincione%
}{%
Marano%
\ \protect \BOthers {.}}{%
{\protect \APACyear {2024}}%
}]{%
marano2024generative}
\APACinsertmetastar {%
marano2024generative}%
\begin{APACrefauthors}%
Marano, G\BPBI C.%
, Rosso, M\BPBI M.%
, Aloisio, A.%
\BCBL {}\ \BBA {} Cirrincione, G.%
\end{APACrefauthors}%
\unskip\
\newblock
\APACrefYearMonthDay{2024}{}{}.
\newblock
{\BBOQ}\APACrefatitle {{Generative adversarial networks review in
  earthquake-related engineering fields}} {{Generative adversarial networks
  review in earthquake-related engineering fields}}.{\BBCQ}
\newblock
\APACjournalVolNumPages{Bulletin of Earthquake Engineering}{22}{7}{3511--3562}.
\newblock
\begin{APACrefDOI} \doi{10.1007/s10518-023-01645-7} \end{APACrefDOI}
\PrintBackRefs{\CurrentBib}

\bibitem [\protect \citeauthoryear {%
Matsumoto%
, Yaoyama%
, Lee%
, Hida%
\BCBL {}\ \BBA {} Itoi%
}{%
Matsumoto%
\ \protect \BOthers {.}}{%
{\protect \APACyear {2024}}%
}]{%
matsumoto2024GAN}
\APACinsertmetastar {%
matsumoto2024GAN}%
\begin{APACrefauthors}%
Matsumoto, Y.%
, Yaoyama, T.%
, Lee, S.%
, Hida, T.%
\BCBL {}\ \BBA {} Itoi, T.%
\end{APACrefauthors}%
\unskip\
\newblock
\APACrefYearMonthDay{2024}{}{}.
\newblock
{\BBOQ}\APACrefatitle {{{Generative Adversarial Networks‐Based
  Ground‐Motion Model for Crustal Earthquakes in Japan Considering Detailed
  Site Conditions}}} {{{Generative Adversarial Networks‐Based Ground‐Motion
  Model for Crustal Earthquakes in Japan Considering Detailed Site
  Conditions}}}.{\BBCQ}
\newblock
\APACjournalVolNumPages{Bulletin of the Seismological Society of
  America}{}{}{}.
\newblock
\begin{APACrefDOI} \doi{10.1785/0120240070} \end{APACrefDOI}
\PrintBackRefs{\CurrentBib}

\bibitem [\protect \citeauthoryear {%
Mousavi%
, Sheng%
, Zhu%
\BCBL {}\ \BBA {} Beroza%
}{%
Mousavi%
\ \protect \BOthers {.}}{%
{\protect \APACyear {2019}}%
}]{%
mousavi2019stanford}
\APACinsertmetastar {%
mousavi2019stanford}%
\begin{APACrefauthors}%
Mousavi, S\BPBI M.%
, Sheng, Y.%
, Zhu, W.%
\BCBL {}\ \BBA {} Beroza, G\BPBI C.%
\end{APACrefauthors}%
\unskip\
\newblock
\APACrefYearMonthDay{2019}{}{}.
\newblock
{\BBOQ}\APACrefatitle {{STanford EArthquake Dataset (STEAD): A Global Data Set
  of Seismic Signals for AI}} {{STanford EArthquake Dataset (STEAD): A Global
  Data Set of Seismic Signals for AI}}.{\BBCQ}
\newblock
\APACjournalVolNumPages{IEEE Access}{}{}{}.
\newblock
\begin{APACrefDOI} \doi{10.1109/ACCESS.2019.2947848} \end{APACrefDOI}
\PrintBackRefs{\CurrentBib}

\bibitem [\protect \citeauthoryear {%
Naoi%
, Tamaribuchi%
, Shimojo%
, Katoh%
\BCBL {}\ \BBA {} Ohyanagi%
}{%
Naoi%
\ \protect \BOthers {.}}{%
{\protect \APACyear {2024}}%
}]{%
naoi2024neural}
\APACinsertmetastar {%
naoi2024neural}%
\begin{APACrefauthors}%
Naoi, M.%
, Tamaribuchi, K.%
, Shimojo, K.%
, Katoh, S.%
\BCBL {}\ \BBA {} Ohyanagi, S.%
\end{APACrefauthors}%
\unskip\
\newblock
\APACrefYearMonthDay{2024}{}{}.
\newblock
{\BBOQ}\APACrefatitle {{Neural phase picker trained on the Japan meteorological
  agency unified earthquake catalog}} {{Neural phase picker trained on the
  Japan meteorological agency unified earthquake catalog}}.{\BBCQ}
\newblock
\APACjournalVolNumPages{Earth, Planets and Space}{76}{1}{150}.
\newblock
\begin{APACrefDOI} \doi{10.1186/s40623-024-02091-8} \end{APACrefDOI}
\PrintBackRefs{\CurrentBib}

\bibitem [\protect \citeauthoryear {%
Nichol%
\ \BBA {} Dhariwal%
}{%
Nichol%
\ \BBA {} Dhariwal%
}{%
{\protect \APACyear {2021}}%
}]{%
nichol2021improved}
\APACinsertmetastar {%
nichol2021improved}%
\begin{APACrefauthors}%
Nichol, A\BPBI Q.%
\BCBT {}\ \BBA {} Dhariwal, P.%
\end{APACrefauthors}%
\unskip\
\newblock
\APACrefYearMonthDay{2021}{}{}.
\newblock
{\BBOQ}\APACrefatitle {{Improved denoising diffusion probabilistic models}}
  {{Improved denoising diffusion probabilistic models}}.{\BBCQ}
\newblock
\BIn{} \APACrefbtitle {International conference on machine learning}
  {International conference on machine learning}\ (\BPGS\ 8162--8171).
\newblock
\begin{APACrefDOI} \doi{10.48550/arXiv.2102.09672} \end{APACrefDOI}
\PrintBackRefs{\CurrentBib}

\bibitem [\protect \citeauthoryear {%
NIED%
}{%
NIED%
}{%
{\protect \APACyear {2019}}%
}]{%
knet2019}
\APACinsertmetastar {%
knet2019}%
\begin{APACrefauthors}%
NIED.%
\end{APACrefauthors}%
\unskip\
\newblock
\APACrefYearMonthDay{2019}{}{}.
\newblock
\APACrefbtitle {{K-NET, KiK-net, National Research Institute for Earth Science
  and Disaster Resilience}.} {{K-NET, KiK-net, National Research Institute for
  Earth Science and Disaster Resilience}.}
\newblock
\begin{APACrefDOI} \doi{10.17598/NIED.0004} \end{APACrefDOI}
\PrintBackRefs{\CurrentBib}

\bibitem [\protect \citeauthoryear {%
Okazaki%
\ \protect \BOthers {.}}{%
Okazaki%
\ \protect \BOthers {.}}{%
{\protect \APACyear {2021}}%
}]{%
okazaki2021simulation}
\APACinsertmetastar {%
okazaki2021simulation}%
\begin{APACrefauthors}%
Okazaki, T.%
, Hachiya, H.%
, Iwaki, A.%
, Maeda, T.%
, Fujiwara, H.%
\BCBL {}\ \BBA {} Ueda, N.%
\end{APACrefauthors}%
\unskip\
\newblock
\APACrefYearMonthDay{2021}{}{}.
\newblock
{\BBOQ}\APACrefatitle {{Simulation of broad-band ground motions with consistent
  long-period and short-period components using the Wasserstein interpolation
  of acceleration envelopes}} {{Simulation of broad-band ground motions with
  consistent long-period and short-period components using the Wasserstein
  interpolation of acceleration envelopes}}.{\BBCQ}
\newblock
\APACjournalVolNumPages{Geophysical Journal International}{227}{1}{333--349}.
\newblock
\begin{APACrefDOI} \doi{10.1093/gji/ggab225} \end{APACrefDOI}
\PrintBackRefs{\CurrentBib}

\bibitem [\protect \citeauthoryear {%
Olsen%
\ \BBA {} Takedatsu%
}{%
Olsen%
\ \BBA {} Takedatsu%
}{%
{\protect \APACyear {2015}}%
}]{%
olsen2015sdsu}
\APACinsertmetastar {%
olsen2015sdsu}%
\begin{APACrefauthors}%
Olsen, K.%
\BCBT {}\ \BBA {} Takedatsu, R.%
\end{APACrefauthors}%
\unskip\
\newblock
\APACrefYearMonthDay{2015}{}{}.
\newblock
{\BBOQ}\APACrefatitle {{The SDSU broadband ground-motion generation module
  BBtoolbox version 1.5}} {{The SDSU broadband ground-motion generation module
  BBtoolbox version 1.5}}.{\BBCQ}
\newblock
\APACjournalVolNumPages{Seismological Research Letters}{86}{1}{81--88}.
\newblock
\begin{APACrefDOI} \doi{10.1785/0220140102} \end{APACrefDOI}
\PrintBackRefs{\CurrentBib}

\bibitem [\protect \citeauthoryear {%
Palgunadi%
, Gabriel%
, Garagash%
, Ulrich%
\BCBL {}\ \BBA {} Mai%
}{%
Palgunadi%
\ \protect \BOthers {.}}{%
{\protect \APACyear {2024}}%
}]{%
palgunadi2024rupture}
\APACinsertmetastar {%
palgunadi2024rupture}%
\begin{APACrefauthors}%
Palgunadi, K\BPBI H.%
, Gabriel, A\BHBI A.%
, Garagash, D\BPBI I.%
, Ulrich, T.%
\BCBL {}\ \BBA {} Mai, P\BPBI M.%
\end{APACrefauthors}%
\unskip\
\newblock
\APACrefYearMonthDay{2024}{}{}.
\newblock
{\BBOQ}\APACrefatitle {{Rupture dynamics of cascading earthquakes in a
  multiscale fracture network}} {{Rupture dynamics of cascading earthquakes in
  a multiscale fracture network}}.{\BBCQ}
\newblock
\APACjournalVolNumPages{Journal of Geophysical Research: Solid
  Earth}{129}{3}{e2023JB027578}.
\newblock
\begin{APACrefDOI} \doi{10.1029/2023JB027578} \end{APACrefDOI}
\PrintBackRefs{\CurrentBib}

\bibitem [\protect \citeauthoryear {%
Paolucci%
\ \protect \BOthers {.}}{%
Paolucci%
\ \protect \BOthers {.}}{%
{\protect \APACyear {2018}}%
}]{%
paolucci2018broadband}
\APACinsertmetastar {%
paolucci2018broadband}%
\begin{APACrefauthors}%
Paolucci, R.%
, Gatti, F.%
, Infantino, M.%
, Smerzini, C.%
, {\"O}zcebe, A\BPBI G.%
\BCBL {}\ \BBA {} Stupazzini, M.%
\end{APACrefauthors}%
\unskip\
\newblock
\APACrefYearMonthDay{2018}{}{}.
\newblock
{\BBOQ}\APACrefatitle {{Broadband ground motions from 3D physics-based
  numerical simulations using artificial neural networks}} {{Broadband ground
  motions from 3D physics-based numerical simulations using artificial neural
  networks}}.{\BBCQ}
\newblock
\APACjournalVolNumPages{Bulletin of the Seismological Society of
  America}{108}{3A}{1272--1286}.
\newblock
\begin{APACrefDOI} \doi{10.1785/0120170293} \end{APACrefDOI}
\PrintBackRefs{\CurrentBib}

\bibitem [\protect \citeauthoryear {%
Paolucci%
, Mazzieri%
\BCBL {}\ \protect \BOthers {.}}{%
Paolucci%
, Mazzieri%
\BCBL {}\ \protect \BOthers {.}}{%
{\protect \APACyear {2021}}%
}]{%
paolucci_earthquake_2021}
\APACinsertmetastar {%
paolucci_earthquake_2021}%
\begin{APACrefauthors}%
Paolucci, R.%
, Mazzieri, I.%
, Piunno, G.%
, Smerzini, C.%
, Vanini, M.%
\BCBL {}\ \BBA {} Özcebe, A.%
\end{APACrefauthors}%
\unskip\
\newblock
\APACrefYearMonthDay{2021}{}{}.
\newblock
{\BBOQ}\APACrefatitle {Earthquake ground motion modeling of induced seismicity
  in the {Groningen} gas field} {Earthquake ground motion modeling of induced
  seismicity in the {Groningen} gas field}.{\BBCQ}
\newblock
\APACjournalVolNumPages{Earthquake Engineering \& Structural
  Dynamics}{50}{1}{135--154}.
\newblock
\begin{APACrefDOI} \doi{10.1002/eqe.3367} \end{APACrefDOI}
\PrintBackRefs{\CurrentBib}

\bibitem [\protect \citeauthoryear {%
Paolucci%
, Smerzini%
\BCBL {}\ \BBA {} Vanini%
}{%
Paolucci%
, Smerzini%
\BCBL {}\ \BBA {} Vanini%
}{%
{\protect \APACyear {2021}}%
}]{%
paolucci_bbspeedset_2021}
\APACinsertmetastar {%
paolucci_bbspeedset_2021}%
\begin{APACrefauthors}%
Paolucci, R.%
, Smerzini, C.%
\BCBL {}\ \BBA {} Vanini, M.%
\end{APACrefauthors}%
\unskip\
\newblock
\APACrefYearMonthDay{2021}{}{}.
\newblock
{\BBOQ}\APACrefatitle {{BB}‐{SPEEDset}: {A} {Validated} {Dataset} of
  {Broadband} {Near}‐{Source} {Earthquake} {Ground} {Motions} from {3D}
  {Physics}‐{Based} {Numerical} {Simulations}} {{BB}‐{SPEEDset}: {A}
  {Validated} {Dataset} of {Broadband} {Near}‐{Source} {Earthquake} {Ground}
  {Motions} from {3D} {Physics}‐{Based} {Numerical} {Simulations}}.{\BBCQ}
\newblock
\APACjournalVolNumPages{Bulletin of the Seismological Society of
  America}{111}{5}{2527--2545}.
\newblock
\begin{APACrefDOI} \doi{10.1785/0120210089} \end{APACrefDOI}
\PrintBackRefs{\CurrentBib}

\bibitem [\protect \citeauthoryear {%
Perraudin%
, Balazs%
\BCBL {}\ \BBA {} Søndergaard%
}{%
Perraudin%
\ \protect \BOthers {.}}{%
{\protect \APACyear {2013}}%
}]{%
perraudin2013fast}
\APACinsertmetastar {%
perraudin2013fast}%
\begin{APACrefauthors}%
Perraudin, N.%
, Balazs, P.%
\BCBL {}\ \BBA {} Søndergaard, P\BPBI L.%
\end{APACrefauthors}%
\unskip\
\newblock
\APACrefYearMonthDay{2013}{}{}.
\newblock
{\BBOQ}\APACrefatitle {A fast {G}riffin-{L}im algorithm} {A fast
  {G}riffin-{L}im algorithm}.{\BBCQ}
\newblock
\BIn{} \APACrefbtitle {{2013 IEEE Workshop on Applications of Signal Processing
  to Audio and Acoustics}} {{2013 IEEE Workshop on Applications of Signal
  Processing to Audio and Acoustics}}\ (\BPGS\ 1--4).
\newblock
\begin{APACrefDOI} \doi{10.1109/WASPAA.2013.6701851} \end{APACrefDOI}
\PrintBackRefs{\CurrentBib}

\bibitem [\protect \citeauthoryear {%
Rodgers%
, Pitarka%
, Pankajakshan%
, Sj{\"o}green%
\BCBL {}\ \BBA {} Petersson%
}{%
Rodgers%
\ \protect \BOthers {.}}{%
{\protect \APACyear {2020}}%
}]{%
rodgers2020regional}
\APACinsertmetastar {%
rodgers2020regional}%
\begin{APACrefauthors}%
Rodgers, A\BPBI J.%
, Pitarka, A.%
, Pankajakshan, R.%
, Sj{\"o}green, B.%
\BCBL {}\ \BBA {} Petersson, N\BPBI A.%
\end{APACrefauthors}%
\unskip\
\newblock
\APACrefYearMonthDay{2020}{}{}.
\newblock
{\BBOQ}\APACrefatitle {{Regional-scale 3D ground-motion simulations of Mw 7
  earthquakes on the Hayward fault, northern California resolving frequencies
  0--10 Hz and including site-response corrections}} {{Regional-scale 3D
  ground-motion simulations of Mw 7 earthquakes on the Hayward fault, northern
  California resolving frequencies 0--10 Hz and including site-response
  corrections}}.{\BBCQ}
\newblock
\APACjournalVolNumPages{Bulletin of the Seismological Society of
  America}{110}{6}{2862--2881}.
\newblock
\begin{APACrefDOI} \doi{10.1785/0120200147} \end{APACrefDOI}
\PrintBackRefs{\CurrentBib}

\bibitem [\protect \citeauthoryear {%
Rombach%
, Blattmann%
, Lorenz%
, Esser%
\BCBL {}\ \BBA {} Ommer%
}{%
Rombach%
\ \protect \BOthers {.}}{%
{\protect \APACyear {2022}}%
}]{%
rombach2022highresolution}
\APACinsertmetastar {%
rombach2022highresolution}%
\begin{APACrefauthors}%
Rombach, R.%
, Blattmann, A.%
, Lorenz, D.%
, Esser, P.%
\BCBL {}\ \BBA {} Ommer, B.%
\end{APACrefauthors}%
\unskip\
\newblock
\APACrefYearMonthDay{2022}{}{}.
\newblock
{\BBOQ}\APACrefatitle {{High-Resolution Image Synthesis With Latent Diffusion
  Models}} {{High-Resolution Image Synthesis With Latent Diffusion
  Models}}.{\BBCQ}
\newblock
\BIn{} \APACrefbtitle {{Proceedings of the IEEE/CVF Conference on Computer
  Vision and Pattern Recognition}.} {{Proceedings of the IEEE/CVF Conference on
  Computer Vision and Pattern Recognition}.}
\newblock
\begin{APACrefDOI} \doi{10.48550/arXiv.2112.10752} \end{APACrefDOI}
\PrintBackRefs{\CurrentBib}

\bibitem [\protect \citeauthoryear {%
Saikia%
\ \BBA {} Somerville%
}{%
Saikia%
\ \BBA {} Somerville%
}{%
{\protect \APACyear {1997}}%
}]{%
saikia1997simulated}
\APACinsertmetastar {%
saikia1997simulated}%
\begin{APACrefauthors}%
Saikia, C\BPBI K.%
\BCBT {}\ \BBA {} Somerville, P.%
\end{APACrefauthors}%
\unskip\
\newblock
\APACrefYearMonthDay{1997}{}{}.
\newblock
{\BBOQ}\APACrefatitle {{Simulated hard-rock motions in Saint Louis, Missouri,
  from large New Madrid earthquakes (Mw$\geq$ 6.5)}} {{Simulated hard-rock
  motions in Saint Louis, Missouri, from large New Madrid earthquakes (Mw$\geq$
  6.5)}}.{\BBCQ}
\newblock
\APACjournalVolNumPages{Bulletin of the Seismological Society of
  America}{87}{1}{123--139}.
\newblock
\begin{APACrefDOI} \doi{10.1785/BSSA0870010123} \end{APACrefDOI}
\PrintBackRefs{\CurrentBib}

\bibitem [\protect \citeauthoryear {%
Savran%
\ \BBA {} Olsen%
}{%
Savran%
\ \BBA {} Olsen%
}{%
{\protect \APACyear {2019}}%
}]{%
savran2019ground}
\APACinsertmetastar {%
savran2019ground}%
\begin{APACrefauthors}%
Savran, W.%
\BCBT {}\ \BBA {} Olsen, K.%
\end{APACrefauthors}%
\unskip\
\newblock
\APACrefYearMonthDay{2019}{}{}.
\newblock
{\BBOQ}\APACrefatitle {{Ground motion simulation and validation of the 2008
  Chino Hills earthquake in scattering media}} {{Ground motion simulation and
  validation of the 2008 Chino Hills earthquake in scattering media}}.{\BBCQ}
\newblock
\APACjournalVolNumPages{Geophysical Journal International}{219}{3}{1836--1850}.
\newblock
\begin{APACrefDOI} \doi{10.1093/gji/ggz399} \end{APACrefDOI}
\PrintBackRefs{\CurrentBib}

\bibitem [\protect \citeauthoryear {%
Sevilla%
}{%
Sevilla%
}{%
{\protect \APACyear {2022}}%
}]{%
Sevilla_STEAD_2022}
\APACinsertmetastar {%
Sevilla_STEAD_2022}%
\begin{APACrefauthors}%
Sevilla, I.%
\end{APACrefauthors}%
\unskip\
\newblock
\APACrefYearMonthDay{2022}{}{}.
\newblock
\APACrefbtitle {Stanford Earthquake Dataset (STEAD).} {Stanford earthquake
  dataset (stead).}
\newblock
\begin{APACrefURL}
  \url{https://www.kaggle.com/datasets/isevilla/stanford-earthquake-dataset-stead}
  \end{APACrefURL}
\newblock
\APACrefnote{Accessed 2025-05-31}
\PrintBackRefs{\CurrentBib}

\bibitem [\protect \citeauthoryear {%
Shi%
, Lavrentiadis%
, Asimaki%
, Ross%
\BCBL {}\ \BBA {} Azizzadenesheli%
}{%
Shi%
\ \protect \BOthers {.}}{%
{\protect \APACyear {2024}}%
}]{%
shi2024broadband}
\APACinsertmetastar {%
shi2024broadband}%
\begin{APACrefauthors}%
Shi, Y.%
, Lavrentiadis, G.%
, Asimaki, D.%
, Ross, Z\BPBI E.%
\BCBL {}\ \BBA {} Azizzadenesheli, K.%
\end{APACrefauthors}%
\unskip\
\newblock
\APACrefYearMonthDay{2024}{}{}.
\newblock
{\BBOQ}\APACrefatitle {{Broadband Ground-Motion Synthesis via Generative
  Adversarial Neural Operators: Development and Validation}} {{Broadband
  Ground-Motion Synthesis via Generative Adversarial Neural Operators:
  Development and Validation}}.{\BBCQ}
\newblock
\APACjournalVolNumPages{Bulletin of the Seismological Society of
  America}{114}{4}{2151--2171}.
\newblock
\begin{APACrefDOI} \doi{10.1785/0120230207} \end{APACrefDOI}
\PrintBackRefs{\CurrentBib}

\bibitem [\protect \citeauthoryear {%
Smerzini%
, Amendola%
, Paolucci%
\BCBL {}\ \BBA {} Bazrafshan%
}{%
Smerzini%
\ \protect \BOthers {.}}{%
{\protect \APACyear {2024}}%
}]{%
smerzini_engineering_2024}
\APACinsertmetastar {%
smerzini_engineering_2024}%
\begin{APACrefauthors}%
Smerzini, C.%
, Amendola, C.%
, Paolucci, R.%
\BCBL {}\ \BBA {} Bazrafshan, A.%
\end{APACrefauthors}%
\unskip\
\newblock
\APACrefYearMonthDay{2024}{}{}.
\newblock
{\BBOQ}\APACrefatitle {{Engineering validation of {BB}-{SPEEDset}, a data set
  of near-source physics-based simulated accelerograms}} {{Engineering
  validation of {BB}-{SPEEDset}, a data set of near-source physics-based
  simulated accelerograms}}.{\BBCQ}
\newblock
\APACjournalVolNumPages{Earthquake Spectra}{40}{1}{420--445}.
\newblock
\begin{APACrefDOI} \doi{10.1177/87552930231206766} \end{APACrefDOI}
\PrintBackRefs{\CurrentBib}

\bibitem [\protect \citeauthoryear {%
Song%
, Dhariwal%
, Chen%
\BCBL {}\ \BBA {} Sutskever%
}{%
Song%
\ \protect \BOthers {.}}{%
{\protect \APACyear {2023}}%
}]{%
song23consistency}
\APACinsertmetastar {%
song23consistency}%
\begin{APACrefauthors}%
Song, Y.%
, Dhariwal, P.%
, Chen, M.%
\BCBL {}\ \BBA {} Sutskever, I.%
\end{APACrefauthors}%
\unskip\
\newblock
\APACrefYearMonthDay{2023}{}{}.
\newblock
{\BBOQ}\APACrefatitle {{Consistency Models}} {{Consistency Models}}.{\BBCQ}
\newblock
\BIn{} \APACrefbtitle {Proceedings of the 40th International Conference on
  Machine Learning.} {Proceedings of the 40th international conference on
  machine learning.}
\newblock
\begin{APACrefDOI} \doi{10.5555/3618408.3619743} \end{APACrefDOI}
\PrintBackRefs{\CurrentBib}

\bibitem [\protect \citeauthoryear {%
Song%
\ \BBA {} Ermon%
}{%
Song%
\ \BBA {} Ermon%
}{%
{\protect \APACyear {2019}}%
}]{%
song2019generative}
\APACinsertmetastar {%
song2019generative}%
\begin{APACrefauthors}%
Song, Y.%
\BCBT {}\ \BBA {} Ermon, S.%
\end{APACrefauthors}%
\unskip\
\newblock
\APACrefYearMonthDay{2019}{}{}.
\newblock
{\BBOQ}\APACrefatitle {{Generative Modeling by Estimating Gradients of the Data
  Distribution}} {{Generative Modeling by Estimating Gradients of the Data
  Distribution}}.{\BBCQ}
\newblock
\BIn{} \APACrefbtitle {Advances in Neural Information Processing Systems.}
  {Advances in neural information processing systems.}
\newblock
\begin{APACrefDOI} \doi{10.48550/arXiv.1907.05600} \end{APACrefDOI}
\PrintBackRefs{\CurrentBib}

\bibitem [\protect \citeauthoryear {%
Song%
\ \protect \BOthers {.}}{%
Song%
\ \protect \BOthers {.}}{%
{\protect \APACyear {2021}}%
}]{%
song2020scorebased}
\APACinsertmetastar {%
song2020scorebased}%
\begin{APACrefauthors}%
Song, Y.%
, Sohl-Dickstein, J.%
, Kingma, D\BPBI P.%
, Kumar, A.%
, Ermon, S.%
\BCBL {}\ \BBA {} Poole, B.%
\end{APACrefauthors}%
\unskip\
\newblock
\APACrefYearMonthDay{2021}{}{}.
\newblock
{\BBOQ}\APACrefatitle {{Score-Based Generative Modeling through Stochastic
  Differential Equations}} {{Score-Based Generative Modeling through Stochastic
  Differential Equations}}.{\BBCQ}
\newblock
\BIn{} \APACrefbtitle {{International Conference on Learning Representations}.}
  {{International Conference on Learning Representations}.}
\newblock
\begin{APACrefDOI} \doi{10.48550/arXiv.2011.13456} \end{APACrefDOI}
\PrintBackRefs{\CurrentBib}

\bibitem [\protect \citeauthoryear {%
Tang%
\ \BBA {} Mai%
}{%
Tang%
\ \BBA {} Mai%
}{%
{\protect \APACyear {2023}}%
}]{%
tang2023stochastic}
\APACinsertmetastar {%
tang2023stochastic}%
\begin{APACrefauthors}%
Tang, Y.%
\BCBT {}\ \BBA {} Mai, P\BPBI M.%
\end{APACrefauthors}%
\unskip\
\newblock
\APACrefYearMonthDay{2023}{}{}.
\newblock
{\BBOQ}\APACrefatitle {{Stochastic Ground-Motion Simulation of the 2021 Mw 5.9
  Woods Point Earthquake: Facilitating Local Probabilistic Seismic Hazard
  Analysis in Australia}} {{Stochastic Ground-Motion Simulation of the 2021 Mw
  5.9 Woods Point Earthquake: Facilitating Local Probabilistic Seismic Hazard
  Analysis in Australia}}.{\BBCQ}
\newblock
\APACjournalVolNumPages{Bulletin of the Seismological Society of
  America}{113}{5}{2119--2143}.
\newblock
\begin{APACrefDOI} \doi{10.1785/0120220260} \end{APACrefDOI}
\PrintBackRefs{\CurrentBib}

\bibitem [\protect \citeauthoryear {%
{thispersondoesnotexist.com}%
}{%
{thispersondoesnotexist.com}%
}{%
{\protect \APACyear {2023}}%
}]{%
thispersondoesnotexist}
\APACinsertmetastar {%
thispersondoesnotexist}%
\begin{APACrefauthors}%
{thispersondoesnotexist.com}.%
\end{APACrefauthors}%
\unskip\
\newblock
\APACrefYearMonthDay{2023}{}{}.
\newblock
\APACrefbtitle {This Person Does Not Exist.} {This person does not exist.}
\newblock
\APAChowpublished {\url{https://thispersondoesnotexist.com}}.
\newblock
\APACrefnote{Accessed: 2024-10-16}
\PrintBackRefs{\CurrentBib}

\bibitem [\protect \citeauthoryear {%
Touhami%
\ \protect \BOthers {.}}{%
Touhami%
\ \protect \BOthers {.}}{%
{\protect \APACyear {2022}}%
}]{%
touhami2022sem3d}
\APACinsertmetastar {%
touhami2022sem3d}%
\begin{APACrefauthors}%
Touhami, S.%
, Gatti, F.%
, Lopez-Caballero, F.%
, Cottereau, R.%
, de Abreu~Corr{\^e}a, L.%
, Aubry, L.%
\BCBL {}\ \BBA {} Clouteau, D.%
\end{APACrefauthors}%
\unskip\
\newblock
\APACrefYearMonthDay{2022}{}{}.
\newblock
{\BBOQ}\APACrefatitle {{SEM3D: A 3D high-fidelity numerical earthquake
  simulator for broadband (0--10 Hz) seismic response prediction at a regional
  scale}} {{SEM3D: A 3D high-fidelity numerical earthquake simulator for
  broadband (0--10 Hz) seismic response prediction at a regional
  scale}}.{\BBCQ}
\newblock
\APACjournalVolNumPages{Geosciences}{12}{3}{112}.
\newblock
\begin{APACrefDOI} \doi{10.3390/geosciences12030112} \end{APACrefDOI}
\PrintBackRefs{\CurrentBib}

\bibitem [\protect \citeauthoryear {%
van Ede%
\ \protect \BOthers {.}}{%
van Ede%
\ \protect \BOthers {.}}{%
{\protect \APACyear {2020}}%
}]{%
van2020hybrid}
\APACinsertmetastar {%
van2020hybrid}%
\begin{APACrefauthors}%
van Ede, M\BPBI C.%
, Molinari, I.%
, Imperatori, W.%
, Kissling, E.%
, Baron, J.%
\BCBL {}\ \BBA {} Morelli, A.%
\end{APACrefauthors}%
\unskip\
\newblock
\APACrefYearMonthDay{2020}{}{}.
\newblock
{\BBOQ}\APACrefatitle {{Hybrid broadband seismograms for seismic shaking
  scenarios: An application to the Po Plain sedimentary Basin (Northern
  Italy)}} {{Hybrid broadband seismograms for seismic shaking scenarios: An
  application to the Po Plain sedimentary Basin (Northern Italy)}}.{\BBCQ}
\newblock
\APACjournalVolNumPages{Pure and Applied Geophysics}{177}{5}{2181--2198}.
\newblock
\begin{APACrefDOI} \doi{10.1007/s00024-019-02322-0} \end{APACrefDOI}
\PrintBackRefs{\CurrentBib}

\bibitem [\protect \citeauthoryear {%
Vincent%
}{%
Vincent%
}{%
{\protect \APACyear {2011}}%
}]{%
vincent2011score}
\APACinsertmetastar {%
vincent2011score}%
\begin{APACrefauthors}%
Vincent, P.%
\end{APACrefauthors}%
\unskip\
\newblock
\APACrefYearMonthDay{2011}{}{}.
\newblock
{\BBOQ}\APACrefatitle {{A Connection Between Score Matching and Denoising
  Autoencoders}} {{A Connection Between Score Matching and Denoising
  Autoencoders}}.{\BBCQ}
\newblock
\APACjournalVolNumPages{{Neural Computation}}{23}{7}{1661--1674}.
\newblock
\begin{APACrefDOI} \doi{10.1162/NECO_a_00142} \end{APACrefDOI}
\PrintBackRefs{\CurrentBib}

\bibitem [\protect \citeauthoryear {%
Wang%
, Trugman%
\BCBL {}\ \BBA {} Lin%
}{%
Wang%
\ \protect \BOthers {.}}{%
{\protect \APACyear {2021}}%
}]{%
wang2021seismogen}
\APACinsertmetastar {%
wang2021seismogen}%
\begin{APACrefauthors}%
Wang, T.%
, Trugman, D.%
\BCBL {}\ \BBA {} Lin, Y.%
\end{APACrefauthors}%
\unskip\
\newblock
\APACrefYearMonthDay{2021}{}{}.
\newblock
{\BBOQ}\APACrefatitle {{SeismoGen: Seismic waveform synthesis using GAN with
  application to seismic data augmentation}} {{SeismoGen: Seismic waveform
  synthesis using GAN with application to seismic data augmentation}}.{\BBCQ}
\newblock
\APACjournalVolNumPages{Journal of Geophysical Research: Solid
  Earth}{126}{4}{e2020JB020077}.
\newblock
\begin{APACrefDOI} \doi{10.1029/2020JB020077} \end{APACrefDOI}
\PrintBackRefs{\CurrentBib}

\bibitem [\protect \citeauthoryear {%
Woollam%
\ \protect \BOthers {.}}{%
Woollam%
\ \protect \BOthers {.}}{%
{\protect \APACyear {2022}}%
}]{%
woollam2022seisbench}
\APACinsertmetastar {%
woollam2022seisbench}%
\begin{APACrefauthors}%
Woollam, J.%
, M{\"u}nchmeyer, J.%
, Tilmann, F.%
, Rietbrock, A.%
, Lange, D.%
, Bornstein, T.%
, Diehl, T.%
, Giunchi, C.%
, Haslinger, F.%
, Jozinovi{\'c}, D.%
, Michelini, A.%
, Saul, J.%
\BCBL {}\ \BBA {} Soto, H.%
\end{APACrefauthors}%
\unskip\
\newblock
\APACrefYearMonthDay{2022}{}{}.
\newblock
{\BBOQ}\APACrefatitle {{SeisBench—A toolbox for machine learning in
  seismology}} {{SeisBench—A toolbox for machine learning in
  seismology}}.{\BBCQ}
\newblock
\APACjournalVolNumPages{Seismological Society of America}{93}{3}{1695--1709}.
\newblock
\begin{APACrefDOI} \doi{10.1785/0220210324} \end{APACrefDOI}
\PrintBackRefs{\CurrentBib}

\bibitem [\protect \citeauthoryear {%
Yang%
\ \protect \BOthers {.}}{%
Yang%
\ \protect \BOthers {.}}{%
{\protect \APACyear {2023}}%
}]{%
yang2023diffusion}
\APACinsertmetastar {%
yang2023diffusion}%
\begin{APACrefauthors}%
Yang, L.%
, Zhang, Z.%
, Song, Y.%
, Hong, S.%
, Xu, R.%
, Zhao, Y.%
, Zhang, W.%
, Cui, B.%
\BCBL {}\ \BBA {} Yang, M\BHBI H.%
\end{APACrefauthors}%
\unskip\
\newblock
\APACrefYearMonthDay{2023}{}{}.
\newblock
{\BBOQ}\APACrefatitle {Diffusion models: A comprehensive survey of methods and
  applications} {Diffusion models: A comprehensive survey of methods and
  applications}.{\BBCQ}
\newblock
\APACjournalVolNumPages{ACM computing surveys}{56}{4}{1--39}.
\newblock
\begin{APACrefDOI} \doi{10.1145/3626235} \end{APACrefDOI}
\PrintBackRefs{\CurrentBib}

\bibitem [\protect \citeauthoryear {%
W.~Zhu%
\ \BBA {} Beroza%
}{%
W.~Zhu%
\ \BBA {} Beroza%
}{%
{\protect \APACyear {2019}}%
}]{%
zhu2019phasenet}
\APACinsertmetastar {%
zhu2019phasenet}%
\begin{APACrefauthors}%
Zhu, W.%
\BCBT {}\ \BBA {} Beroza, G\BPBI C.%
\end{APACrefauthors}%
\unskip\
\newblock
\APACrefYearMonthDay{2019}{}{}.
\newblock
{\BBOQ}\APACrefatitle {{PhaseNet: a deep-neural-network-based seismic
  arrival-time picking method}} {{PhaseNet: a deep-neural-network-based seismic
  arrival-time picking method}}.{\BBCQ}
\newblock
\APACjournalVolNumPages{Geophysical Journal International}{216}{1}{261--273}.
\newblock
\begin{APACrefDOI} \doi{10.1093/gji/ggy423} \end{APACrefDOI}
\PrintBackRefs{\CurrentBib}

\bibitem [\protect \citeauthoryear {%
Z.~Zhu%
\ \protect \BOthers {.}}{%
Z.~Zhu%
\ \protect \BOthers {.}}{%
{\protect \APACyear {2023}}%
}]{%
zhu2023diffusion}
\APACinsertmetastar {%
zhu2023diffusion}%
\begin{APACrefauthors}%
Zhu, Z.%
, Zhao, H.%
, He, H.%
, Zhong, Y.%
, Zhang, S.%
, Guo, H.%
, Chen, T.%
\BCBL {}\ \BBA {} Zhang, W.%
\end{APACrefauthors}%
\unskip\
\newblock
\APACrefYearMonthDay{2023}{}{}.
\newblock
{\BBOQ}\APACrefatitle {{Diffusion models for reinforcement learning: A survey}}
  {{Diffusion models for reinforcement learning: A survey}}.{\BBCQ}
\newblock
\APACjournalVolNumPages{arXiv preprint arXiv:2311.01223}{}{}{}.
\newblock
\begin{APACrefDOI} \doi{10.48550/arXiv.2311.01223} \end{APACrefDOI}
\PrintBackRefs{\CurrentBib}

\end{thebibliography}

\clearpage
 \setcounter{figure}{0}
 \renewcommand{\thefigure}{S\arabic{figure}}

%
%

\title{Supporting Information for "High Resolution Seismic Waveform Generation using Denoising Diffusion"}
%
%

%
%

\authors{Kadek Hendrawan Palgunadi\affil{1, \oplus}, Andreas Bergmeister\affil{2,\dagger, \oplus},  Andrea Bosisio\affil{3}, Laura Ermert\affil{1, \ddagger}, Maria Koroni\affil{1}, Nathanaël Perraudin\affil{2,\dagger}, Simon Dirmeier\affil{2}, Men-Andrin Meier\affil{4,*}}

\affiliation{1}{Swiss Seismological Service (SED), ETH Zürich, Switzerland}
\affiliation{2}{Swiss Data Science Center (SDSC), ETH Zürich, Switzerland}
\affiliation{3}{Politecnico di Milano, Italy}
\affiliation{4}{Earth and Planetary Science Department, ETH Zürich, Switzerland}
\affiliation{\dagger}{Work conducted while being employed at SDSC}
\affiliation{\ddagger}{Now at: Univ. Grenoble Alpes, Univ. Savoie Mont Blanc, CNRS, IRD, Univ. Gustave Eiffel, ISTerre, Grenoble, France}
\affiliation{\oplus}{equal contributions}

\begin{figure}[!htp]
    \centering
    \includegraphics[width=0.49\textwidth]{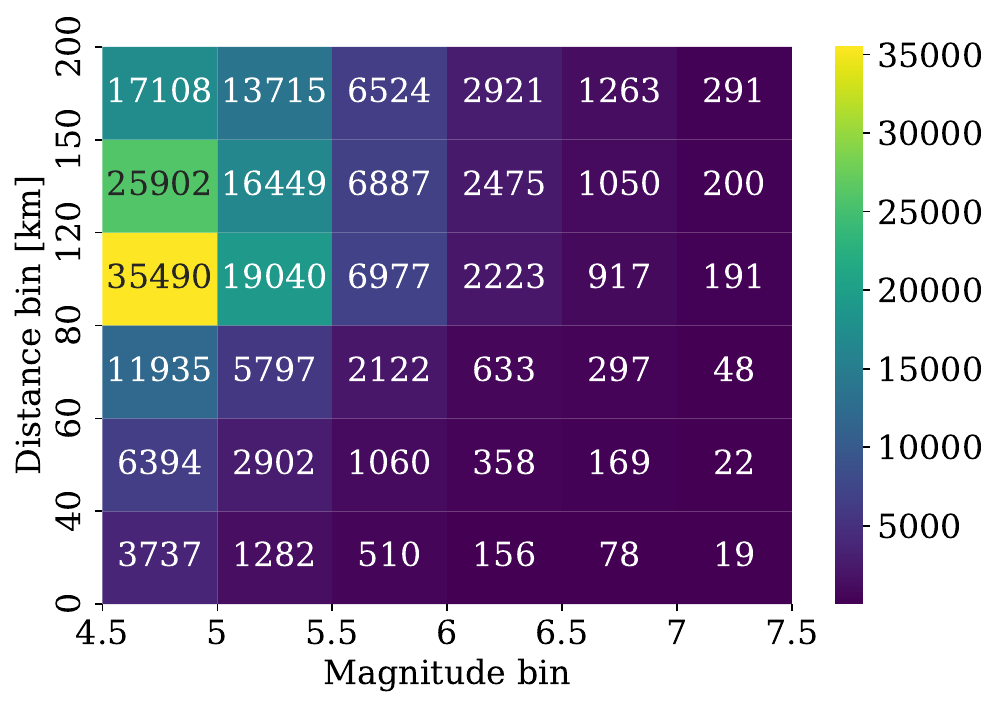}
    \caption{Number of samples in each magnitude-distance bin for all of the following bin plots. Predicted and target denote the generative waveform model (GWM) and real data.}
    \label{fig:supp_grid_plot}
\end{figure}

\begin{figure}[!htp]
    \centering
    \includegraphics[width=0.6\textwidth]{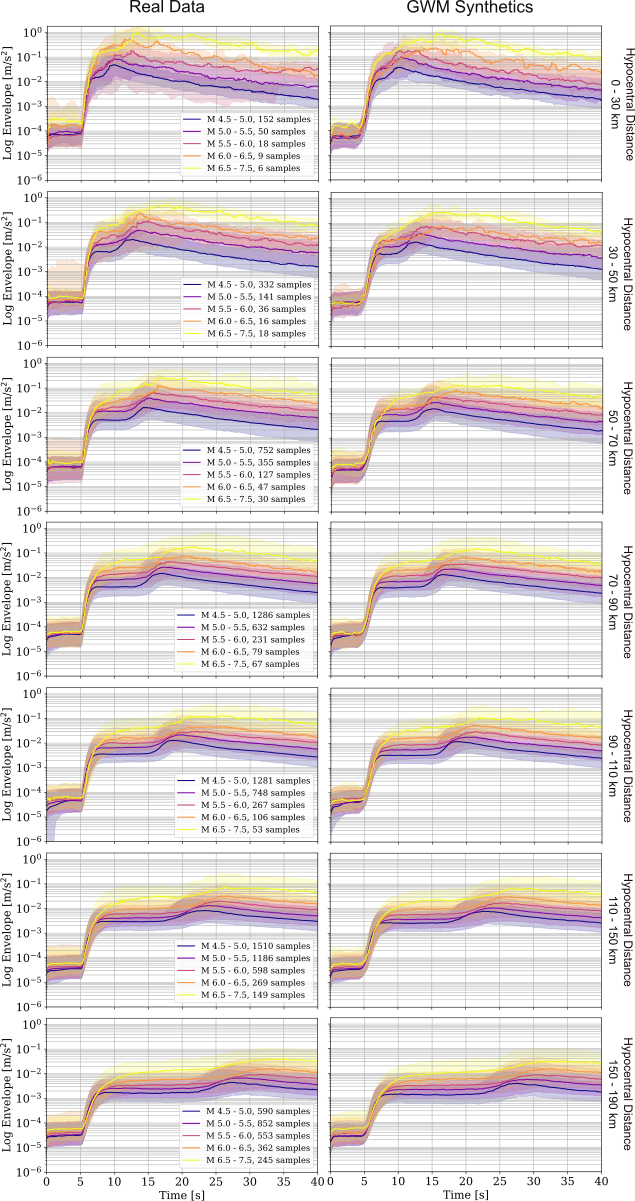}
    \caption{Distribution of time-domain envelopes for Radial-component seismograms in different magnitude and distance bins. Predicted and target denote the generative waveform model (GWM) and real data.}
    \label{fig:supp_envelope_0}
\end{figure}

\begin{figure}[!htp]
    \centering
    \includegraphics[width=0.6\textwidth]{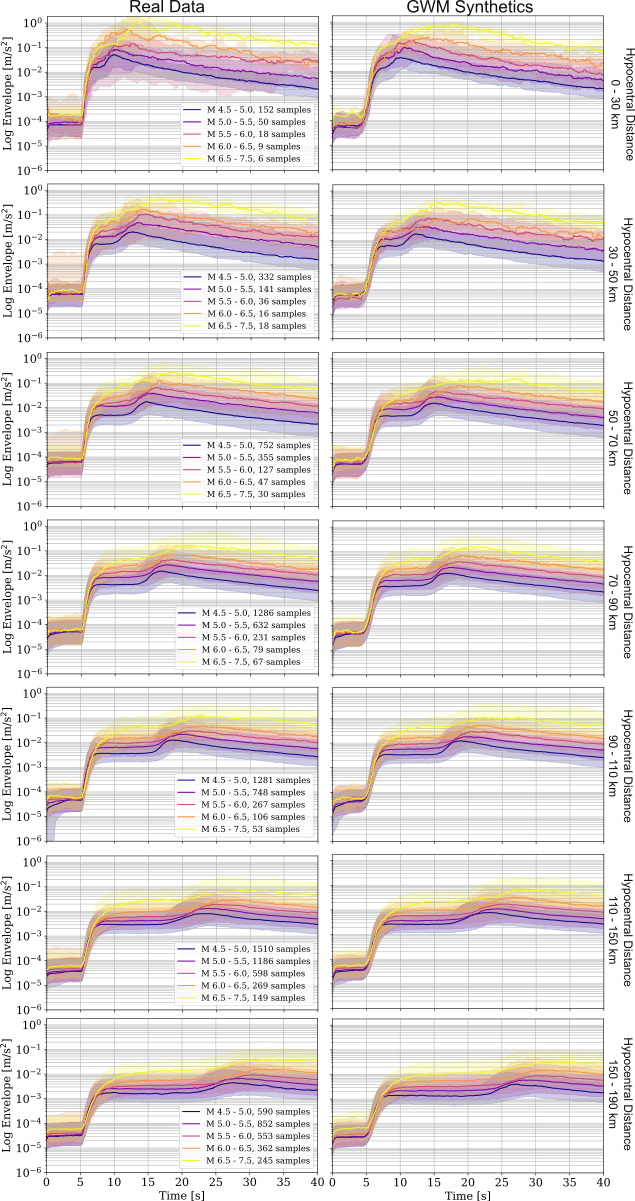}
    \caption{Distribution of time-domain envelopes for transverse-component seismograms in different magnitude and distance bins. Predicted and target denote the generative waveform model (GWM) and real data.}
    \label{fig:supp_envelope_1}
\end{figure}

\begin{figure}[!htp]
    \centering
    \includegraphics[width=0.6\textwidth]{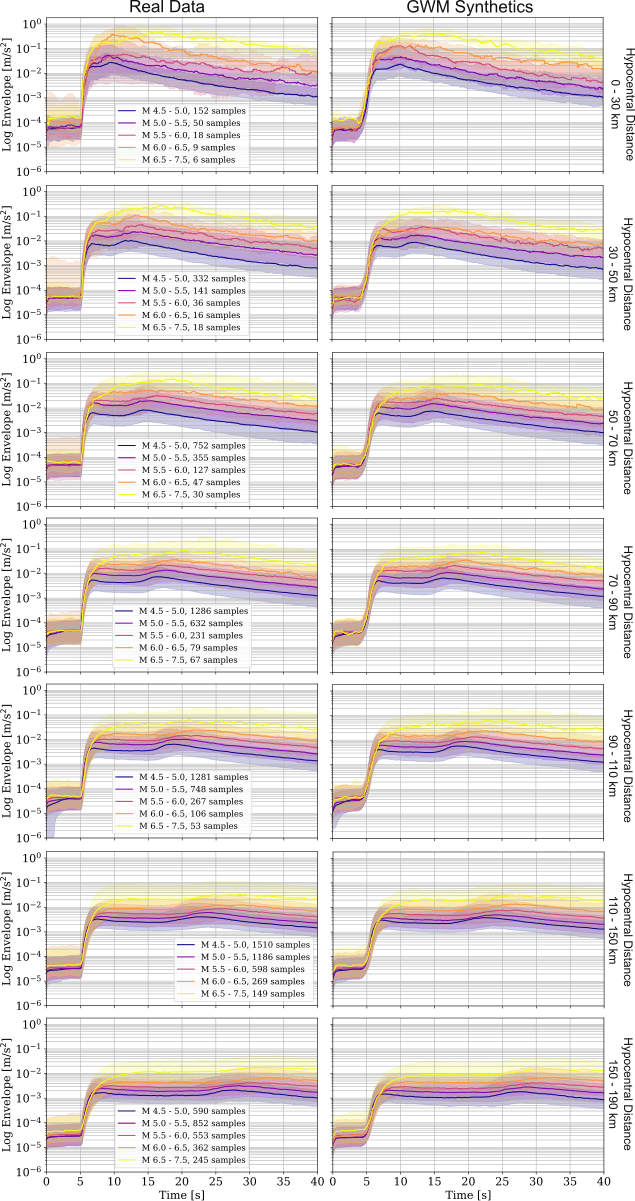}
    \caption{Distribution of time-domain envelopes for vertical-component seismograms in different magnitude and distance bins. Predicted and target denote the generative waveform model (GWM) and real data.}
    \label{fig:supp_envelope_2}
\end{figure}

\begin{figure}[!htp]
    \centering
    \includegraphics[width=0.6\textwidth]{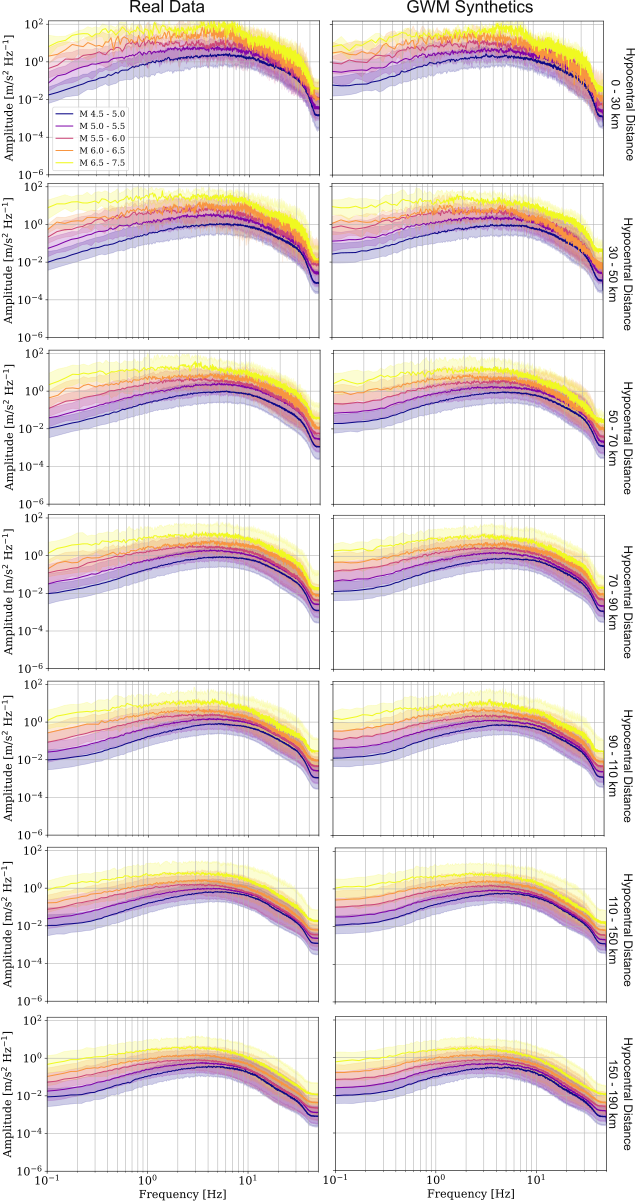}
    \caption{Distribution of Fourier spectra log-amplitudes for radial-component seismograms in different magnitude and distance bins. Predicted and target denote the generative waveform model (GWM) and real data.}
    \label{fig:supp_asd_0}
\end{figure}

\begin{figure}[!htp]
    \centering
    \includegraphics[width=0.6\textwidth]{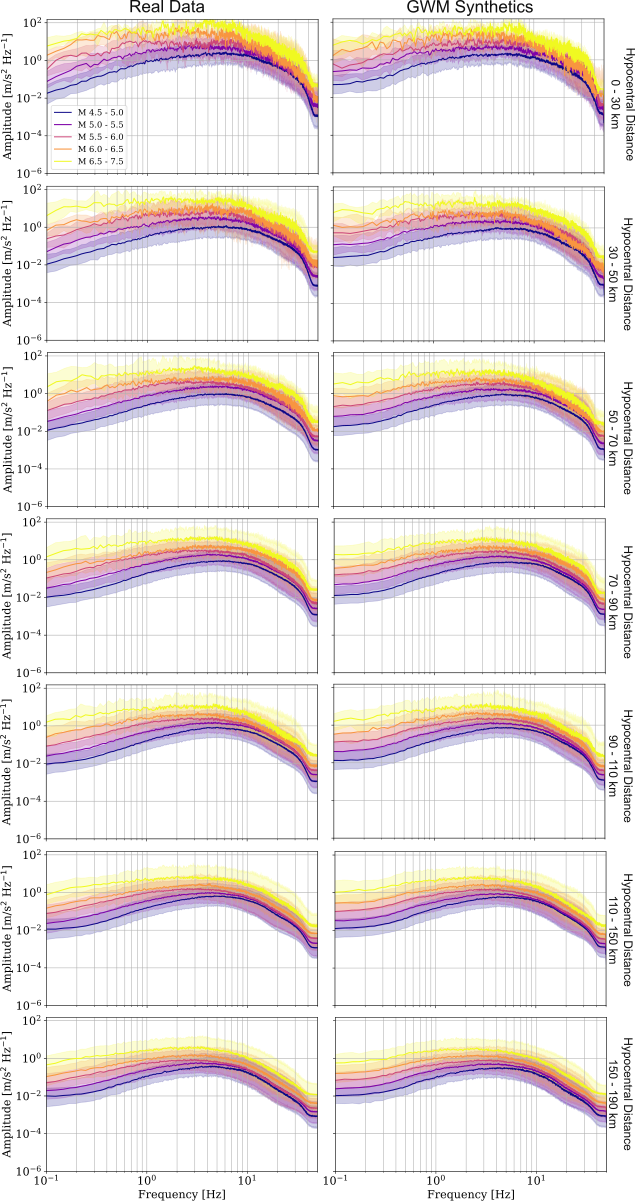}
    \caption{Distribution of Fourier spectra log-amplitudes for tranverse-component seismograms in different magnitude and distance bins. Predicted and target denote the generative waveform model (GWM) and real data.}
    \label{fig:supp_asd_1}
\end{figure}

\begin{figure}[!htp]
    \centering
    \includegraphics[width=0.6\textwidth]{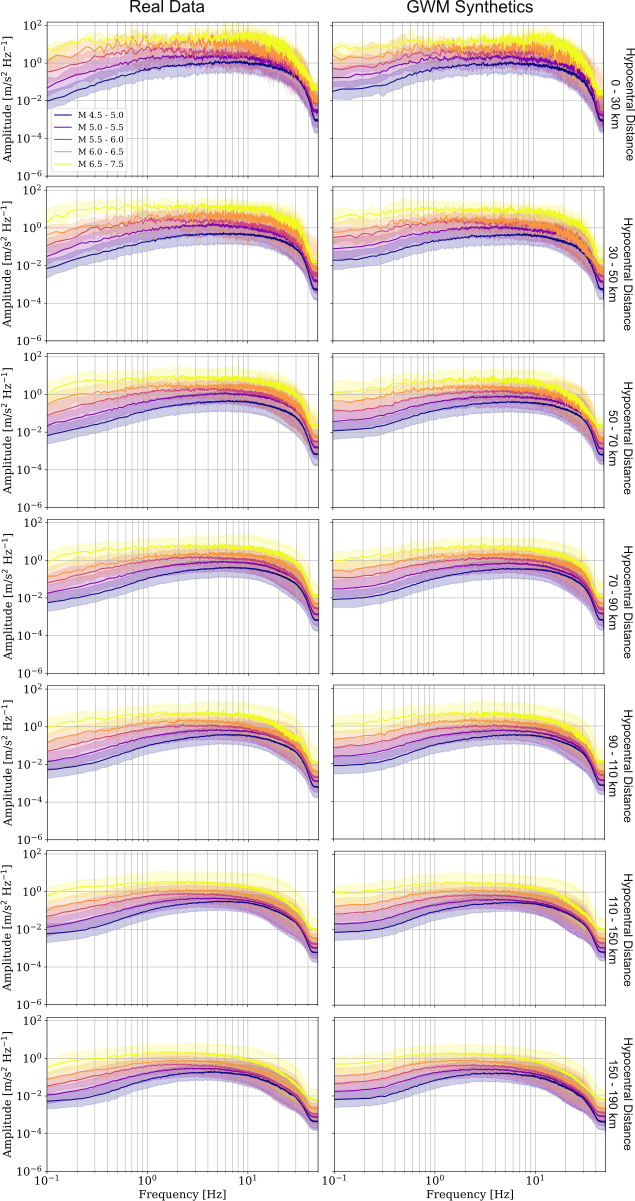}
    \caption{Distribution of Fourier spectra log-amplitudes for vertical-component seismograms in different magnitude and distance bins. Predicted and target denote the generative waveform model (GWM) and real data.}
    \label{fig:supp_asd_2}
\end{figure}

\begin{figure}[!htp]
    \centering
    \includegraphics[width=0.9\textwidth]{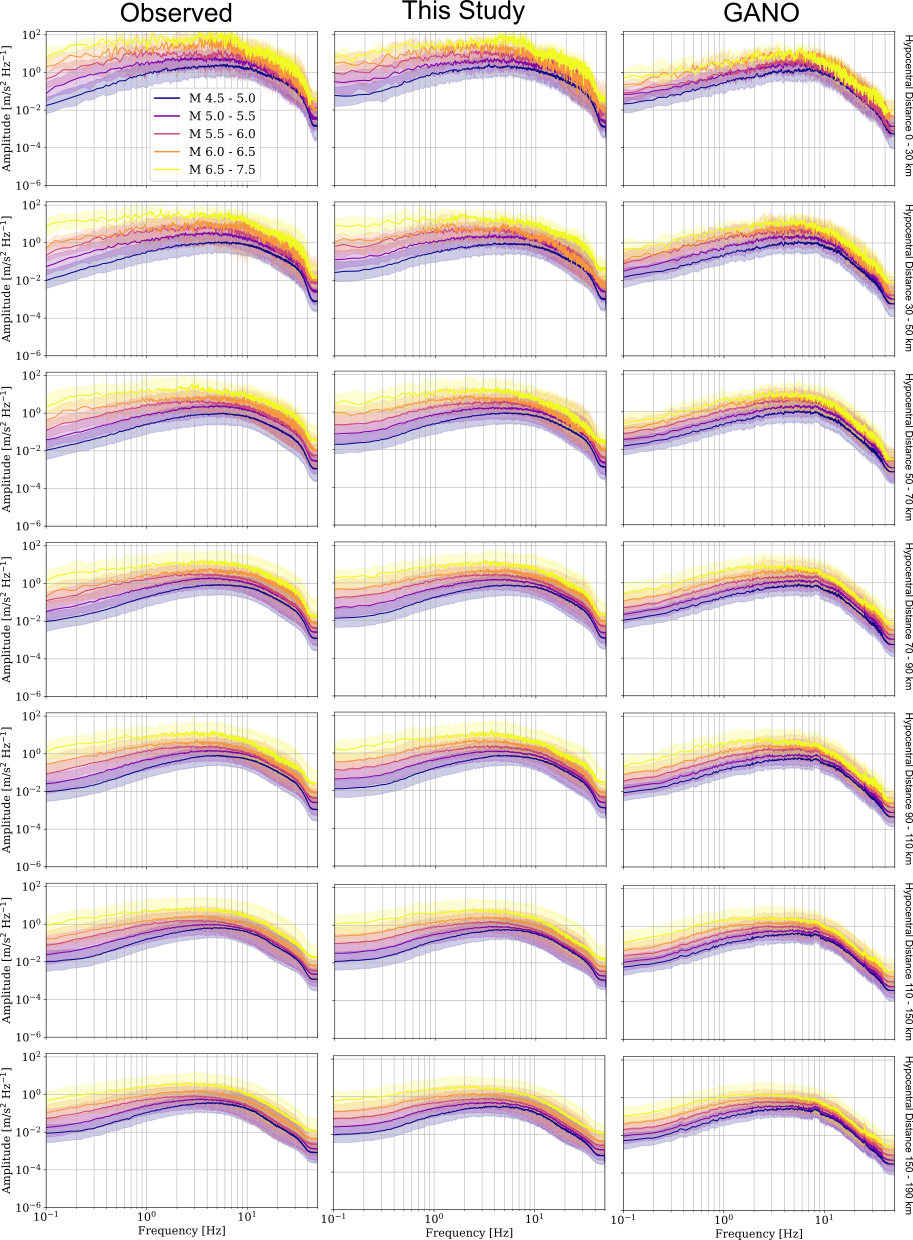}
    \caption{Distribution of Fourier spectra log-amplitudes for Radial-component seismograms in different magnitude and distance bins. Observed, this study, and GANO denote real data, our GWM, and GANO.}
    \label{fig:supp_gano_0}
\end{figure}

\begin{figure}[!htp]
    \centering
    \includegraphics[width=0.9\textwidth]{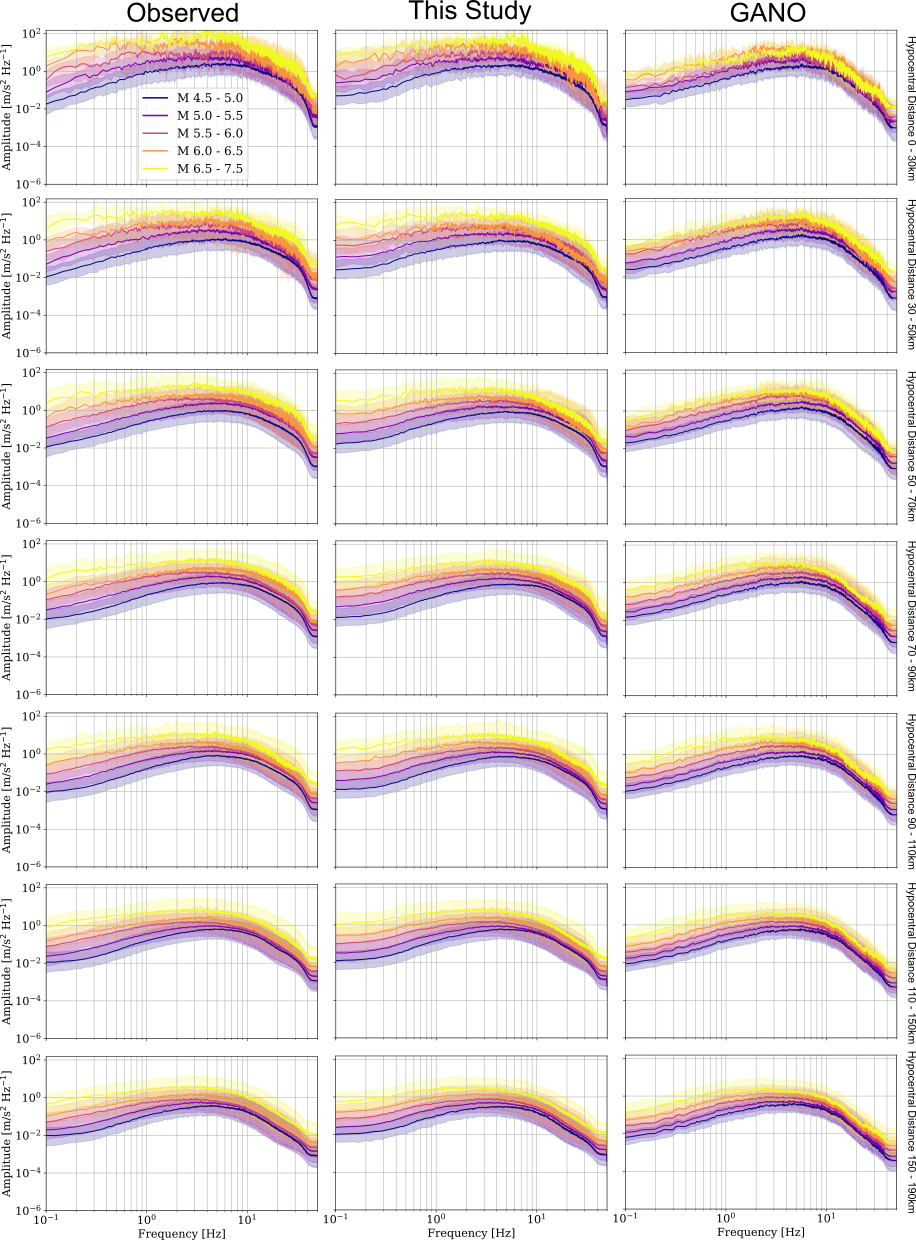}
    \caption{Distribution of Fourier spectra log-amplitudes for Transverse-component seismograms in different magnitude and distance bins. Observed, this study, and GANO denote real data, our GWM, and GANO.}
    \label{fig:supp_gano_1}
\end{figure}

\begin{figure}[!htp]
    \centering
    \includegraphics[width=0.9\textwidth]{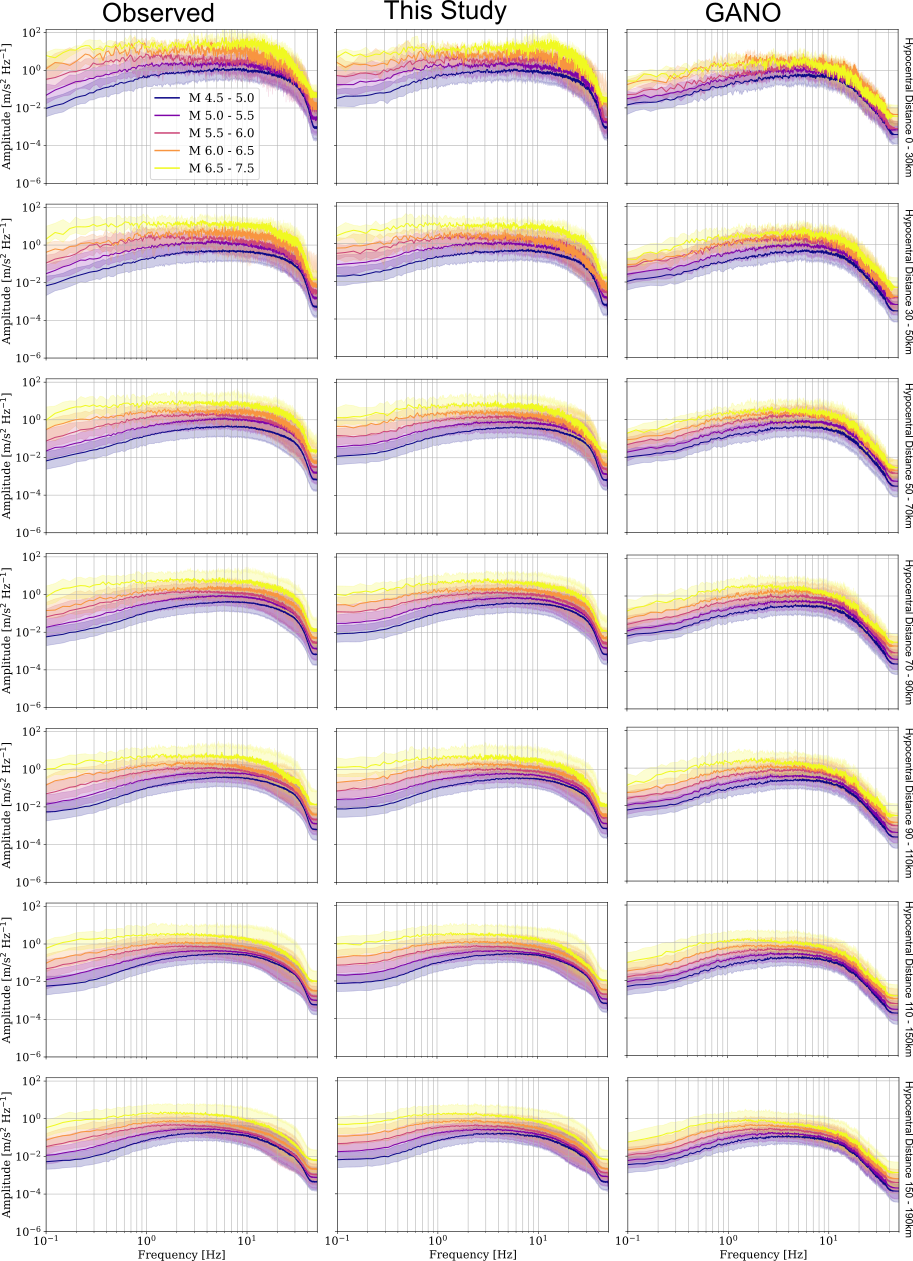}
    \caption{Distribution of Fourier spectra log-amplitudes for Vertical-component seismograms in different magnitude and distance bins. Observed, this study, and GANO denote real data, our GWM, and GANO.}
    \label{fig:supp_gano_2}
\end{figure}

\begin{figure}[!htp]
    \centering
    \includegraphics[width=0.7\linewidth]{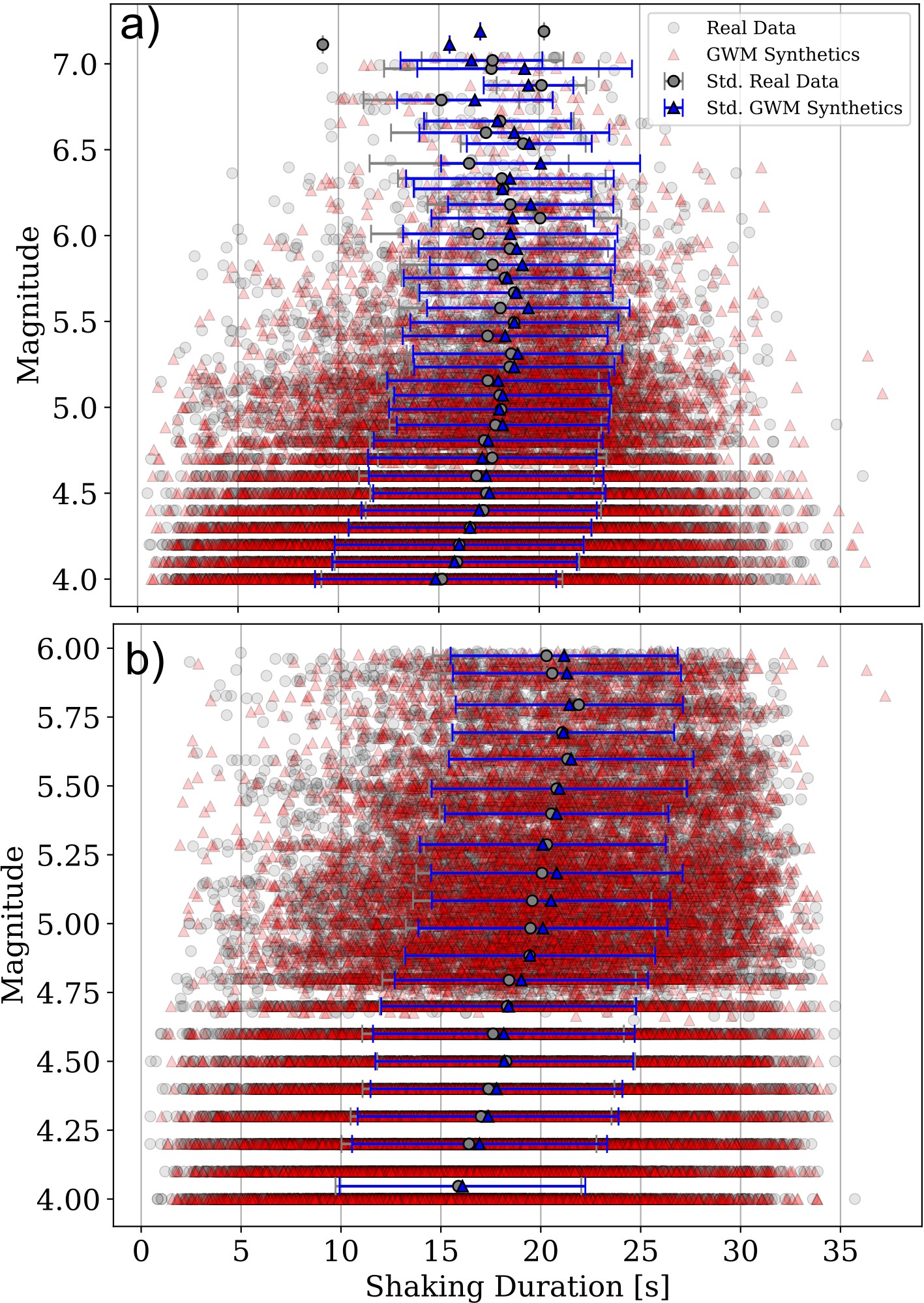}
    \caption{Shaking duration for all GWM synthetics (red triangles) using one realization and real data (grey circles) with corresponding conditioning parameters. For each magnitude bin (every 0.08) from $M$ 4.5 - 9.0, grey dots and lines show the mean and standard deviation of the real data, while blue triangles and lines show the mean and standard deviation of the GWM synthetics. a) Shaking duration for all test datasets. b) Shaking duration of filtered test datasets for M $\leq 6$ and hypocentral distance $< 100$ km.}
    \label{fig:shaking_duration}
\end{figure}

\begin{figure}[!htp]
    \centering
    \includegraphics[width=0.99\textwidth]{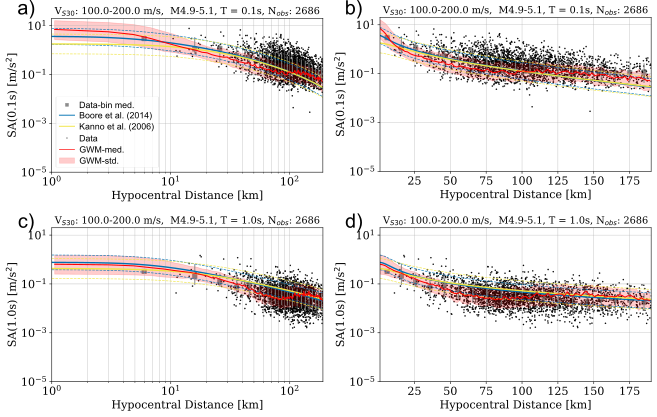}
    \caption{RotD50 pseudo-spectral acceleration (SA) with a damping factor 5\% versus hypocentral distance for periods ($T$) of 0.1 s and 1.0 s. Median prediction of the GWM (red line) and standard deviation (red shaded area), along with median prediction (solid lines) and standard deviation (dashed lines) of the \protect\citeA{boore2014nga} GMM (blue), and the \protect\citeA{kanno2006new} GMM (green), using M5 and \(V_{S30} = 150\) m/s. The data are sampled from narrow magnitude and $V_{S30}$ bins, as written in the figure titles, and shown by their median (green squares) and standard deviations (green lines).}
    \label{fig:SA_GMM_5.0_150}
\end{figure}

\begin{figure}[!htp]
    \centering
    \includegraphics[width=0.99\textwidth]{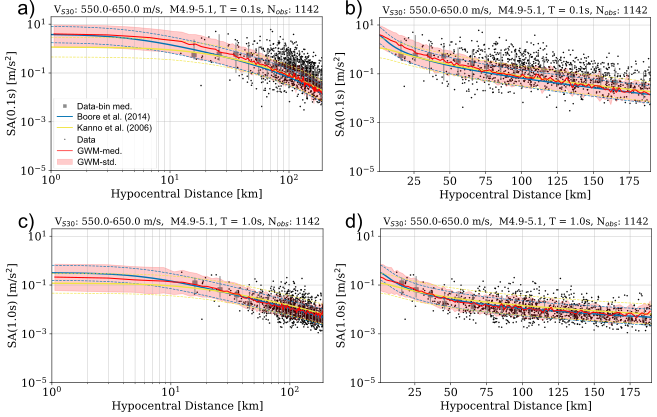}
    \caption{Same as Figure \ref{fig:SA_GMM_5.0_150} but for the magnitude bin M5 and \(V_{S30} = 600\) m/s.}
    \label{fig:SA_GMM_5.0_600}
\end{figure}

\begin{figure}[!htp]
    \centering
    \includegraphics[width=0.99\textwidth]{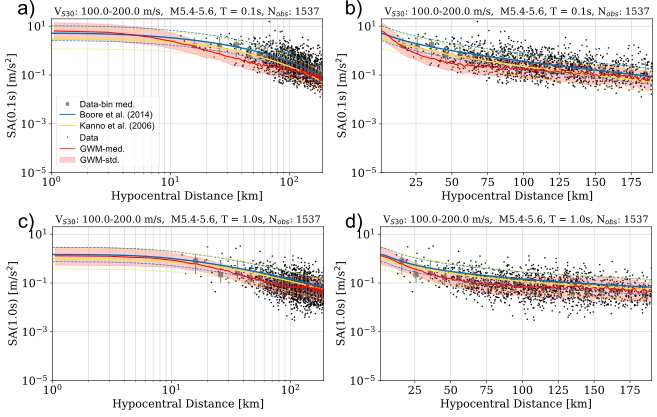}
    \caption{Same as Figure \ref{fig:SA_GMM_5.0_150} but for the magnitude bin M5.5 and \(V_{S30} = 150\) m/s.}
    \label{fig:SA_GMM_5.5_150}
\end{figure}

\begin{figure}[!htp]
    \centering
    \includegraphics[width=0.99\textwidth]{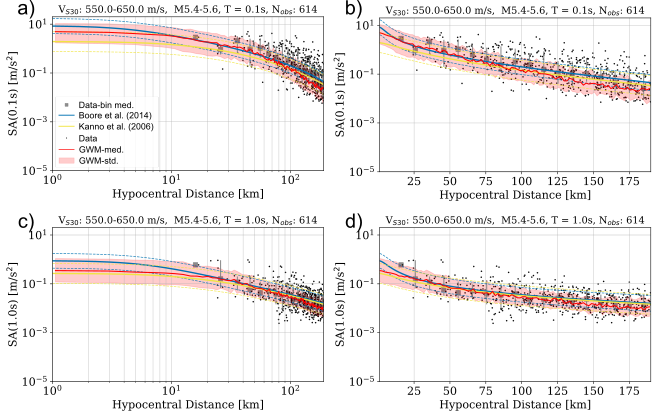}
    \caption{Same as Figure \ref{fig:SA_GMM_5.0_150} but for the magnitude bin M5.5 and \(V_{S30} = 600\) m/s.}
    \label{fig:SA_GMM_5.5_600}
\end{figure}

\begin{figure}[!htp]
    \centering
    \includegraphics[width=0.99\textwidth]{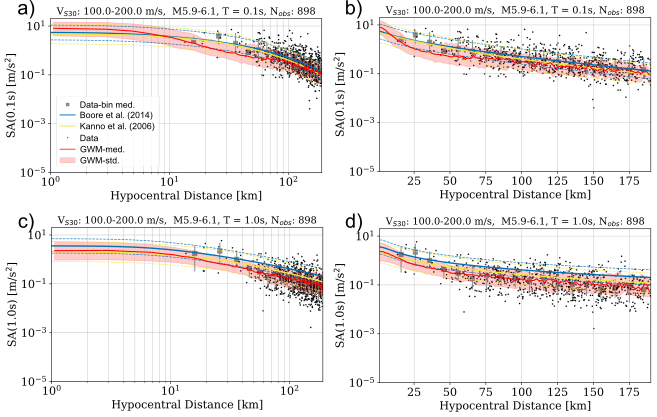}
    \caption{Same as Figure \ref{fig:SA_GMM_5.0_150} but for the magnitude bin M6.0 and \(V_{S30} = 150\) m/s.}
    \label{fig:SA_GMM_6.0_150}
\end{figure}

\begin{figure}[!htp]
    \centering
    \includegraphics[width=0.99\textwidth]{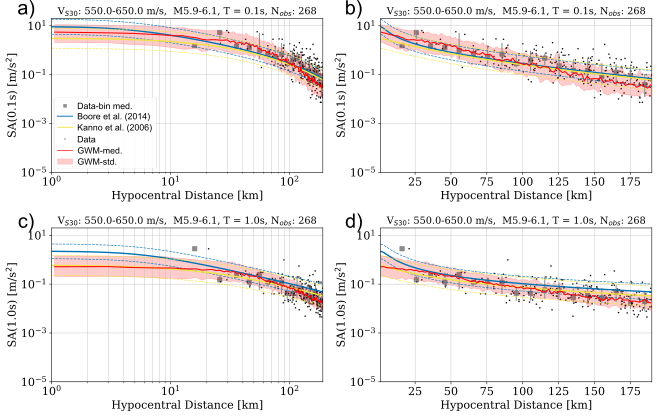}
    \caption{Same as Figure \ref{fig:SA_GMM_5.0_150} but for the magnitude bin M6.0 and \(V_{S30} = 600\) m/s.}
    \label{fig:SA_GMM_6.0_600}
\end{figure}

\begin{figure}[!htp]
    \centering
    \includegraphics[width=0.99\textwidth]{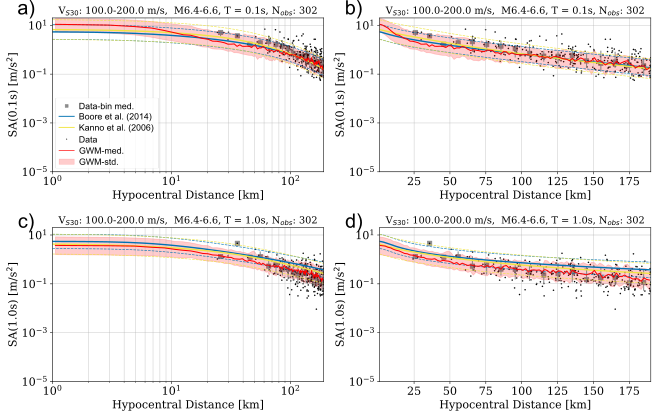}
    \caption{Same as Figure \ref{fig:SA_GMM_5.0_150} but for the magnitude bin M6.5 and \(V_{S30} = 150\) m/s.}
    \label{fig:SA_GMM_6.5_150}
\end{figure}

\begin{figure}[!htp]
    \centering
    \includegraphics[width=0.99\textwidth]{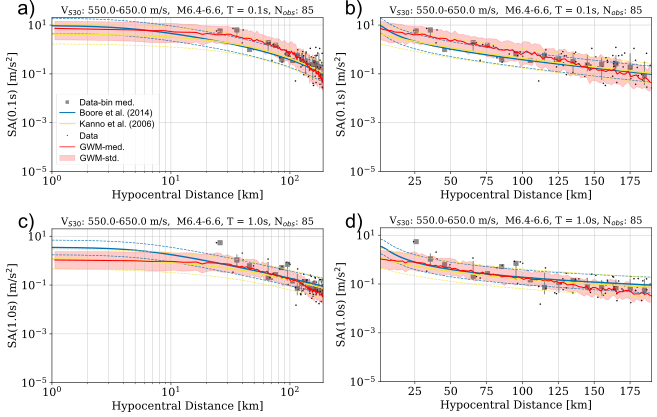}
    \caption{Same as Figure \ref{fig:SA_GMM_5.0_150} but for the magnitude bin M6.5 and \(V_{S30} = 600\) m/s.}
    \label{fig:SA_GMM_6.5_600}
\end{figure}

\begin{figure}[!htp]
    \centering
    \includegraphics[width=0.99\textwidth]{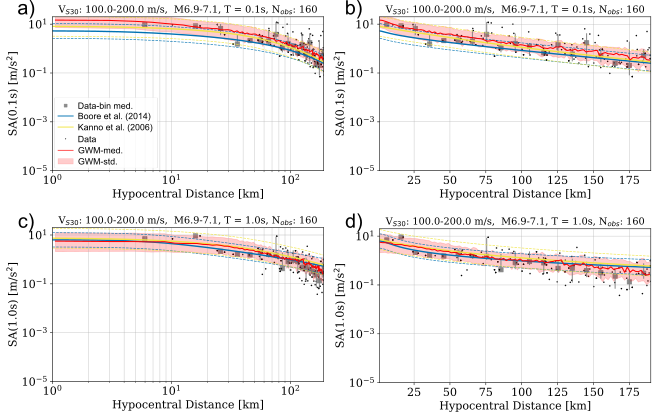}
    \caption{Same as Figure \ref{fig:SA_GMM_5.0_150} but for the magnitude bin M7.0 and \(V_{S30} = 150\) m/s.}
    \label{fig:SA_GMM_7.0_150}
\end{figure}

\begin{figure}[!htp]
    \centering
    \includegraphics[width=0.99\textwidth]{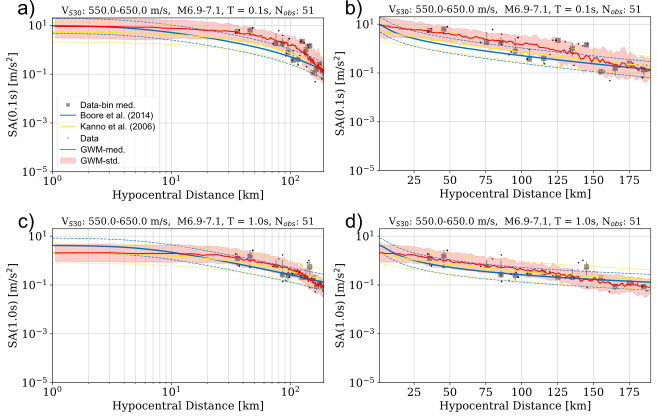}
    \caption{Same as Figure \ref{fig:SA_GMM_5.0_150} but for the magnitude bin M7.0 and \(V_{S30} = 600\) m/s.}
    \label{fig:SA_GMM_7.0_600}
\end{figure}

\begin{figure}[!htp]
    \centering
    \includegraphics[width=0.9\textwidth]{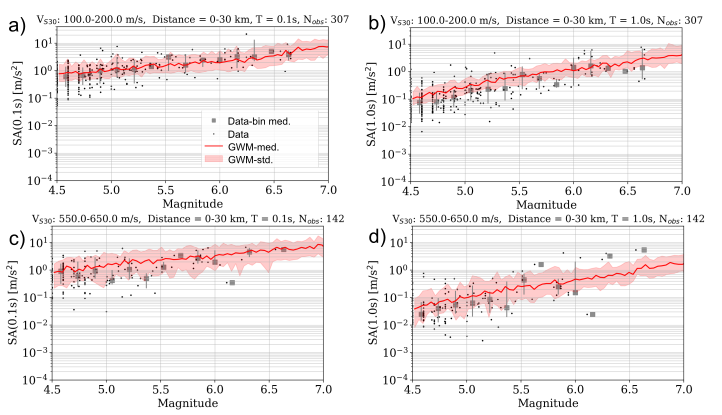}
    \caption{RotD50 pseudo-spectral acceleration (SA) with a damping factor of 5\% versus magnitude. Median of the GWM prediction (red lines) and its standard deviation (red shaded areas), using $R = 15$ km. Panels a) and c) show RotD50 pseudo-spectral acceleration for \(V_{S30} = 150\) m/s at periods of 0.1 s and 1.0 s, respectively. Panels b) and d) show RotD50 pseudo-spectral acceleration for \(V_{S30} = 600\) m/s at periods of 0.1 s and 1.0 s, respectively. The data (grey dots) are sampled from narrow magnitude, $R$, and $V_{S30}$ bins, as written in the figure titles.}
    \label{fig:SA_magnitude_0-30}
\end{figure}

\begin{figure}[!htp]
    \centering
    \includegraphics[width=0.9\textwidth]{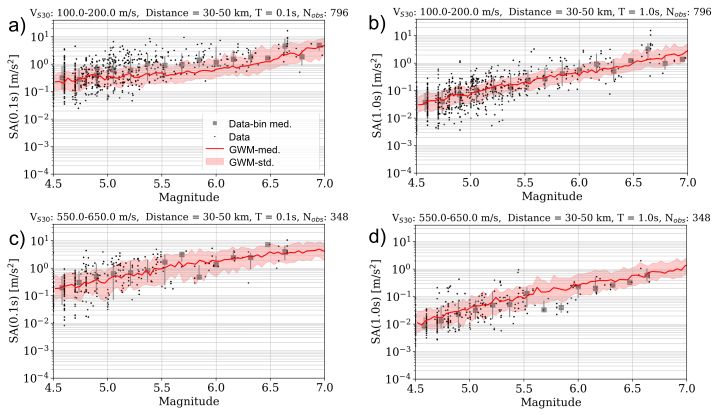}
    \caption{Same as Figure \ref{fig:SA_magnitude_0-30} a distance bin of 40 km.}
    \label{fig:SA_magnitude_30-50}
\end{figure}

\begin{figure}[!htp]
    \centering
    \includegraphics[width=0.9\textwidth]{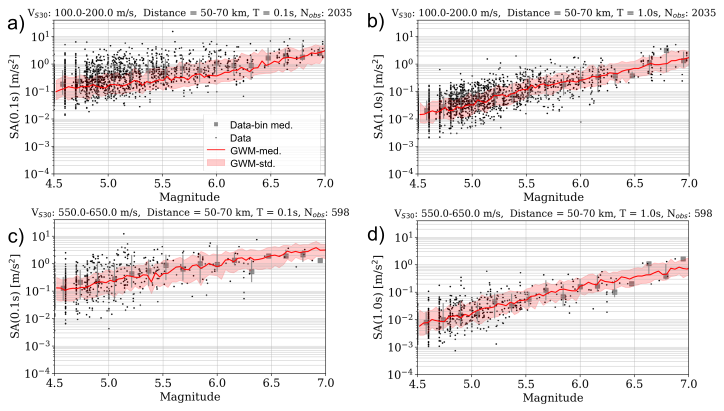}
    \caption{Same as Figure \ref{fig:SA_magnitude_0-30} a distance bin of 60 km.}
    \label{fig:SA_magnitude_50-70}
\end{figure}

\begin{figure}[!htp]
    \centering
    \includegraphics[width=0.9\textwidth]{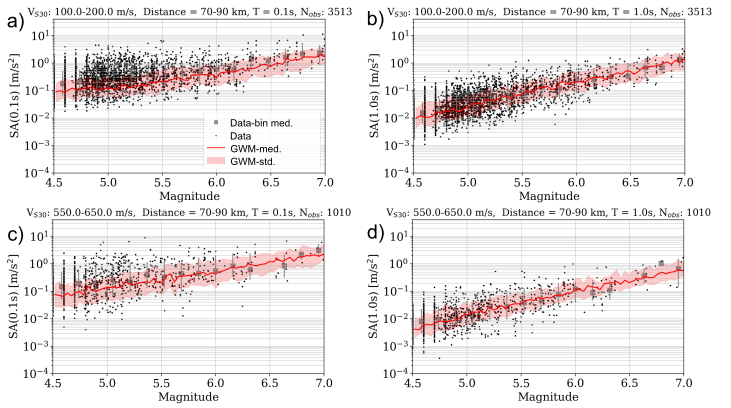}
    \caption{Same as Figure \ref{fig:SA_magnitude_0-30} a distance bin of 80 km.}
    \label{fig:SA_magnitude_70-90}
\end{figure}

\begin{figure}[!htp]
    \centering
    \includegraphics[width=0.9\textwidth]{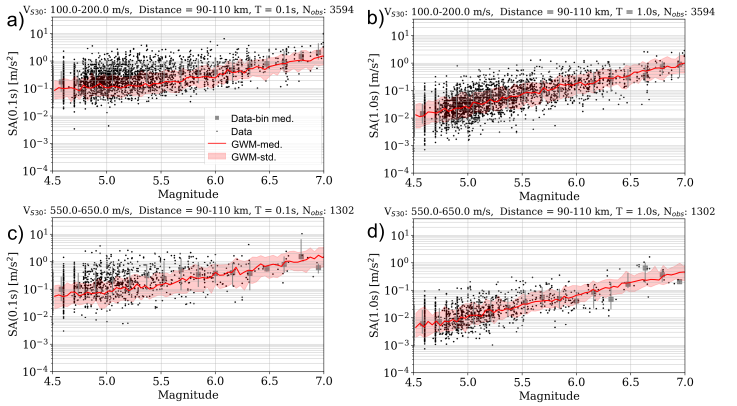}
    \caption{Same as Figure \ref{fig:SA_magnitude_0-30} a distance bin of 100 km.}
    \label{fig:SA_magnitude_90-110}
\end{figure}

\begin{figure}[!htp]
    \centering
    \includegraphics[width=0.9\textwidth]{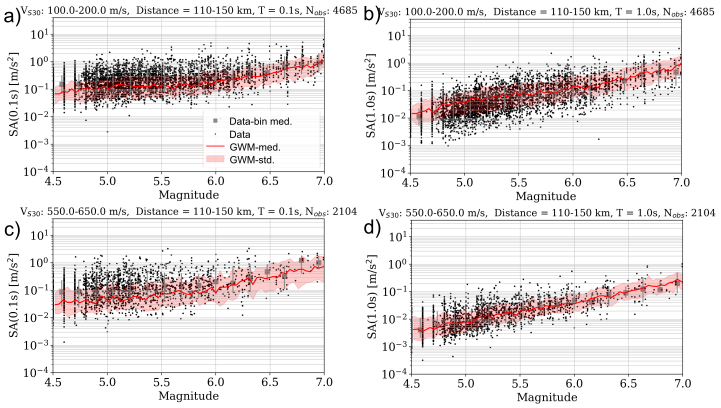}
    \caption{Same as Figure \ref{fig:SA_magnitude_0-30} a distance bin of 130 km.}
    \label{fig:SA_magnitude_110-150}
\end{figure}

\begin{figure}[!htp]
    \centering
    \includegraphics[width=0.99\textwidth]{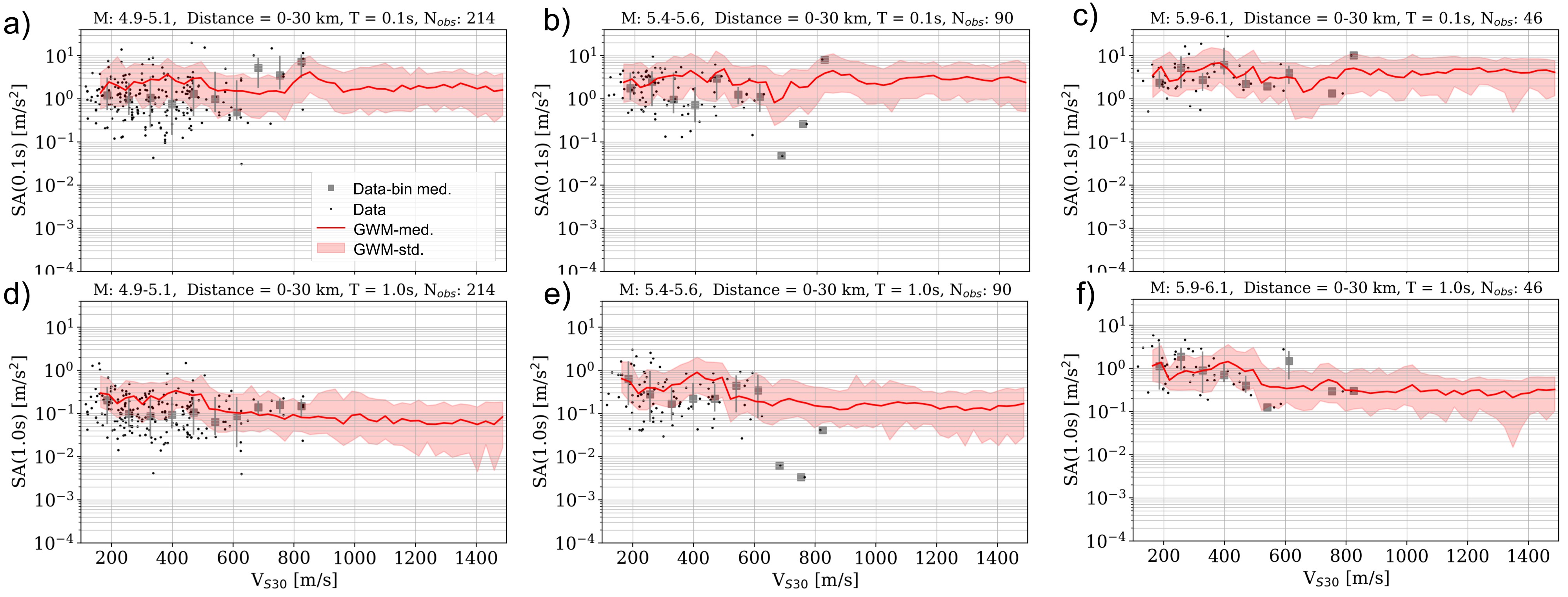}
    \caption{RotD50 pseudo-spectral acceleration (SA) with a damping factor of 5\% versus magnitude. Median of the GWM prediction (red) and its standard deviation (red shaded areas), using $R = 15$ km. Panels a) and d) show RotD50 pseudo-spectral acceleration for \(M5\) m/s at periods of 0.1 s and 1.0 s, respectively. Panels b) and e) show RotD50 pseudo-spectral acceleration for \(M5.5\) m/s at periods of 0.1 s and 1.0 s, respectively. Panels c) and f) show RotD50 pseudo-spectral acceleration for \(M6.0\) m/s at periods of 0.1 s and 1.0 s, respectively. The data (black dots) are sampled from narrow magnitude, $R$, and magnitude bins, as written in the figure titles.}
    \label{fig:SA_vs30_0-30}
\end{figure}

\begin{figure}[!htp]
    \centering
    \includegraphics[width=0.99\textwidth]{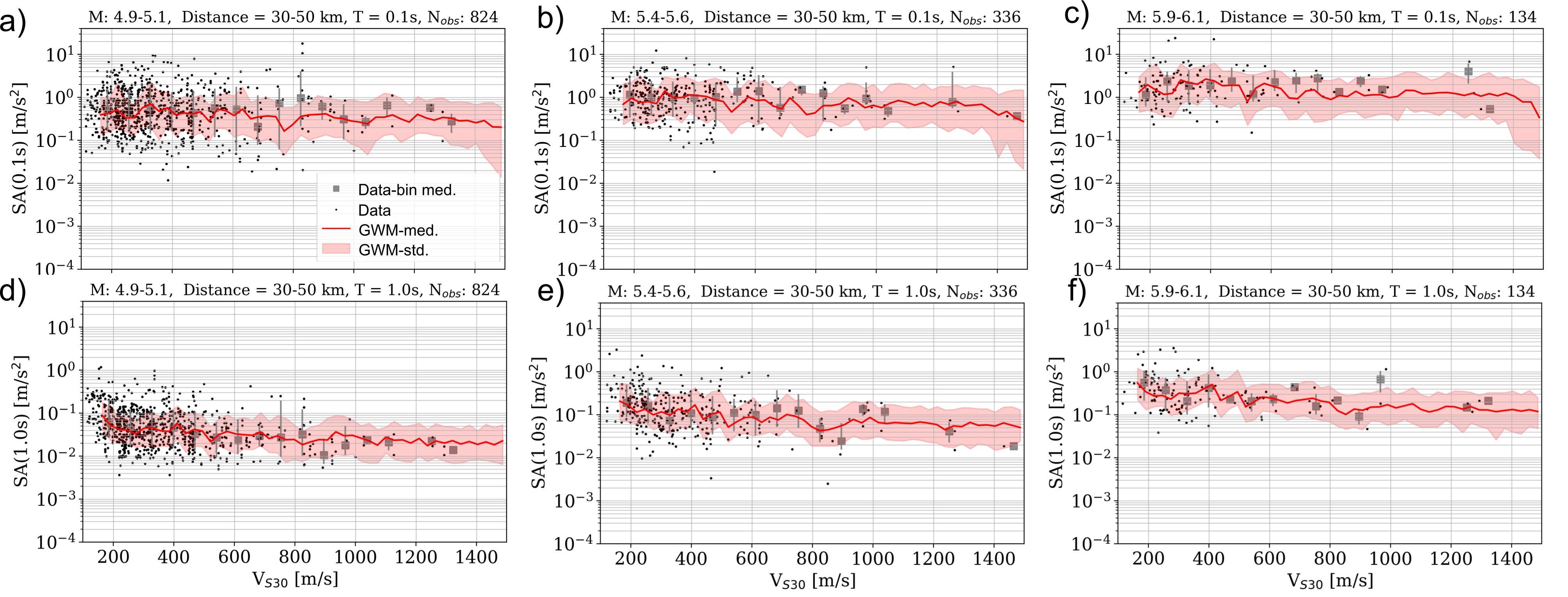}
    \caption{Same as Figure \ref{fig:SA_vs30_0-30} but with a distance bin of 40 km.}
    \label{fig:SA_vs30_30-50}
\end{figure}

\begin{figure}[!htp]
    \centering
    \includegraphics[width=0.99\textwidth]{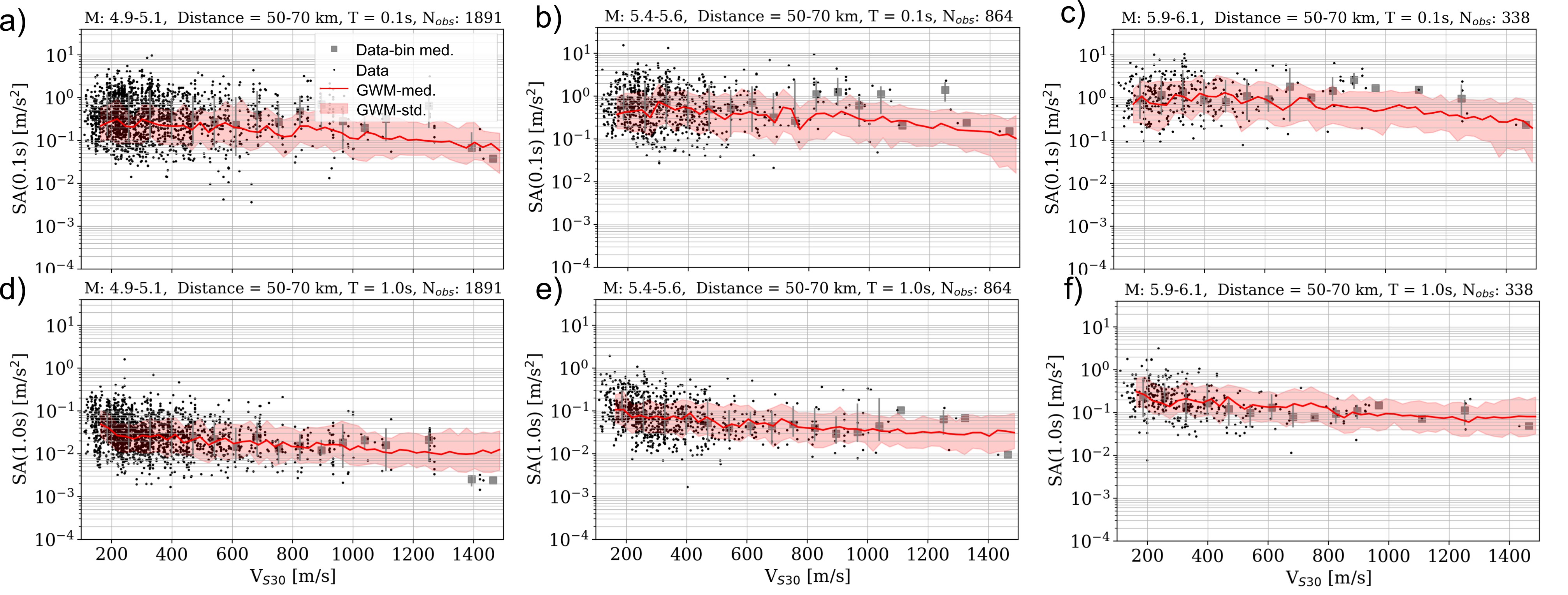}
    \caption{Same as Figure \ref{fig:SA_vs30_0-30} but with a distance bin of 60 km.}
    \label{fig:SA_vs30_50-70}
\end{figure}

\begin{figure}[!htp]
    \centering
    \includegraphics[width=0.99\textwidth]{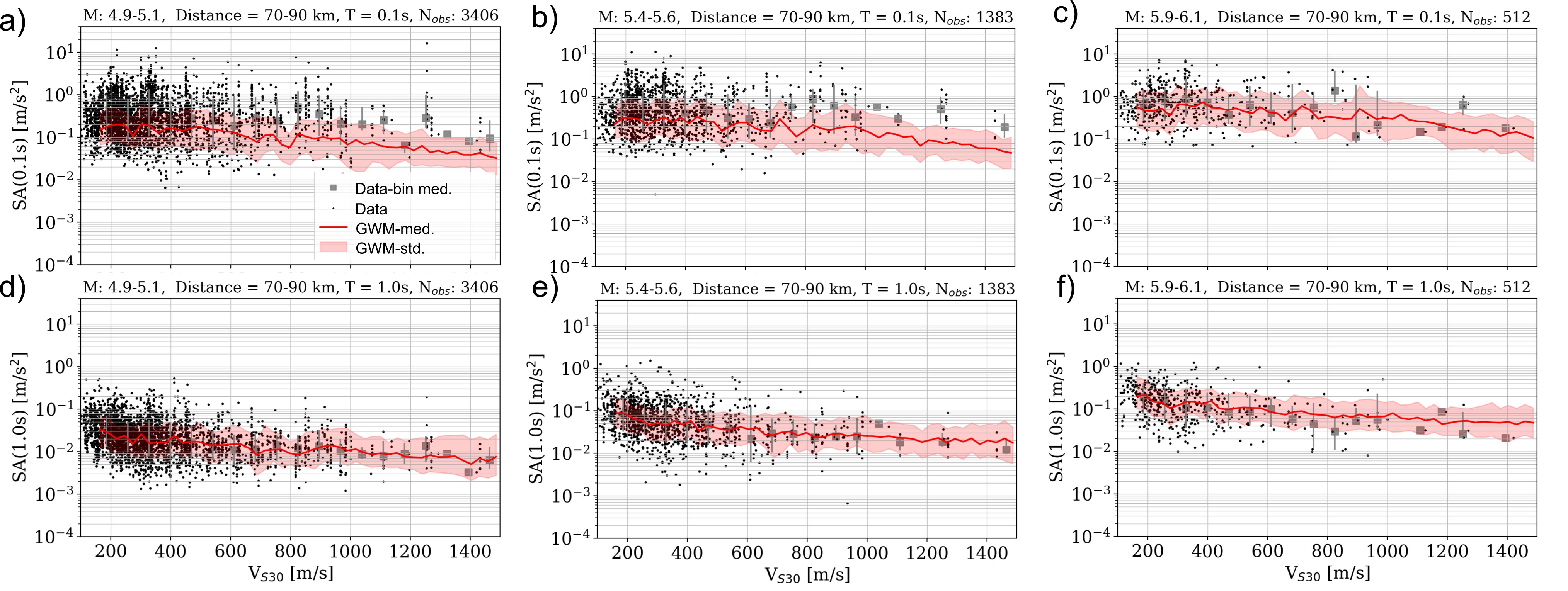}
    \caption{Same as Figure \ref{fig:SA_vs30_0-30} but with a distance bin of 80 km.}
    \label{fig:SA_vs30_70-90}
\end{figure}

\begin{figure}[!htp]
    \centering
    \includegraphics[width=0.99\textwidth]{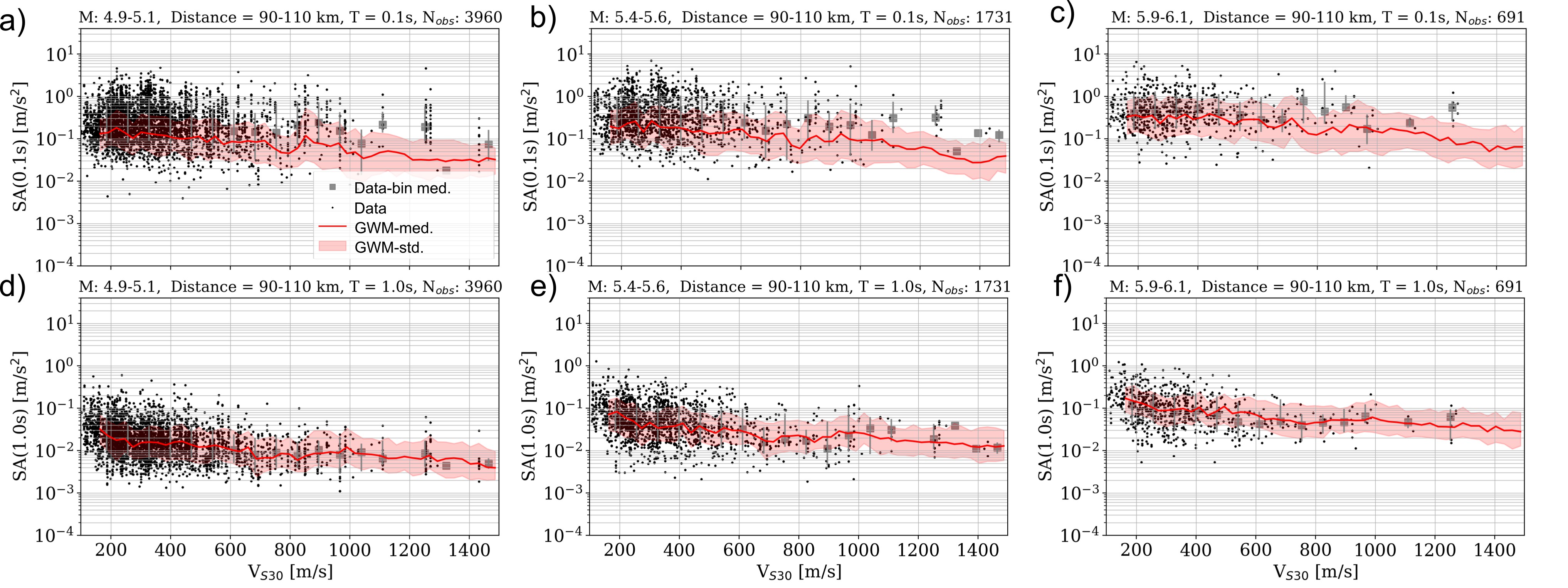}
    \caption{Same as Figure \ref{fig:SA_vs30_0-30} but with a distance bin of 100 km.}
    \label{fig:SA_vs30_90-110}
\end{figure}

\begin{figure}[!htp]
    \centering
    \includegraphics[width=0.99\textwidth]{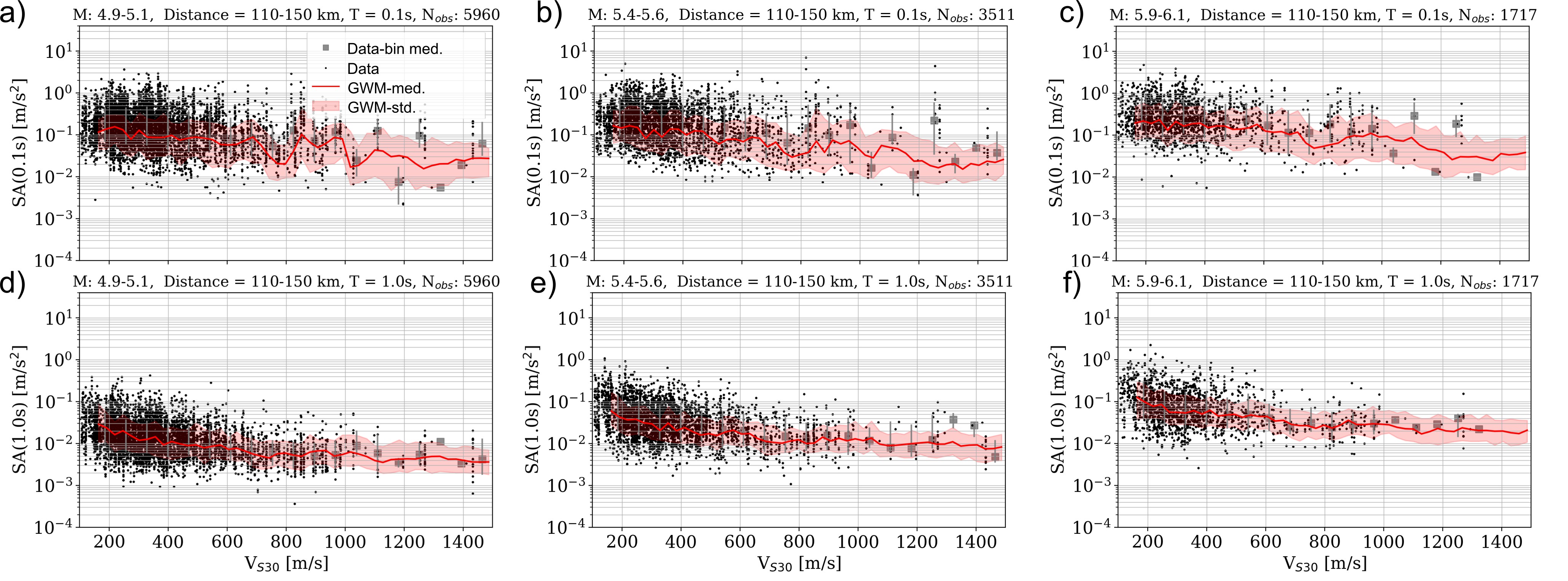}
    \caption{Same as Figure \ref{fig:SA_vs30_0-30} but with a distance bin of 130 km.}
    \label{fig:SA_vs30_110-150}
\end{figure}

\begin{figure}[!htp]
    \centering
    \includegraphics[width=0.99\linewidth]{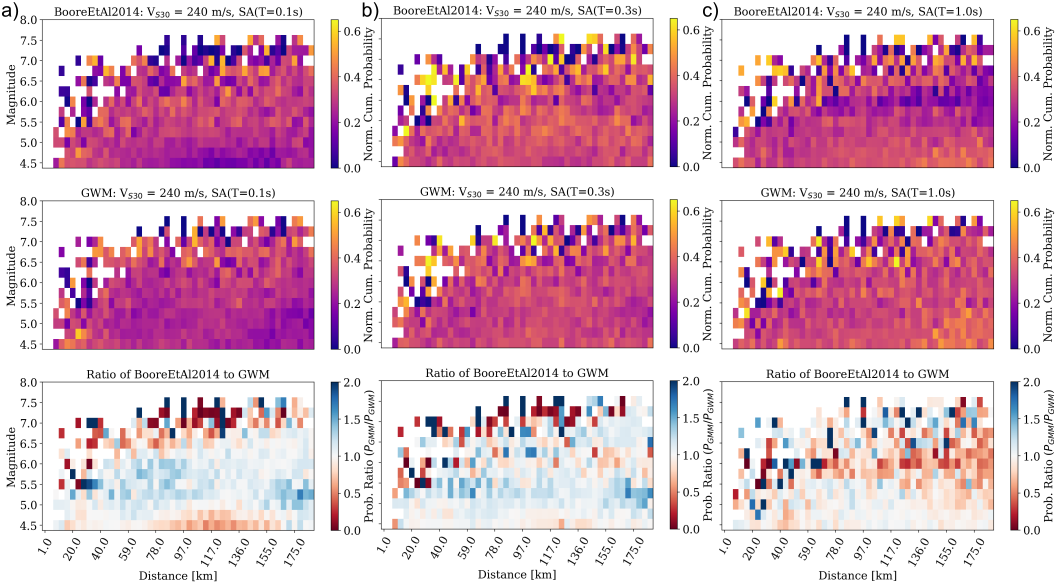}
    \caption{Average model probabilities given the SA data of the ground motion model (GMM) by \protect\citeA{boore2014nga}, generative waveform modeling (GWM), and the ratio between the two distributions given the data as a function of magnitude and recording distance for $V_{S30} = 240$ m/s. Panels a), b), and c) show the model likelihoods and their ratios at $T = 0.1$ s, $T = 0.3$ s, and $T = 1.0$ s, respectively.}
    \label{fig:prob_GMM_boore_240}
\end{figure}

\begin{figure}[!htp]
    \centering
    \includegraphics[width=0.99\linewidth]{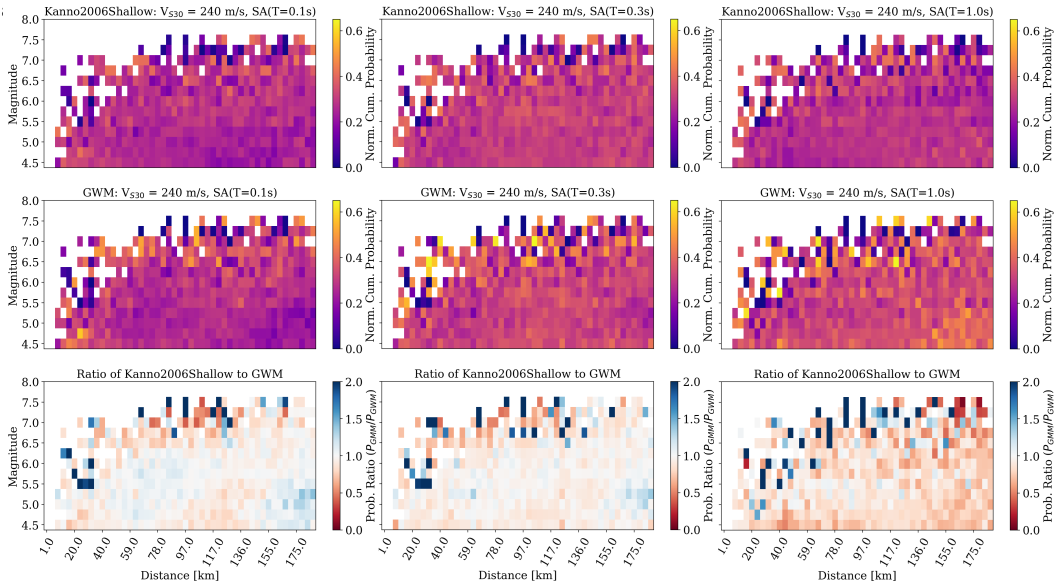}
    \caption{Same as Figure \ref{fig:prob_GMM_boore_240} but for GMM model of \protect\citeA{kanno2006new} for $V_{S30} = 240$ m/s. Panels a), b), and c) show the model likelihoods and their ratios at $T = 0.1$ s, $T = 0.3$ s, and $T = 1.0$ s, respectively.}
    \label{fig:prob_GMM_kanno_240}
\end{figure}

\begin{figure}[!htp]
    \centering
    \includegraphics[width=0.99\linewidth]{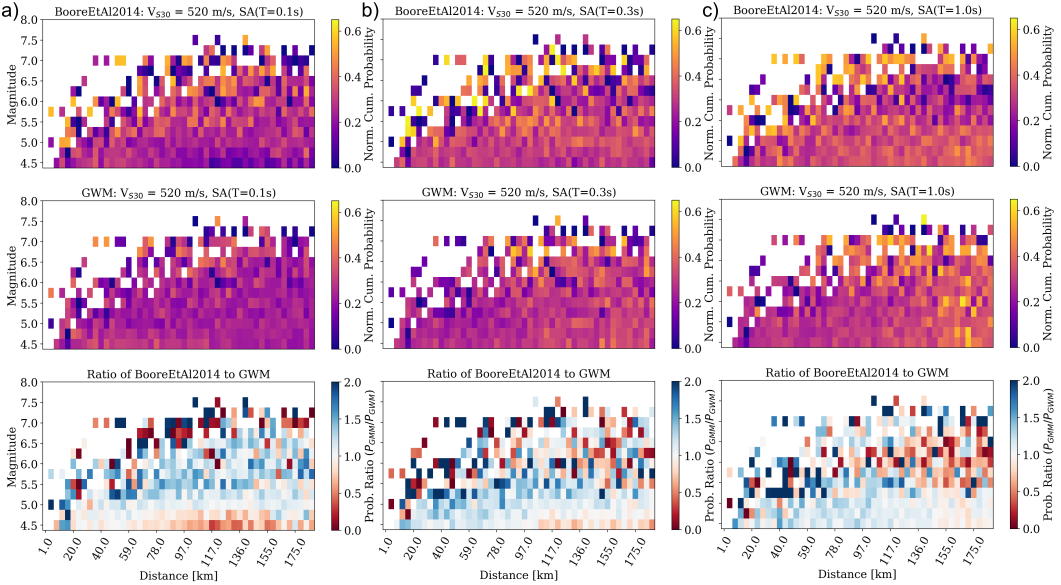}
    \caption{Same as Figure \ref{fig:prob_GMM_boore_240} but for $V_{S30} = 520$ m/s.}
    \label{fig:prob_GMM_boore_520}
\end{figure}

\begin{figure}[!htp]
    \centering
    \includegraphics[width=0.99\linewidth]{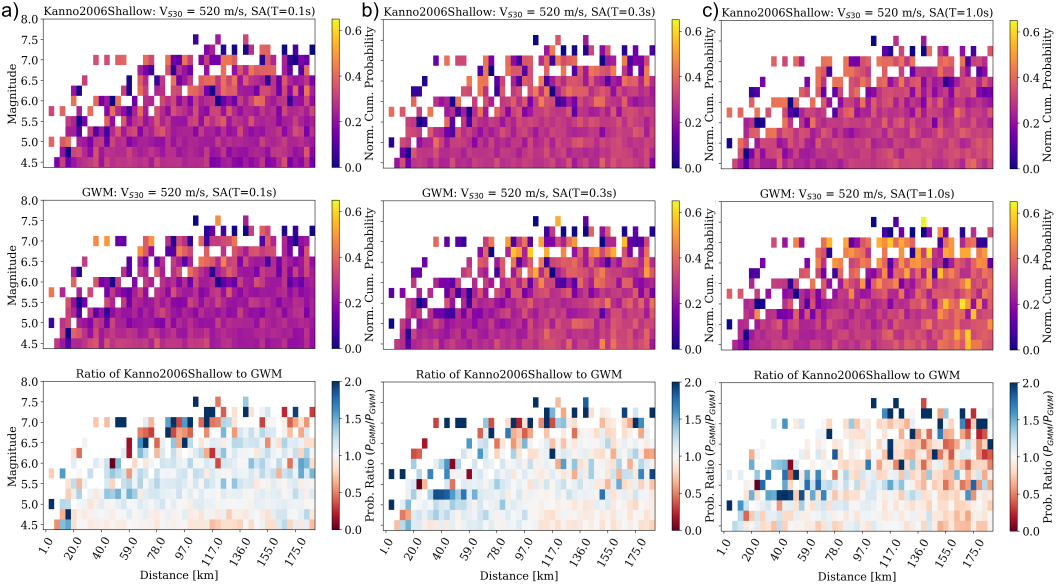}
    \caption{Same as Figure \ref{fig:prob_GMM_kanno_240} but for $V_{S30} = 520$ m/s.}
    \label{fig:prob_GMM_kanno_520}
\end{figure}

\begin{figure}[!htp]
    \centering
    \includegraphics[width=0.99\linewidth]{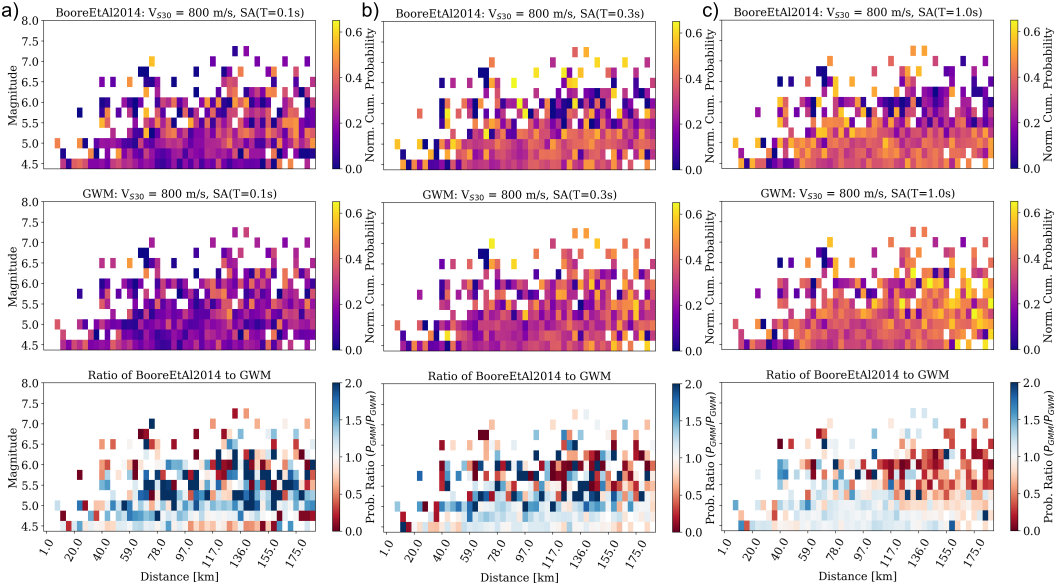}
    \caption{Same as Figure \ref{fig:prob_GMM_boore_240} but for $V_{S30} = 800$ m/s.}
    \label{fig:prob_GMM_boore_800}
\end{figure}

\begin{figure}[!htp]
    \centering
    \includegraphics[width=0.99\linewidth]{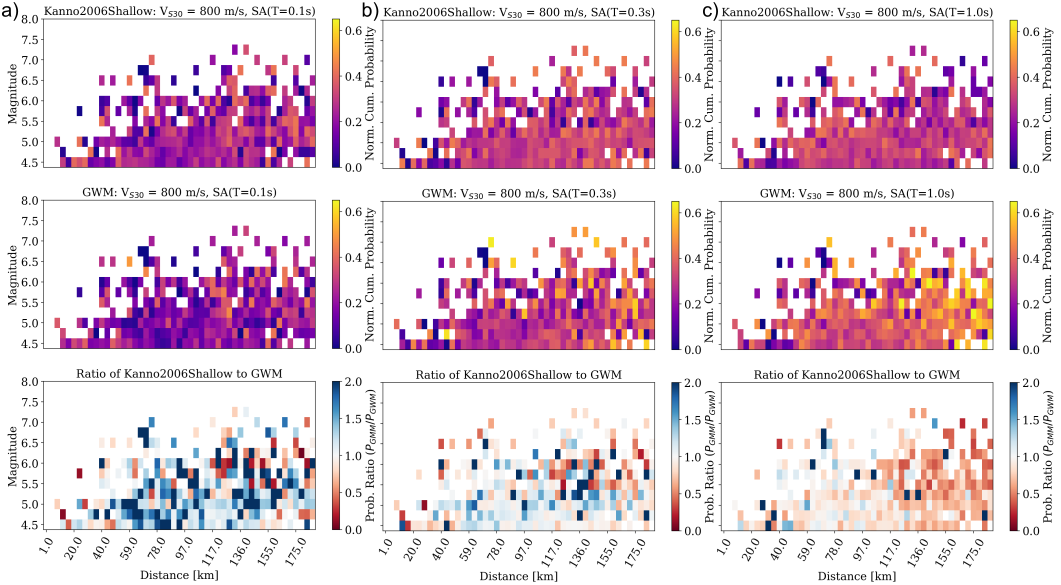}
    \caption{Same as Figure \ref{fig:prob_GMM_kanno_240} but for $V_{S30} = 800$ m/s.}
    \label{fig:prob_GMM_kanno_800}
\end{figure}

\begin{figure}[!htp]
    \centering
    \includegraphics[width=0.99\linewidth]{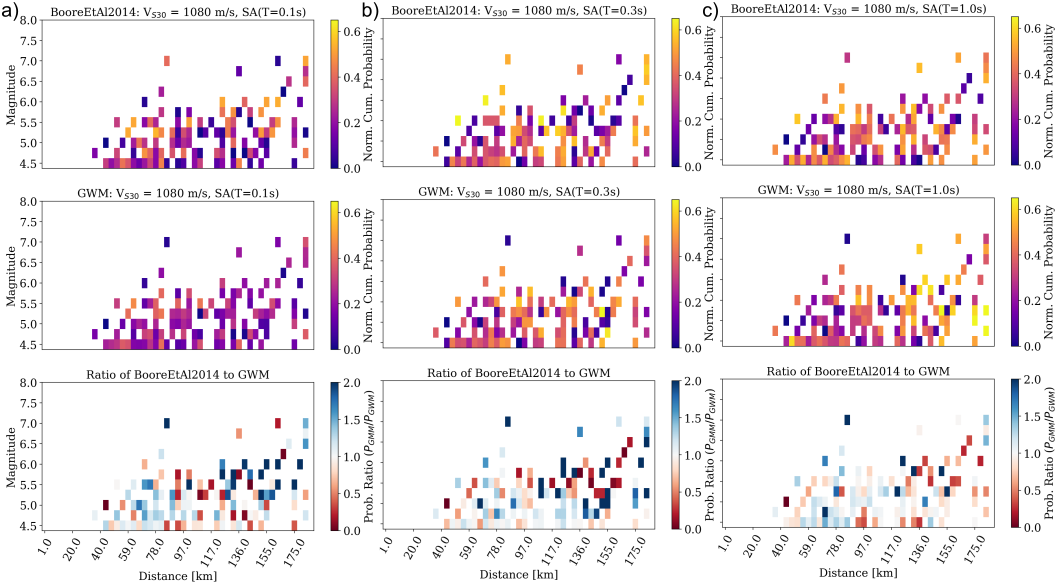}
    \caption{Same as Figure \ref{fig:prob_GMM_boore_240} but for $V_{S30} = 1080$ m/s.}
    \label{fig:prob_GMM_boore_1080}
\end{figure}

\begin{figure}[!htp]
    \centering
    \includegraphics[width=0.99\linewidth]{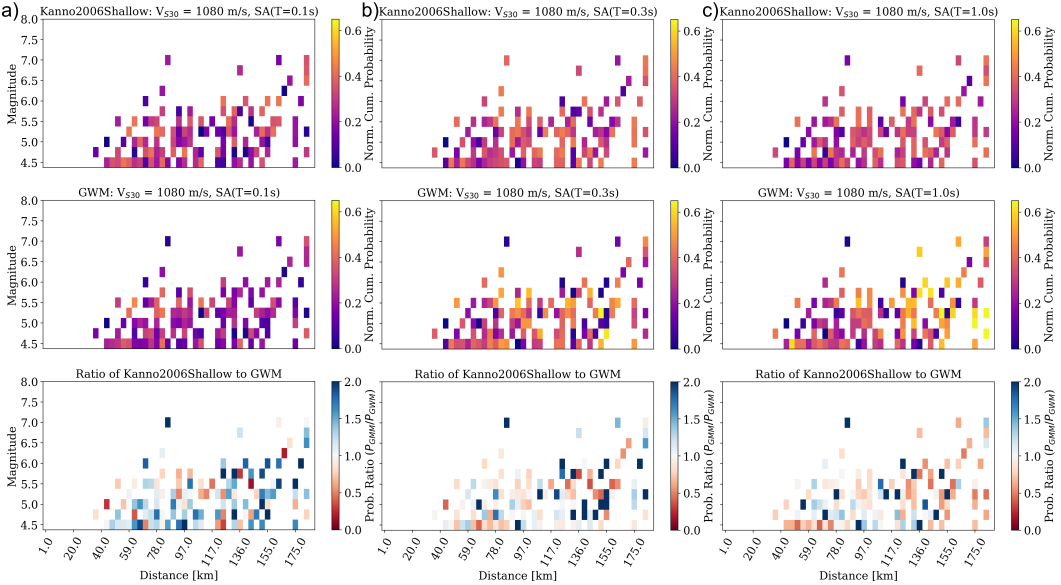}
    \caption{Same as Figure \ref{fig:prob_GMM_kanno_240} but for $V_{S30} = 1080$ m/s.}
    \label{fig:prob_GMM_kanno_1080}
\end{figure}

\begin{figure}[!htp]
    \centering
    \includegraphics[width=0.99\linewidth]{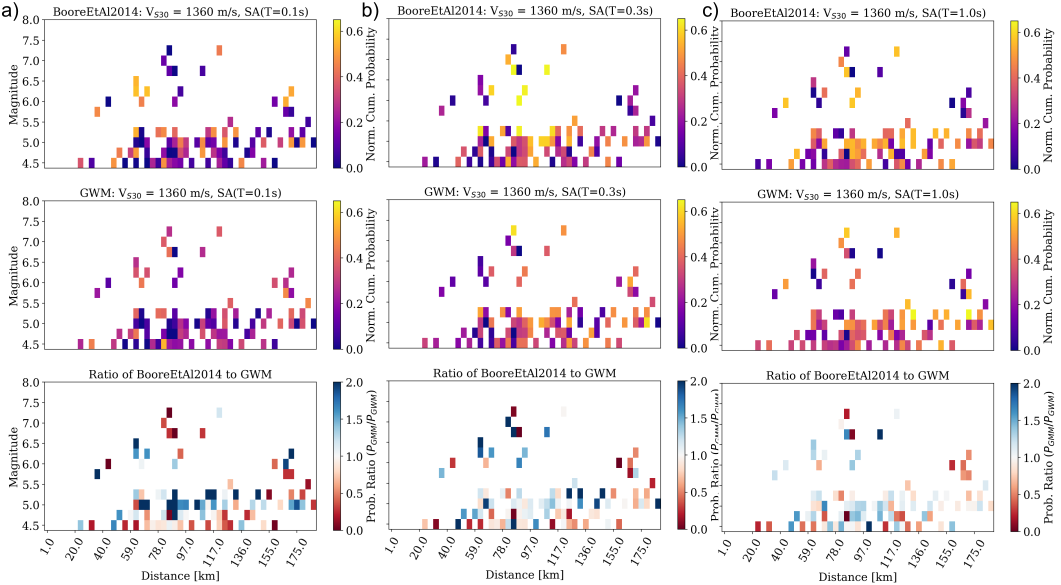}
    \caption{Same as Figure \ref{fig:prob_GMM_boore_240} but for $V_{S30} = 1360$ m/s.}
    \label{fig:prob_GMM_boore_1360}
\end{figure}

\begin{figure}[!htp]
    \centering
    \includegraphics[width=0.99\linewidth]{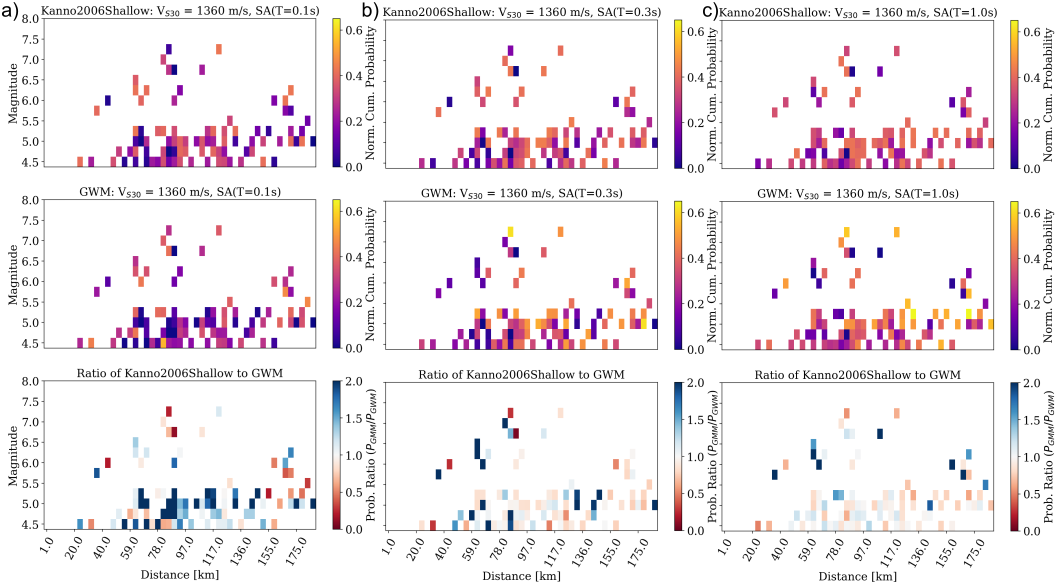}
    \caption{Same as Figure \ref{fig:prob_GMM_kanno_240} but for $V_{S30} = 1360$ m/s.}
    \label{fig:prob_GMM_kanno_1360}
\end{figure}

\begin{figure}[!htp]
    \centering
    \includegraphics[width=0.7\linewidth]{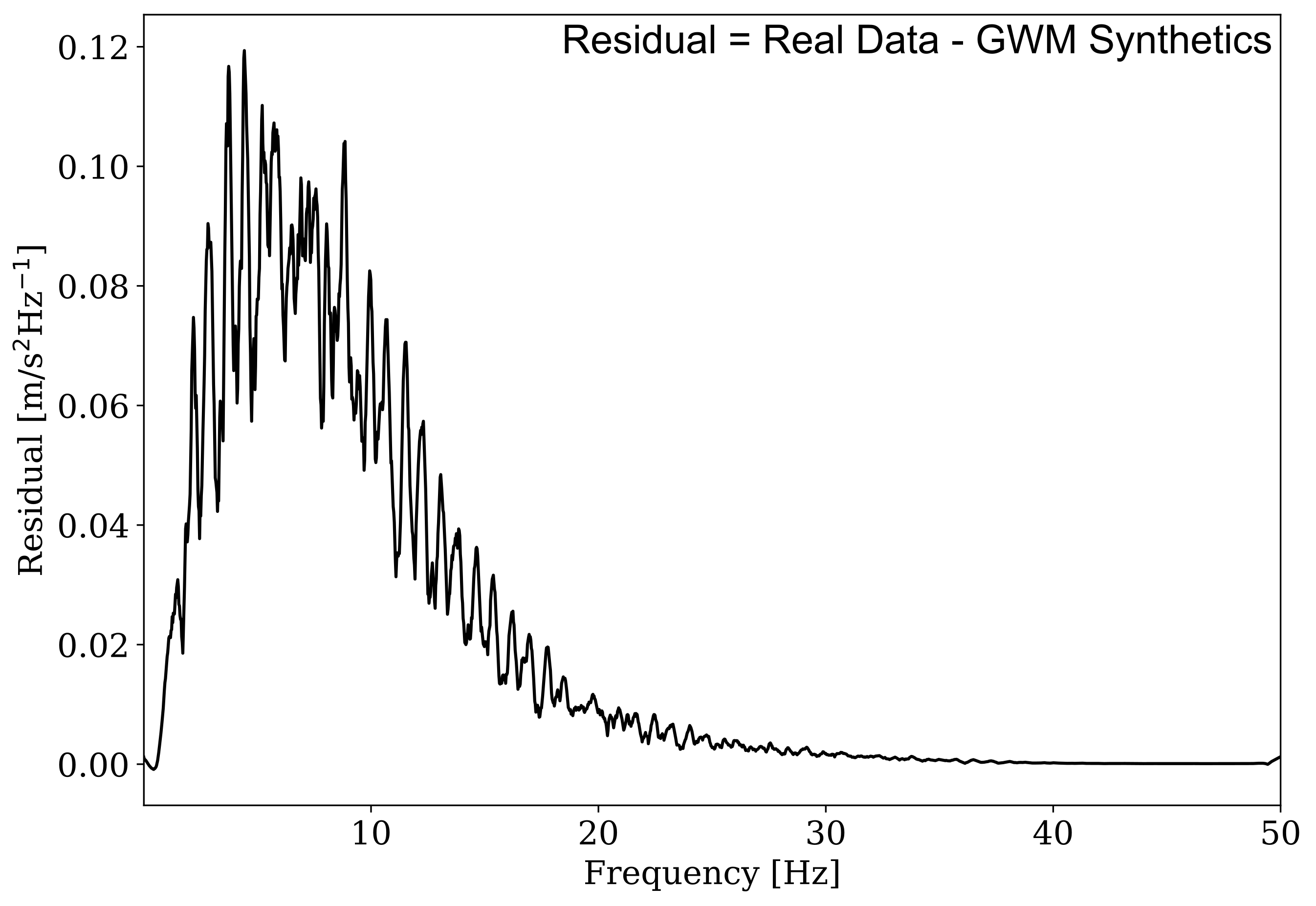}
    \caption{Residual of the spectral mean amplitude between real data and GWM for all records.}
    \label{fig:amp_residual}
\end{figure}

\begin{figure}[!htp]
    \centering
    \begin{subfigure}[b]{0.32\textwidth}
        \includegraphics[width=\textwidth]{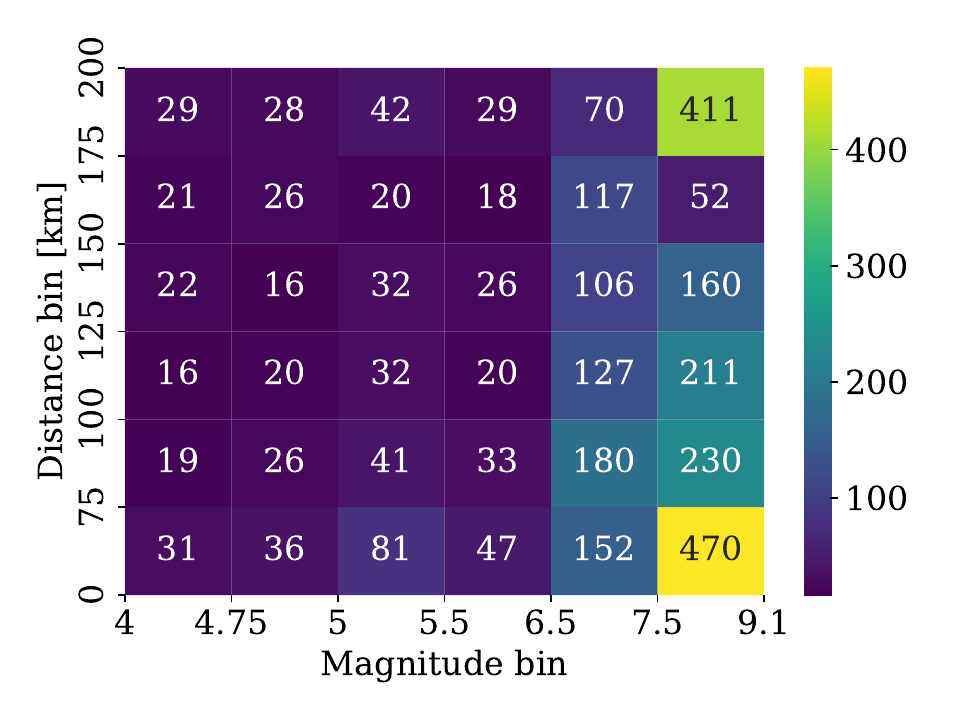}
        \caption{East-West}
    \end{subfigure}
    \begin{subfigure}[b]{0.32\textwidth}
        \includegraphics[width=\textwidth]{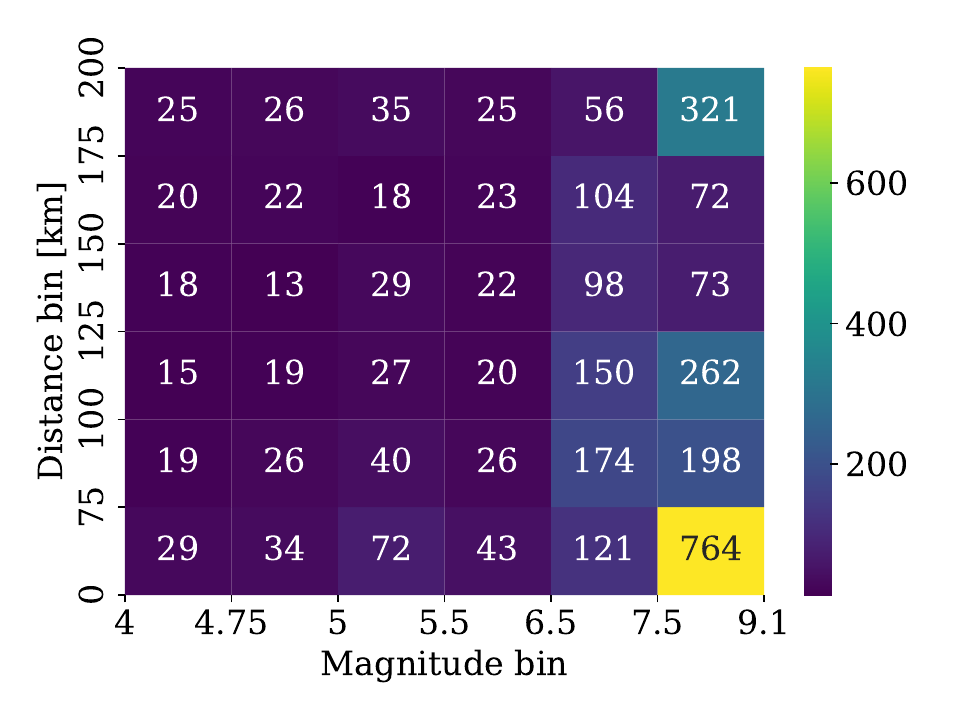}
        \caption{North-South}
    \end{subfigure}
    \begin{subfigure}[b]{0.32\textwidth}
        \includegraphics[width=\textwidth]{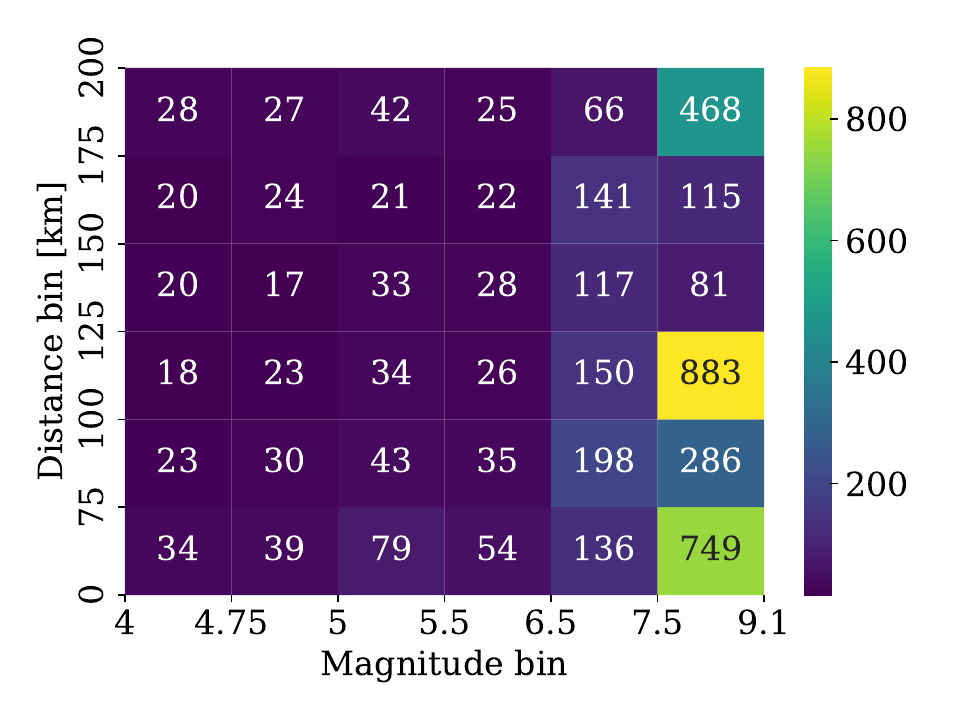}
        \caption{Vertical}
    \end{subfigure}
    \caption{Log-amplitude Fourier spectra Fréchet Distance heatmaps for all three components in different magnitude and distance bins.}
    \label{fig:supp_asd}
\end{figure}

\noindent\textbf{The content of the supplementary figures is listed as follows:}
\begin{enumerate}
\item Figures \ref{fig:supp_grid_plot} to \ref{fig:supp_asd} for supplementary of the maintext.
\item Figures \ref{fig:remove_vs30_rhyp} to \ref{fig:station_vs_vs30} for Text S1.
\item Figures \ref{fig:statistics_onset} to \ref{fig:SNR_total} for Text S2.
\item Figures \ref{fig:spectral_envelope} to \ref{fig:waveform_stead_bad} for Text S3.
\end{enumerate}

\section*{Introduction}
This supplementary document expands on the main paper by providing additional information and figures that deepen understanding and further validate the results. It offers visual evidence that supports and extends the findings discussed in the main text.

\subsection*{Additional Evaluation Figures}

Side-by-side comparisons of the generative waveform model (GWM) and the
observed data are provided, binned by magnitude, hypocentral distance,
\(V_{S30}\), hypocentral depth, and azimuthal gap:

\begin{enumerate}
  \item Time-domain signal envelopes
  \item Fourier amplitude spectra
  \item Shaking-duration statistics for all magnitudes
  \item Pseudo-spectral acceleration at periods \(T = 0.1,\ 0.3,\) and \(1.0\ \text{s}\) versus distance
  \item Pseudo-spectral acceleration versus magnitude
  \item Pseudo-spectral acceleration versus \(V_{S30}\)
  \item Average model probabilities across magnitude–distance bins
  \item Residuals of mean spectral amplitude
\end{enumerate}

\subsection*{Additional Texts}

The supplement also contains four short sections that describe:
\begin{itemize}
  \item Text S1: the preprocessing steps used in the main study.
  \item Text S2: the data distribution.
  \item Text S3: the reproducibility workflow based on the Stanford Earthquake Dataset (STEAD).
  \item Text S4: the comparison of the HighFEM with the GANO model.
\end{itemize}

\clearpage

\section*{Figures:}

\begin{figure}[!htp]
    \centering
    \includegraphics[width=0.9\textwidth]{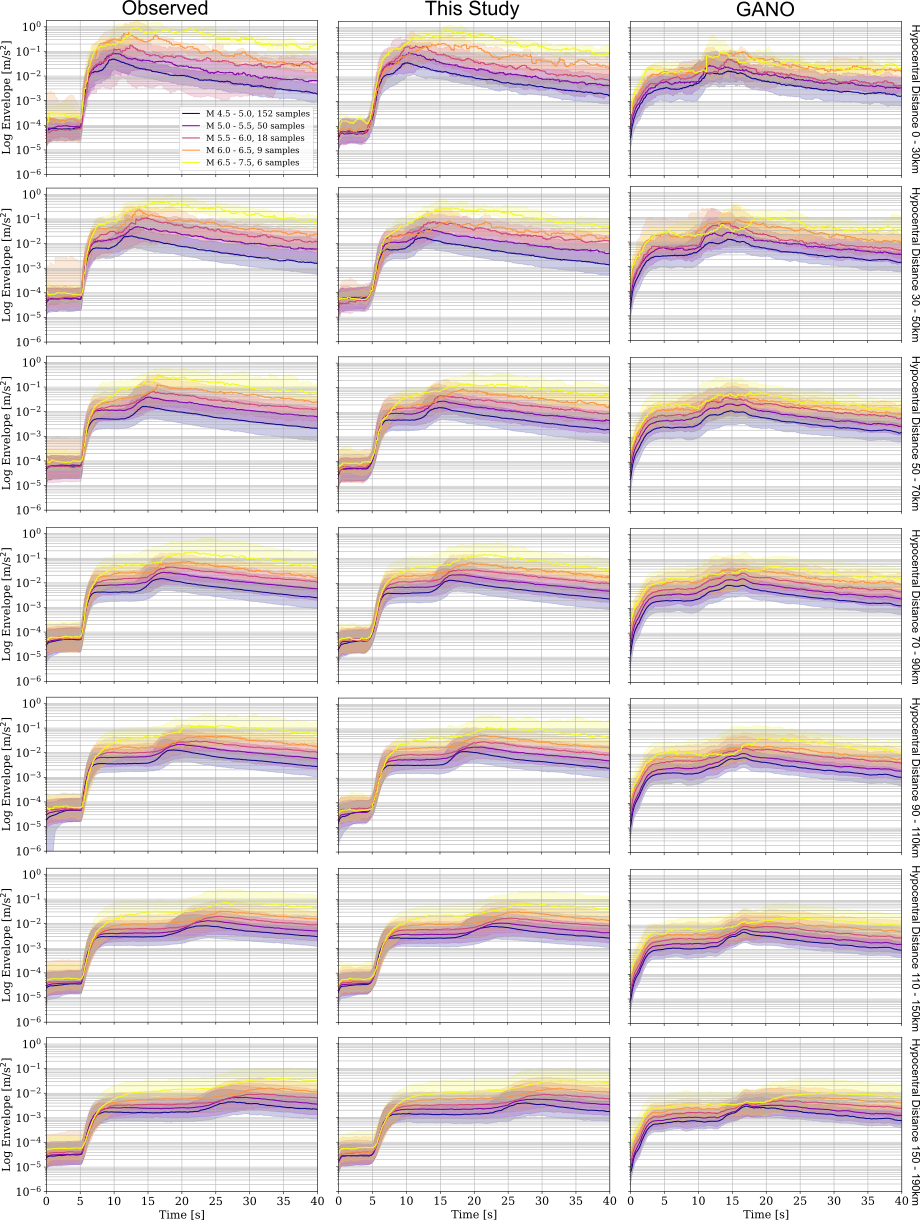}
    \caption{Distribution of time-domain envelopes for Radial-component seismograms in different magnitude and distance bins. Observed, this study, and GANO denote real data, our GWM, and GANO}
    \label{fig:supp_envelope_gano_0}
\end{figure}

\begin{figure}[!htp]
    \centering
    \includegraphics[width=0.9\textwidth]{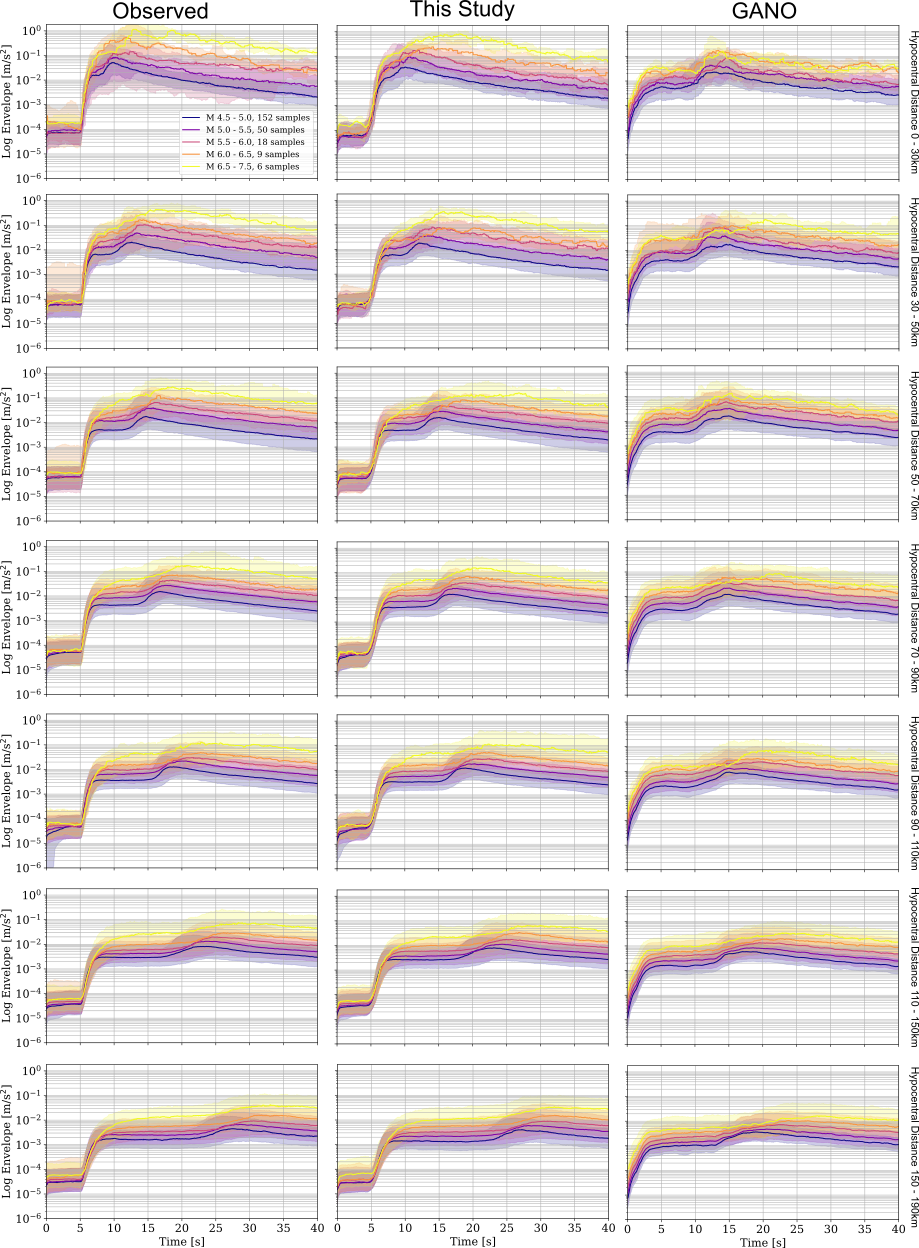}
    \caption{Distribution of time-domain envelopes for Transverse-component seismograms in different magnitude and distance bins. Observed, this study, and GANO denote real data, our GWM, and GANO.}
    \label{fig:supp_envelope_gano_1}
\end{figure}

\begin{figure}[!htp]
    \centering
    \includegraphics[width=0.9\textwidth]{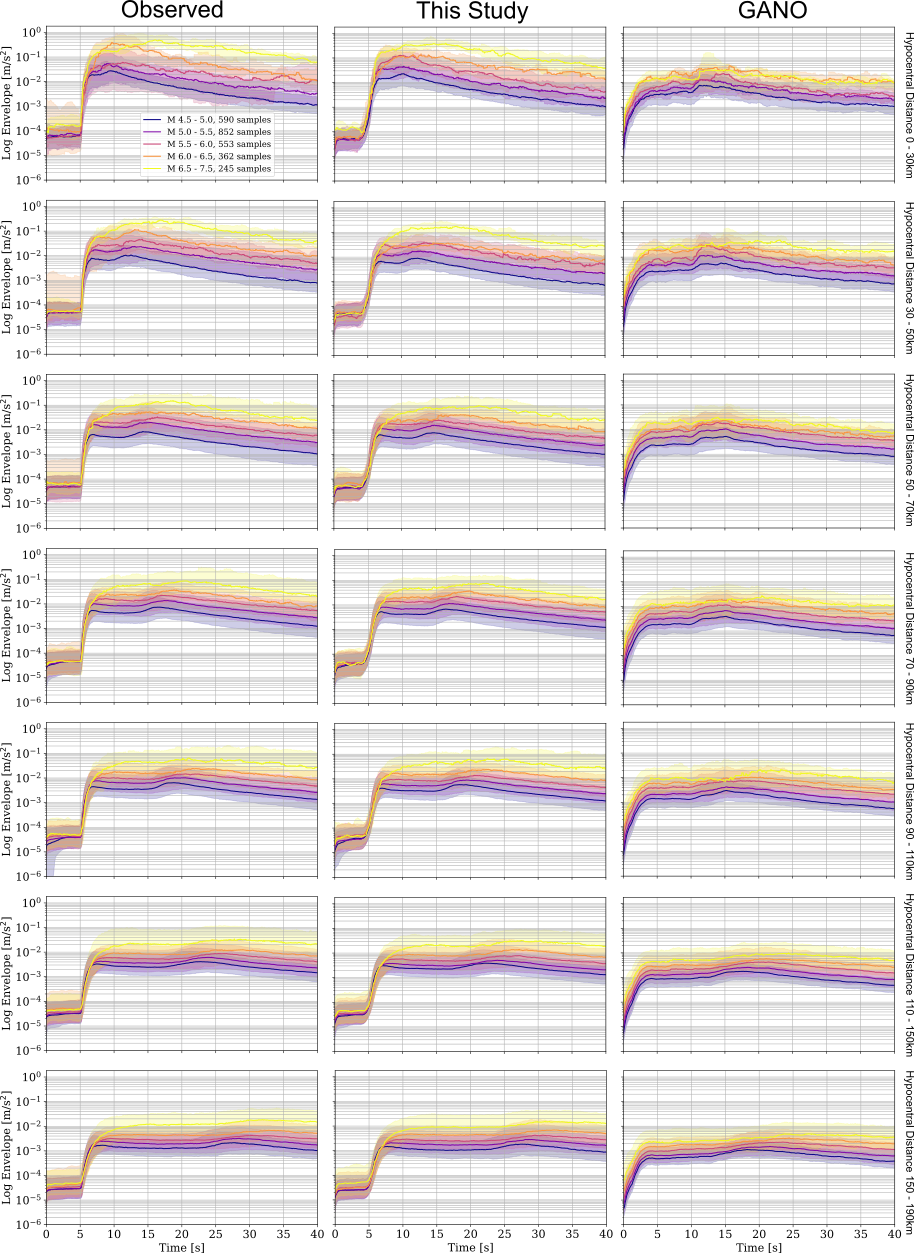}
    \caption{Distribution of time-domain envelopes for Vertical-component seismograms in different magnitude and distance bins. Observed, this study, and GANO denote real data, our GWM, and GANO.}
    \label{fig:supp_envelope_gano_2}
\end{figure}

\begin{figure}[!htp]
    \centering
    \includegraphics[width=0.9\textwidth]{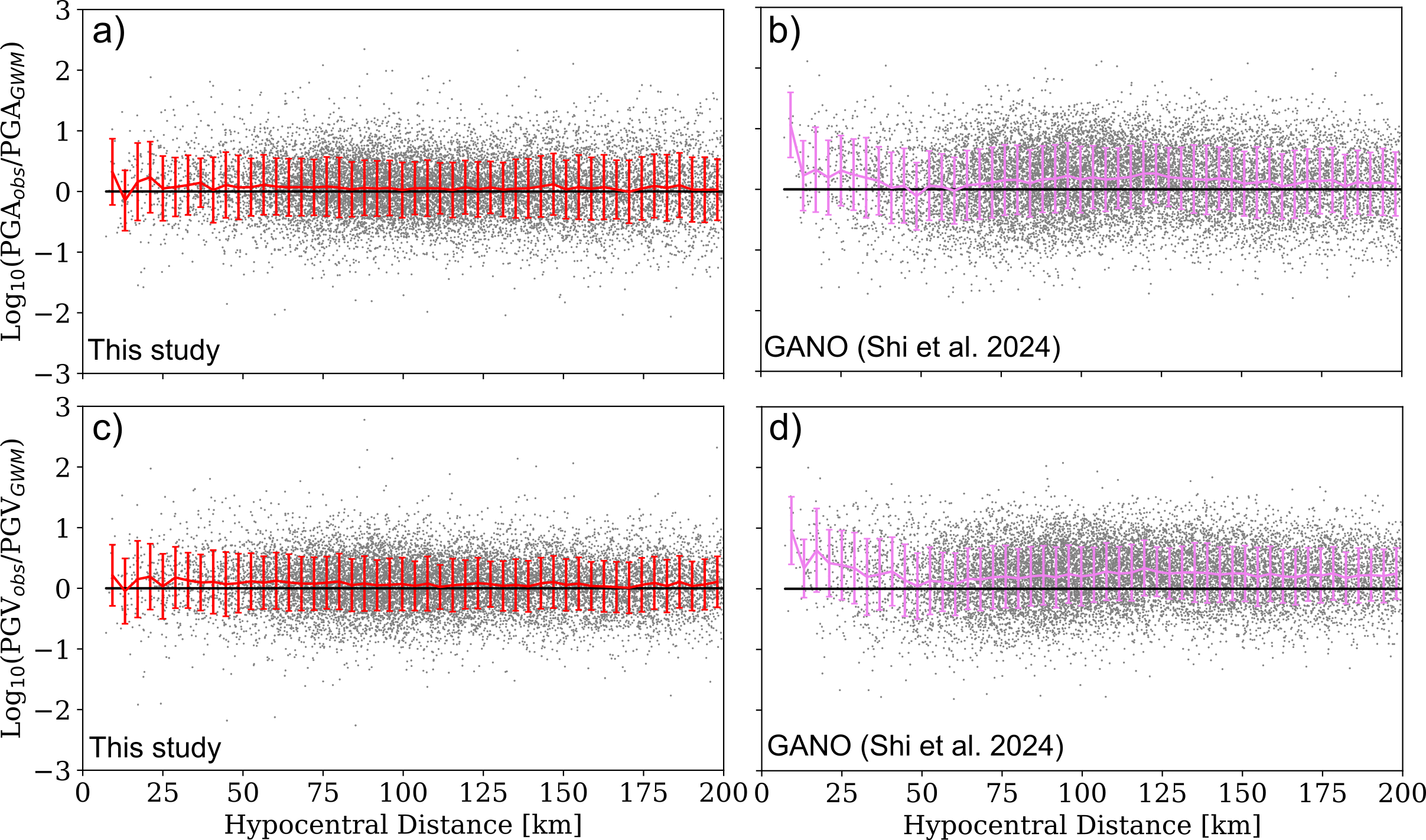}
    \caption{Model bias as a function of hypocentral distance for the generative waveform model (GWM) in this study (red) and GWM from \protect\citeA{shi2024broadband} (violet) for PGA (a and b) and PGV (b and d), with respect to real data. Colored lines represent the mean of the ratio in 50 distance bins of 3.67 km width. The bars represent the standard deviation in each bin.}
    \label{fig:bias_gano}
\end{figure}

\clearpage

\section*{Text S1: Preprocessing Steps for Data Preparation}

A clean dataset is essential for generating accurate models and results representative of the training data. We perform step-by-step data preprocessing and filtering. Below, we detail our approach:

\begin{itemize}
    \item Create a list of records without initial processing, and download the dataset from \url{https://www.kyoshin.bosai.go.jp/}.
    \begin{enumerate}
        \item Save a list of records for each earthquake, matching metadata information from CMT, F-NET, and JMA datasets.
        \item For all events, store all information in a \texttt{*.h5} file.
        \item Automatically pick waveforms from the original data using STA/LTA for P-phase picking.
    \end{enumerate}

    \item Use the Python script \url{01_preprocess.py} to load and preprocess data for all events with existing headers in a \texttt{*.h5} file from the previous step.
    \begin{enumerate}
        \item Selection criteria:
        \begin{itemize}
            \item Minimum hypocentral distance: 0 km
            \item Maximum hypocentral distance: 200 km
            \item Minimum magnitude: 4.0
            \item Maximum magnitude: 9.1
            \item Minimum depth: 0 km
            \item Maximum depth: 100 km
        \end{itemize}
        \item Filter parameters:
        \begin{itemize}
            \item High-pass filter at 0.1 Hz
            \item Causal filter
            \item Filter order: 2
            \item Waveform length: 12,501 samples (125.01 s)
        \end{itemize}
        \item Waveforms are read using the Python module \texttt{obspy} in KNET format and calibrated accordingly.
        \item Each waveform is detrended before filtering. Each event generates a 3D matrix with dimensions $n_t$ time steps, $n_{rec}$ records, and $n_c = 3$ components.
        \item Save all processed data as \texttt{*.h5} files.
    \end{enumerate}

    \item Use \url{02_extractMatFileWaveform.py} to load the \texttt{*.h5} files and prepare data in seisbench format (i.e., \url{metadata.csv} and \url{waveforms.h5}). This step also corrects \texttt{NaN} values. If \texttt{NaN} values exist, perform interpolation as follows:
    \begin{enumerate}
        \item Check the number of \texttt{NaN} values. If the number exceeds half of the data length, discard the waveform from training.
        \item Perform linear interpolation in the frequency domain using NumPy’s \texttt{interp}.
        \item Apply iterative spectral fitting, including:
        \begin{enumerate}
            \item Perform Fast Fourier Transform (FFT) on the interpolated signal.
            \item Identify and focus on the dominant waveform frequency.
            \item Perform inverse FFT.
            \item Enforce data consistency by preserving original non-\texttt{NaN} samples.
            \item Check convergence.
            \item Limit to 100 iterations or stop if tolerance falls below a predefined threshold (i.e., $1 \times 10^{-4}$).
        \end{enumerate}
        \item The total number of waveforms after this process is 384,896.
    \end{enumerate}

    \item Use \url{03_picking_save2training.py} to load waveforms in Python:
    \begin{enumerate}
        \item Pick phases using PhaseNet and a pretrained model from the Japan Meteorological Agency (JMA) \cite{naoi2024neural}.
        \item Read preprocessed data and metadata, then calculate the hypocenter position based on whether the event is within or outside Japan's island boundary, and include this as a new parameter.
        \item For each waveform copy, perform detrending using both demeaning and linear methods. Duplicate waveforms specifically for picking purposes.
        \item Apply bandpass causal filtering with a minimum frequency of 0.1 Hz and a maximum frequency of 30 Hz, as recommended by \citeA{zhu2019phasenet}.
        \item Remove poor-quality waveforms based on:
        \begin{itemize}
            \item Lack of noise before the P-wave onset, likely due to missing initial picks during download. Waveforms exceeding a maximum noise amplitude of $1.5 \times 10^{-4}$ (based on mean noise amplitude within the first 100 samples or 1 second) are automatically discarded, removing approximately 98\% of problematic waveforms.
            \item Instrumental spikes.
            \item Unavailable $V_{S30}$ in the data.
        \end{itemize}
        \item Align all P-phase picks to 5 seconds after the start time and rotate waveforms into radial and transverse components.
        \item Save waveforms in RTZ (radial, tangential, vertical) format.
        \item The total number of waveforms remaining after this process is 295,552.
    \end{enumerate}

    \item Use \url{04_filter_waveforms.py} to remove poor-quality picks:
    \begin{enumerate}
        \item Read the \url{raw_waveforms.h5} file.
        \item Select only records with positive $V_{S30}$ values.
        \item Repick waveforms and discard any waveform with pick-time values less than 2 s or greater than 7 s.
        \item The total number of removed waveforms is 29,436.
    \end{enumerate}

    \item Use \url{05_raw_data_filter_indices.py} to remove poor-quality picks:
    \begin{enumerate}
        \item Read the \url{raw_waveforms_filtered.h5} file.
        \item Append indices of the valid waveforms.
    \end{enumerate}

    \item The total number of high-quality waveforms is 266,216.

    \item The removed waveforms are primarily characterized by large hypocentral distance ($>100$ km), low $V_{S30}$ ($<1000$ m/s), and low magnitude ($<5.5$) (Figures \ref{fig:remove_vs30_rhyp}, \ref{fig:remove_mag_rhyp}, and \ref{fig:remove_vs30_mag}). Poor-quality data mainly originate from the KNET network rather than the KiKnet network, as expected, since KNET is a surface network located near urban areas (Figure \ref{fig:station_vs_vs30}).
    
    \begin{figure}[!htp]
        \centering
        \includegraphics[width=0.4\linewidth]{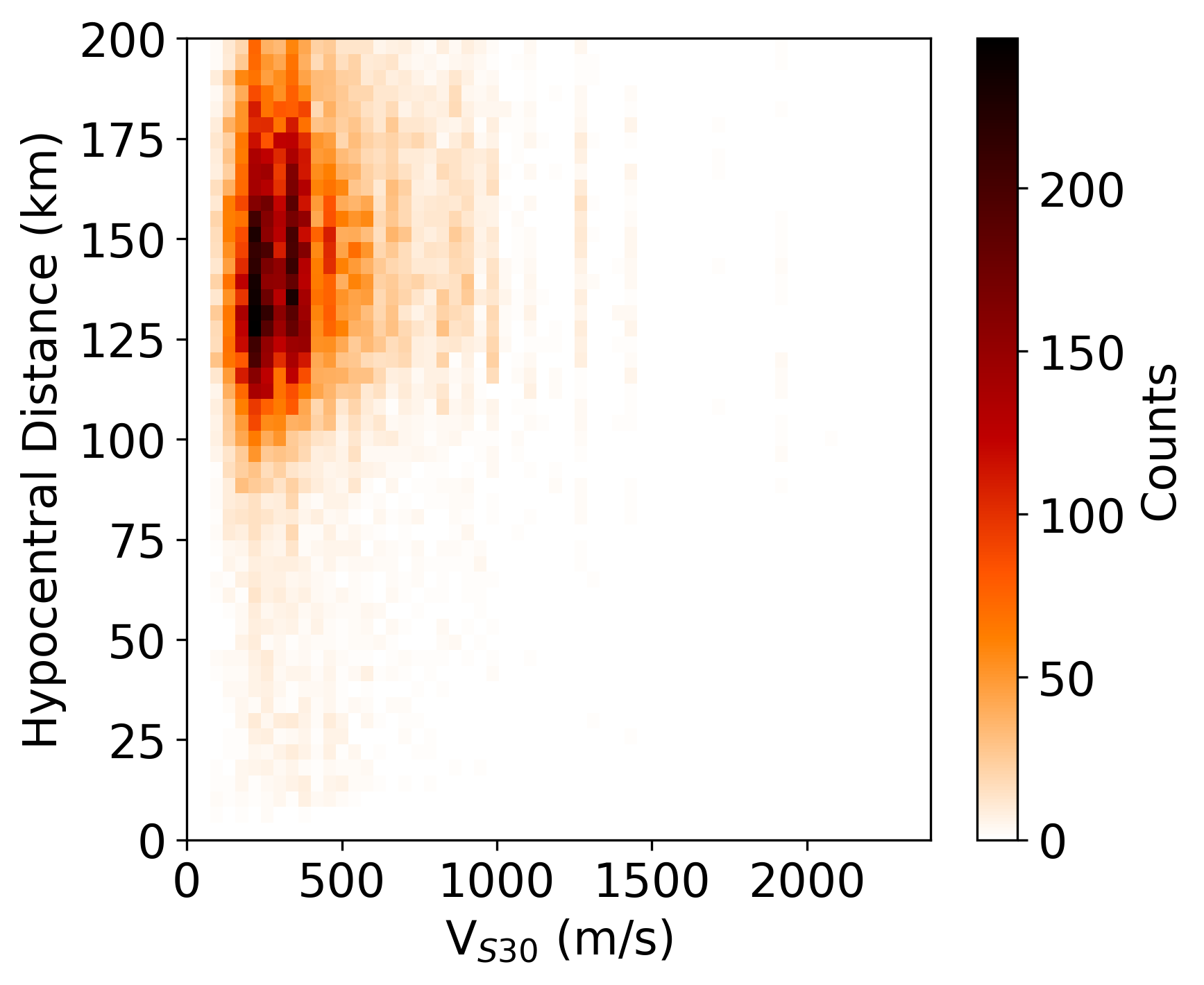}
        \caption{2D histogram showing the distribution of removed waveforms as a function of hypocentral distance (x-axis, in km) and $V_{S30}$ (y-axis, in m/s). Waveforms are removed during the second filtering step (i.e., quality filtering based on noise, instrumental spikes, and availability of $V_{S30}$).}
        \label{fig:remove_vs30_rhyp}
    \end{figure}

    \begin{figure}[!htp]
        \centering
        \includegraphics[width=0.4\linewidth]{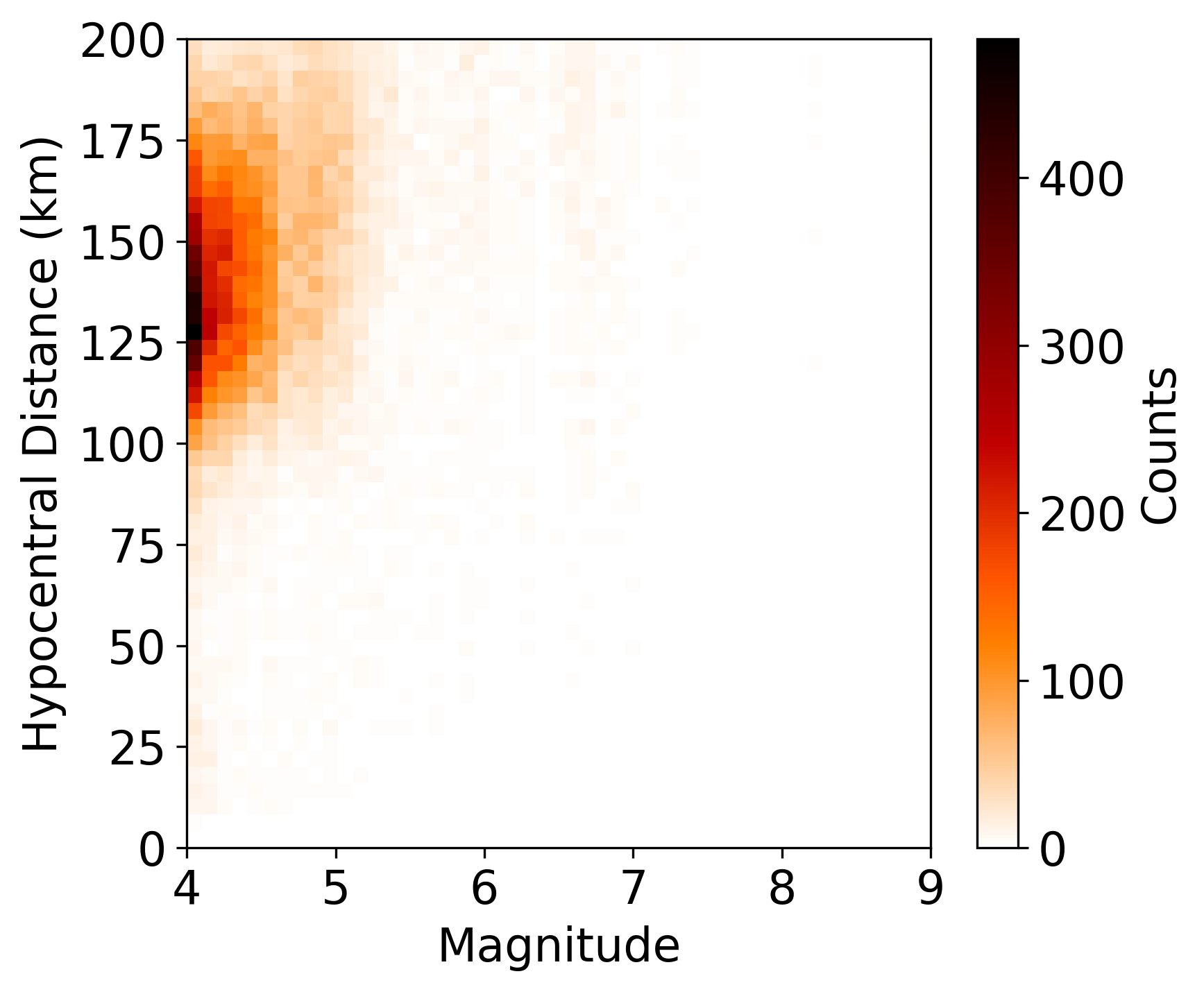}
        \caption{Same as Figure \ref{fig:remove_vs30_rhyp}, but showing the distribution of removed waveforms as a function of hypocentral distance (x-axis, in km) and magnitude (y-axis).}
        \label{fig:remove_mag_rhyp}
    \end{figure}

    \begin{figure}[!htp]
        \centering
        \includegraphics[width=0.4\linewidth]{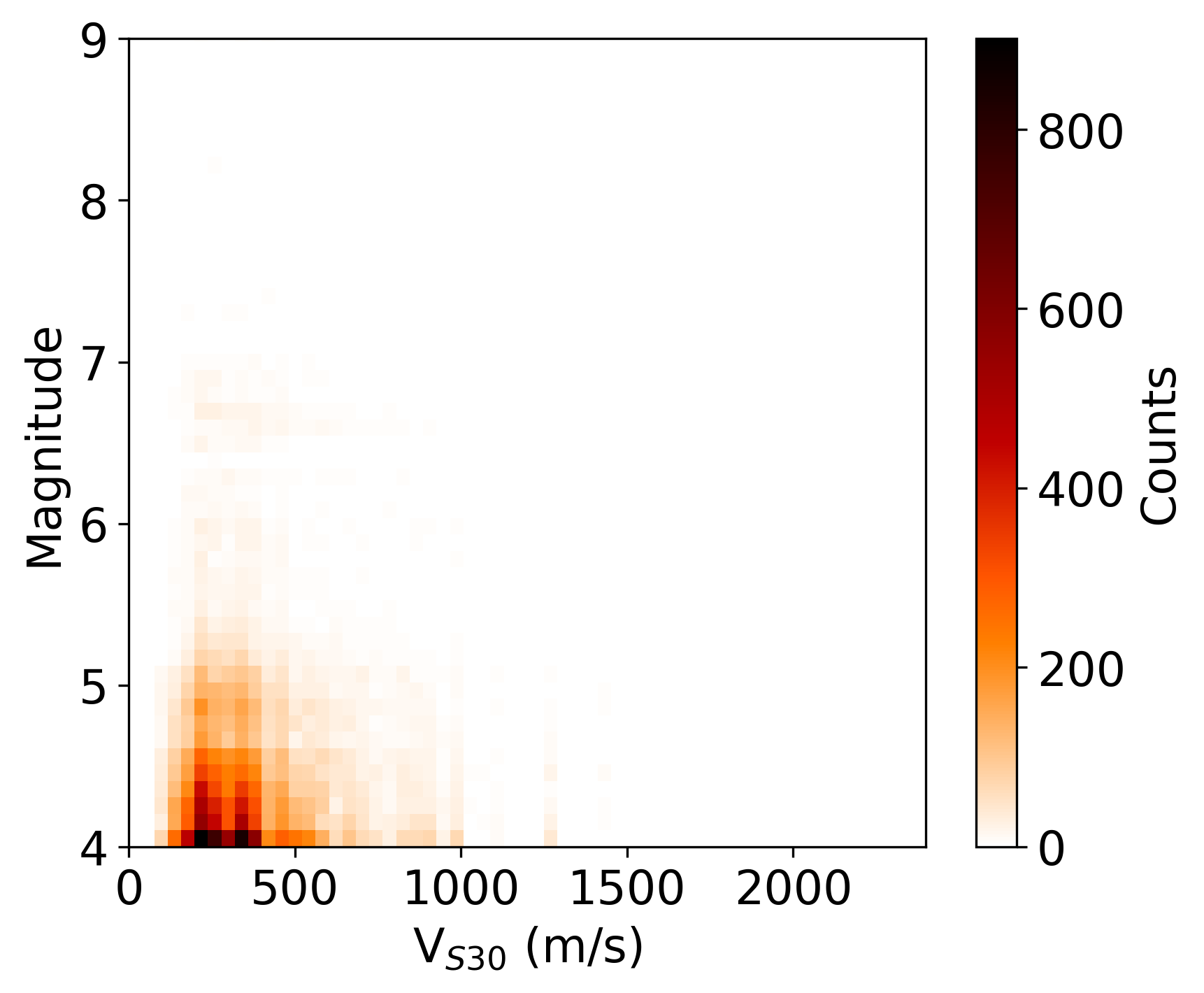}
        \caption{Same as Figure \ref{fig:remove_vs30_rhyp}, but showing the distribution of removed waveforms as a function of magnitude (x-axis) and $V_{S30}$ (y-axis, in m/s).}
        \label{fig:remove_vs30_mag}
    \end{figure}

    \begin{figure}[!htp]
        \centering
        \includegraphics[width=0.4\linewidth]{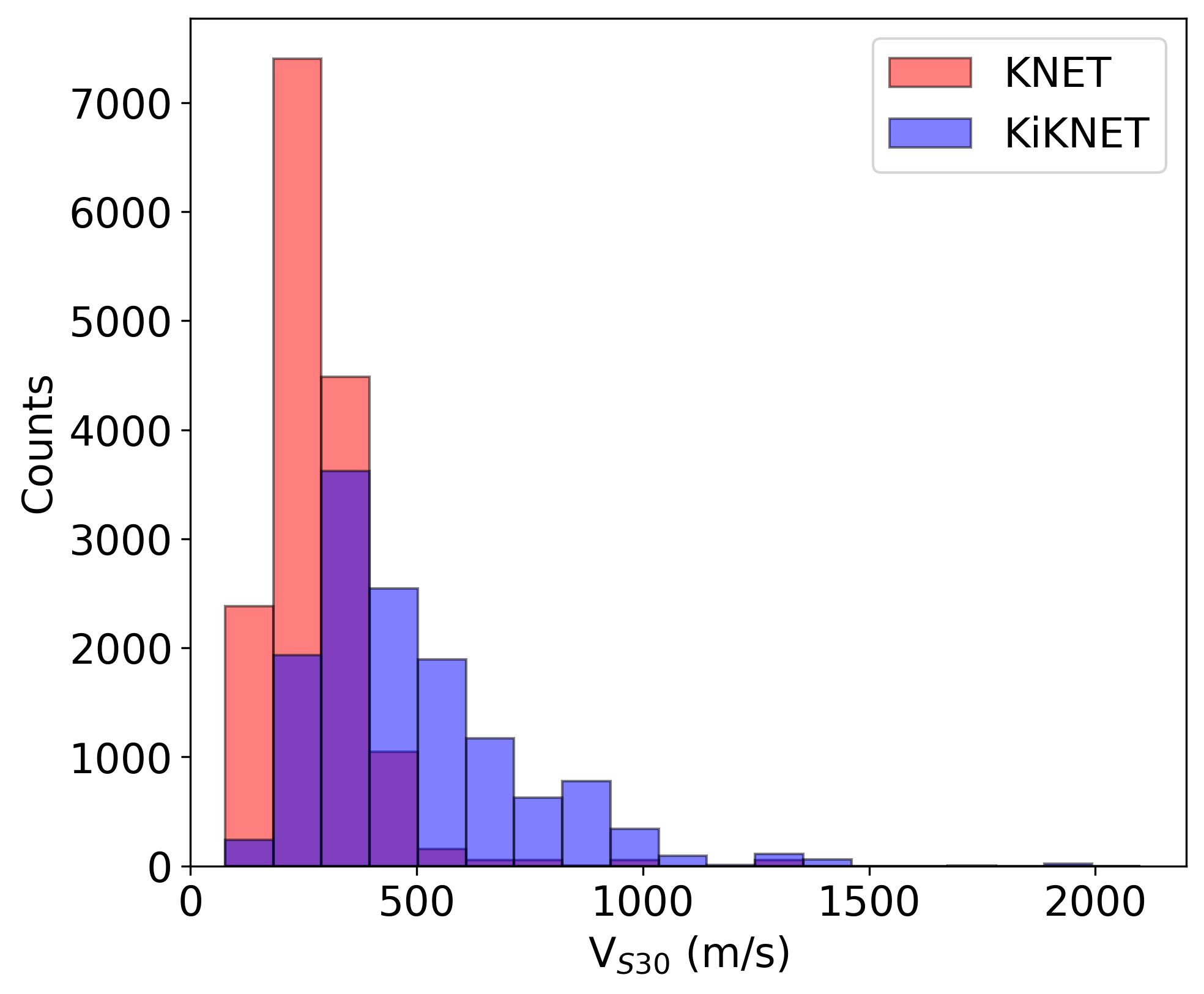}
        \caption{Histogram comparing the number of removed waveforms for two seismic networks: KNET (red bars) and KiKnet (blue bars), categorized by $V_{S30}$ values.}
        \label{fig:station_vs_vs30}
    \end{figure}

\end{itemize}

\clearpage
\section*{Text S2: Statistics of the Final Preprocessed Dataset}

\begin{figure}[!htp]
    \centering
    \includegraphics[width=0.4\linewidth]{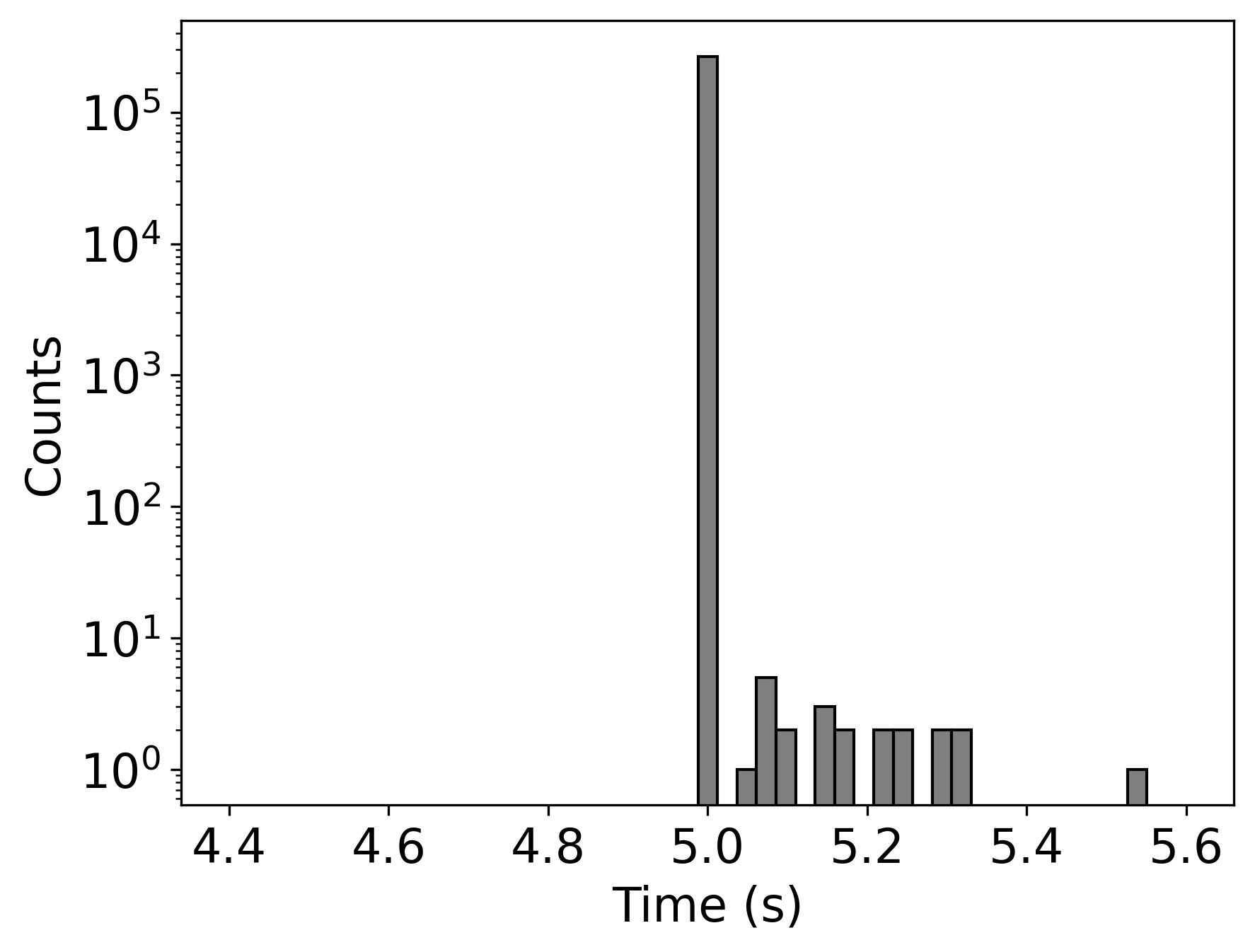}
    \caption{Distribution of the P-wave onset times relative to the waveform starting time.}
    \label{fig:statistics_onset}
\end{figure}

\begin{figure}[!htp]
    \centering
    \includegraphics[width=0.4\linewidth]{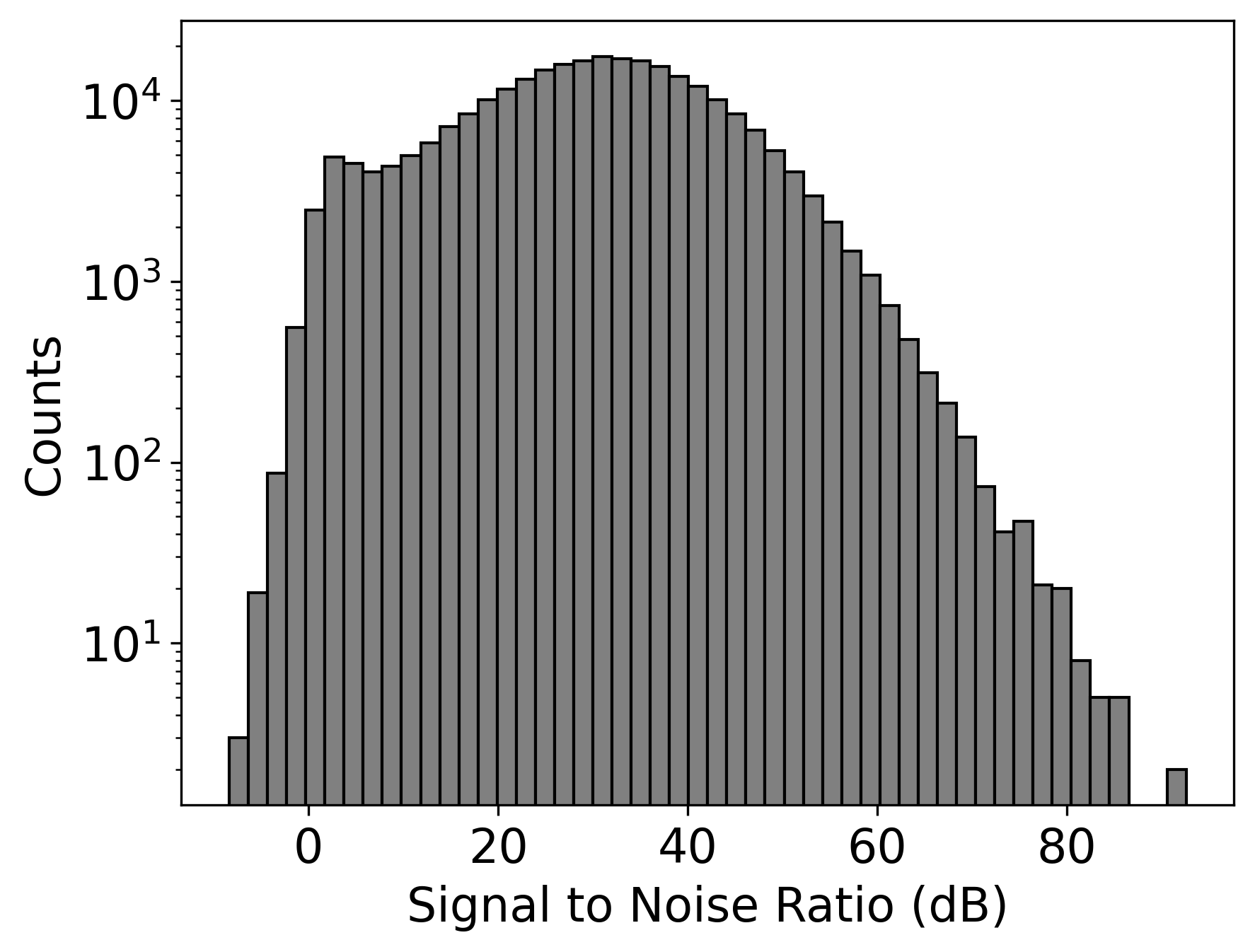}
    \caption{Distribution of the signal-to-noise ratio (SNR) for the final dataset.}
    \label{fig:SNR_total}
\end{figure}

The final “clean” dataset consists of 266,216 high-quality waveforms. Each waveform includes three components rotated into radial, tangential, and vertical directions, with a duration of 125 s sampled at 100 samples per second. The dataset covers magnitudes from 4.0 to 9.1, hypocentral distances between 0 and 200 km, and $V_{S30}$ values ranging from 76 to 2100 m/s. Figures \ref{fig:statistics_onset} and \ref{fig:SNR_total} show the distributions of P-wave onset times and signal-to-noise ratios (SNR), respectively.

\begin{figure}[!htp]
    \centering
    \includegraphics[width=\linewidth]{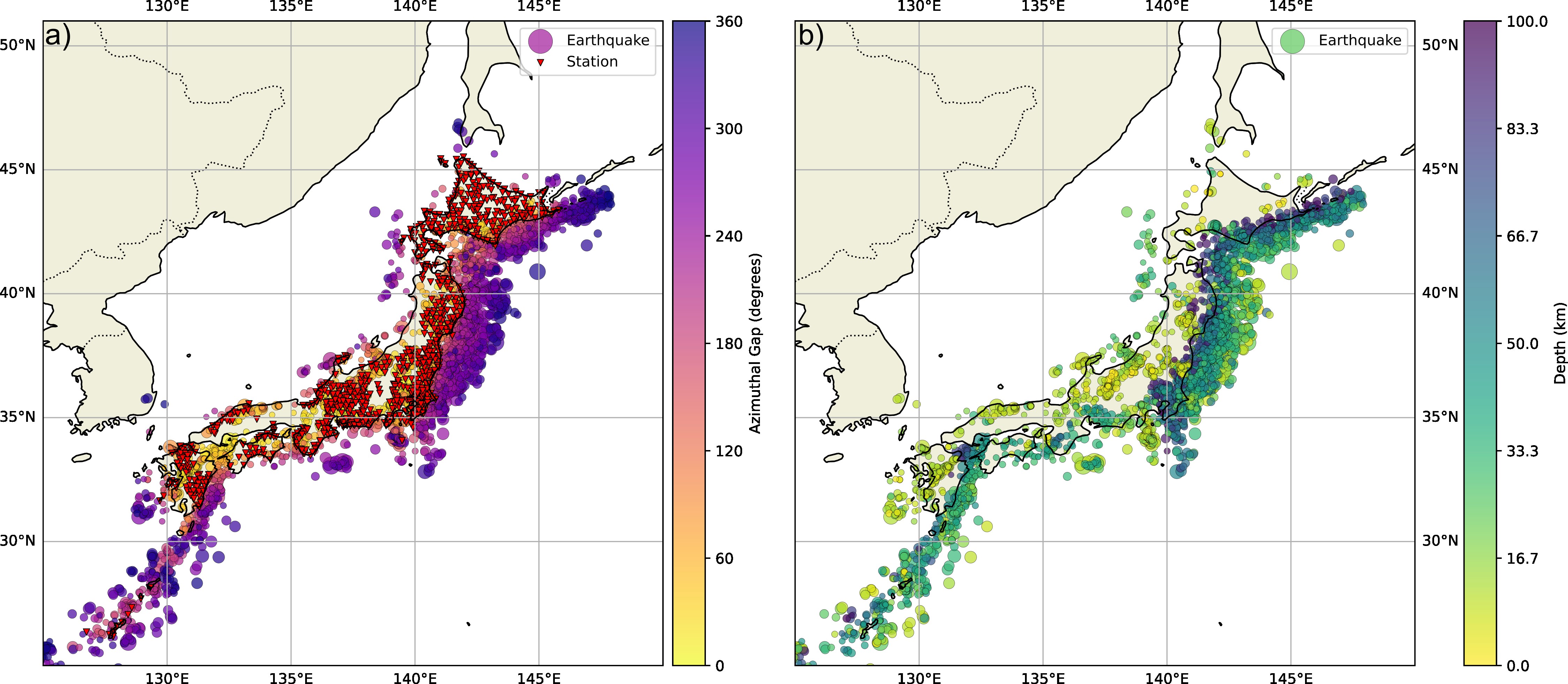}
    \caption{Distribution of events and stations in the region of Japan by (a) azimuthal gap and (b) depth.}
    \label{fig:data_distribution}
\end{figure}

Figure \ref{fig:data_distribution} summarises the dataset: panel (a) shows the azimuthal gap for each event, while panel (b) presents the corresponding depth distribution.

\clearpage
\section*{Text S3: Stanford Earthquake Dataset (STEAD) for reproducibility}

\subsection*{Motivations}
To ensure that our model training and evaluation pipeline can be replicated end-to-end without restrictive licensing, we rerun every experiment on the publicly available Stanford Earthquake Dataset (STEAD) \cite{mousavi2019stanford}.  STEAD contains more than 1.2M labelled three-component waveforms (East-West, North-South, and vertical) spanning a wide magnitude ($M_w$ -0.5 – 7.9), hypocenter depth (-3.49 - 342 km), and distance (0–347 km) range, making it an ideal surrogate for benchmarking ground-motion models, envelope extractors, and signal-based classifiers.  This section describes (i) how the data are obtained and processed, (ii) how the evaluation scripts are adapted for STEAD, and (iii) how the resulting performance is reproduced based on the STEAD datasets.

\subsection*{Data descriptions and preprocessing steps for the STEAD datasets}
The dataset used in this demonstration contains only local earthquakes with moment magnitude $M_{\mathrm{w}} > 4.5$ and hypocentral distance $< 200$ km. After filtering, we retain 3188 three-component (E–N–Z) waveform records.

Our model is conditioned on five parameters—magnitude, hypocentral distance, hypocentral depth, $V_{S30}$, and azimuthal gap. Because STEAD does not report site‐condition values ($V_{S30}$), we assign each record a random value drawn uniformly from 400 m/s to 1500 m/s (random seed = 42 for reproducibility). When fewer than three stations are available, we substitute the epicentral azimuth for the azimuthal gap. Horizontal components are subsequently rotated into radial and transverse components.

The fully pre-processed dataset is archived on Zenodo. To reproduce it from raw STEAD files, use the following scripts provided in this repository:

\begin{enumerate}
\item \textbf{Step 1 — Download}\
\url{stead_download.py}: downloads the STEAD archive from Kaggle \cite{Sevilla_STEAD_2022} ($\sim 98$ GB).
\item \textbf{Step 2 — Pre-process}\
\url{create_dataset_from_STEAD.py}: filters, converts, and stores the waveforms in \url{raw_waveforms.h5}.
\item \textbf{Step 3 — Post-process}\ 
\url{Residual_plot_stead.ipynb}: visualize GWM and evaluate the GWM statistics. 
\end{enumerate}

Both scripts include hard-coded directory paths. These paths must be edited to correspond to your local file-system layout before the code is executed. 

We partitioned the dataset into a 90\% training set and a 10\% test set. The seismological evaluation uses only the test subset, which contains 319 three-component seismic waveforms. Model training was carried out on a single node equipped with four NVIDIA Grace-Hopper GPUs, taking roughly 2 hours for the first-stage autoencoder and an additional 1.5 hour for the second-stage diffusion model.

\subsection*{Results and discussions}
\begin{figure}[!htp]
    \centering
    \includegraphics[width=0.99\linewidth]{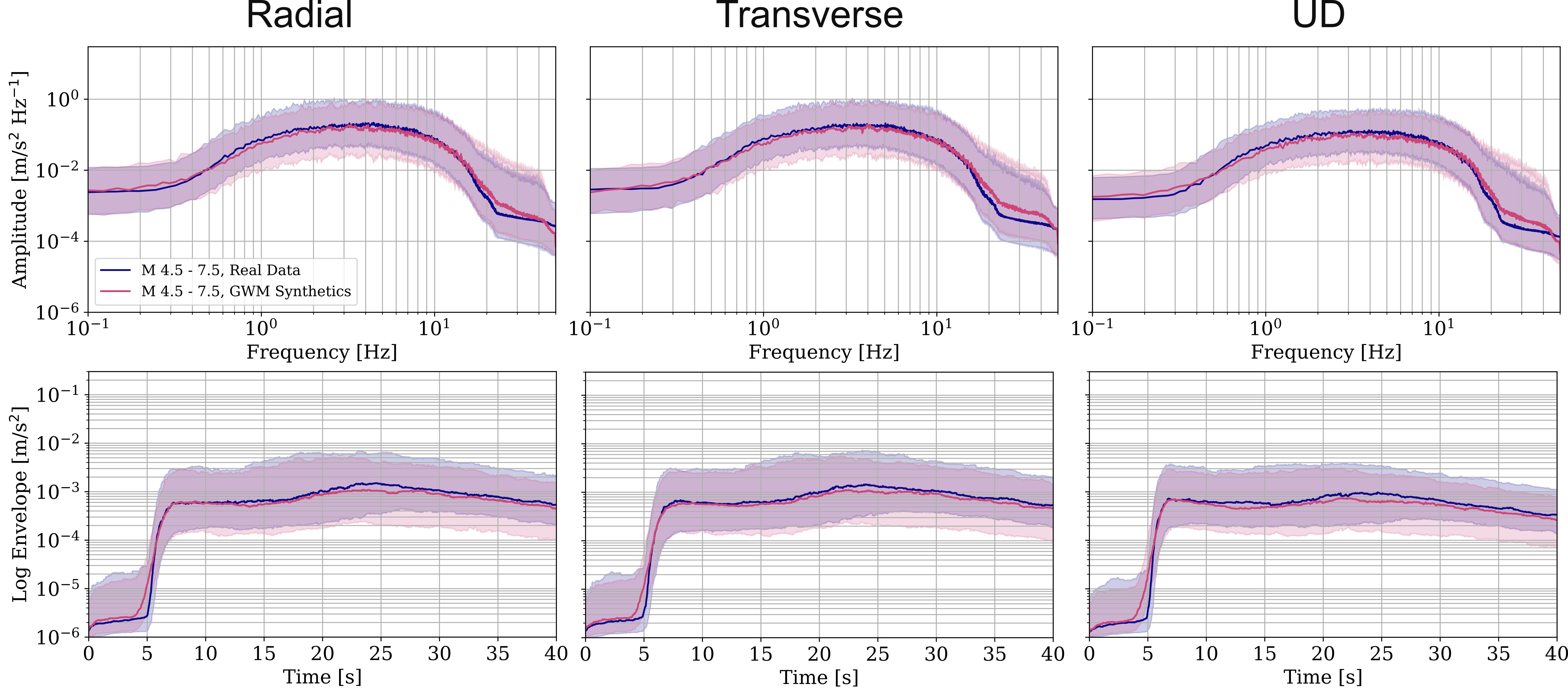}
    \caption{Fourier amplitude spectra (upper panel) and waveform envelope (lower panel) for all the test dataset.}
    \label{fig:spectral_envelope}
\end{figure}

Despite the limited training set, 2869 three-component seismic waveforms, our model accurately reproduces first-order spectral statistics, capturing both the mean and the standard deviation of the frequency content across samples (Figure \ref{fig:spectral_envelope}). The generative waveform model (GWM) also reconstructs the waveform envelope, but it systematically lags the true P-wave onset by about 1 s in all components (Figure \ref{fig:spectral_envelope}).

\begin{figure}[!htp]
    \centering
    \includegraphics[width=0.5\linewidth]{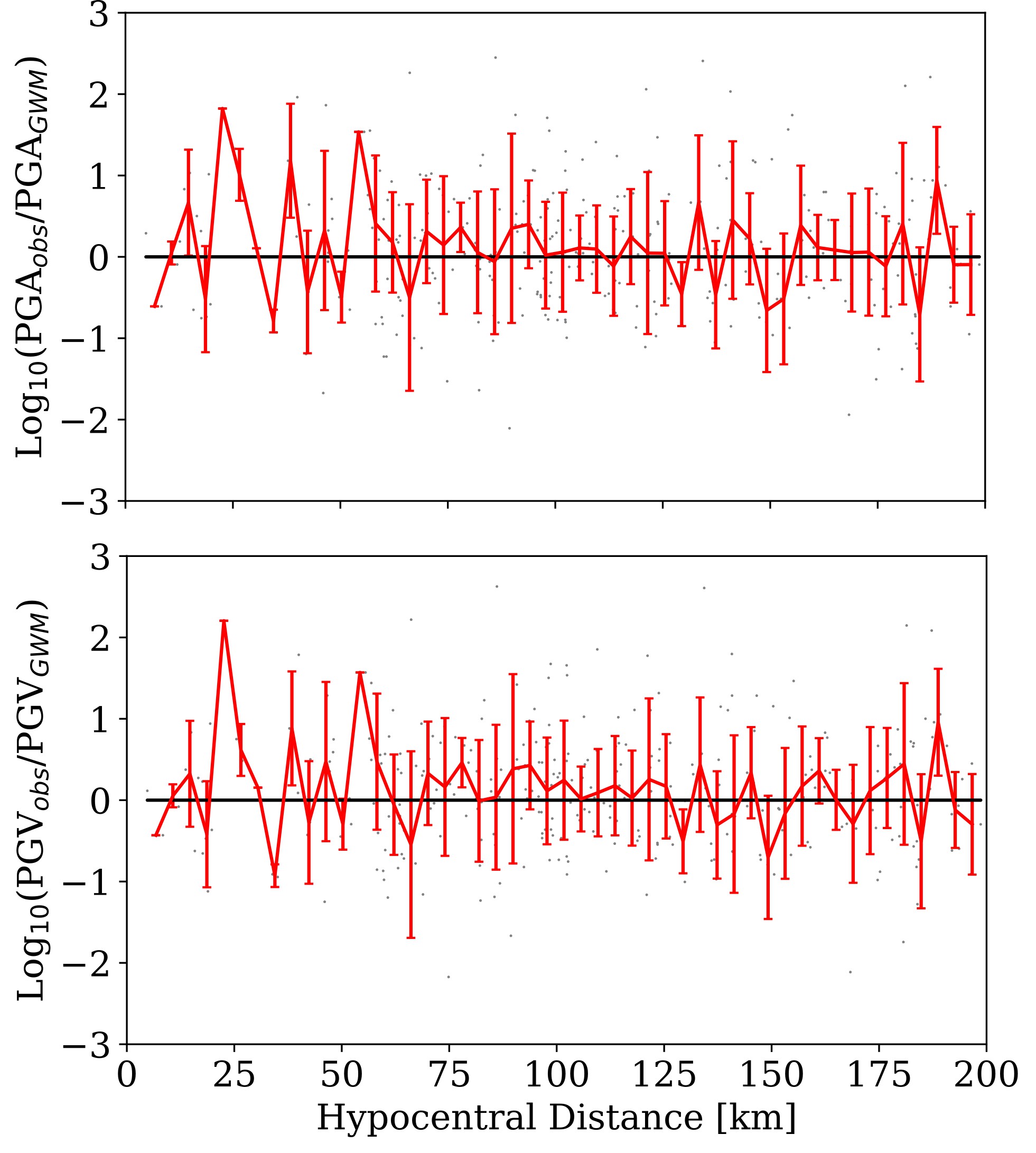}
    \caption{Ratio of peak ground acceleration (PGA, upper panel) and peak ground velocity (PGV, lower panel) predicted by the generative waveform model (GWM) to those observed in the data; both intensity measures are calculated as RotD50 values. On average, the GWM underpredicts PGA by a logarithmic ratio of 0.122 ($\approx~31\%$) and PGV by 0.136 ($\approx~37\%$).}
    \label{fig:bias_stead}
\end{figure}

We quantified the bias of the GWM against the test set for peak ground acceleration (PGA) and peak ground velocity (PGV) (Figure \ref{fig:bias_stead}). On average, the model underpredicts PGA by 0.122 in logarithmic units ($\approx$ 31\%) and PGV by 0.136 in logarithmic units ($\approx$ 37\%).

\begin{figure}[!htp]
    \centering
    \includegraphics[width=0.99\linewidth]{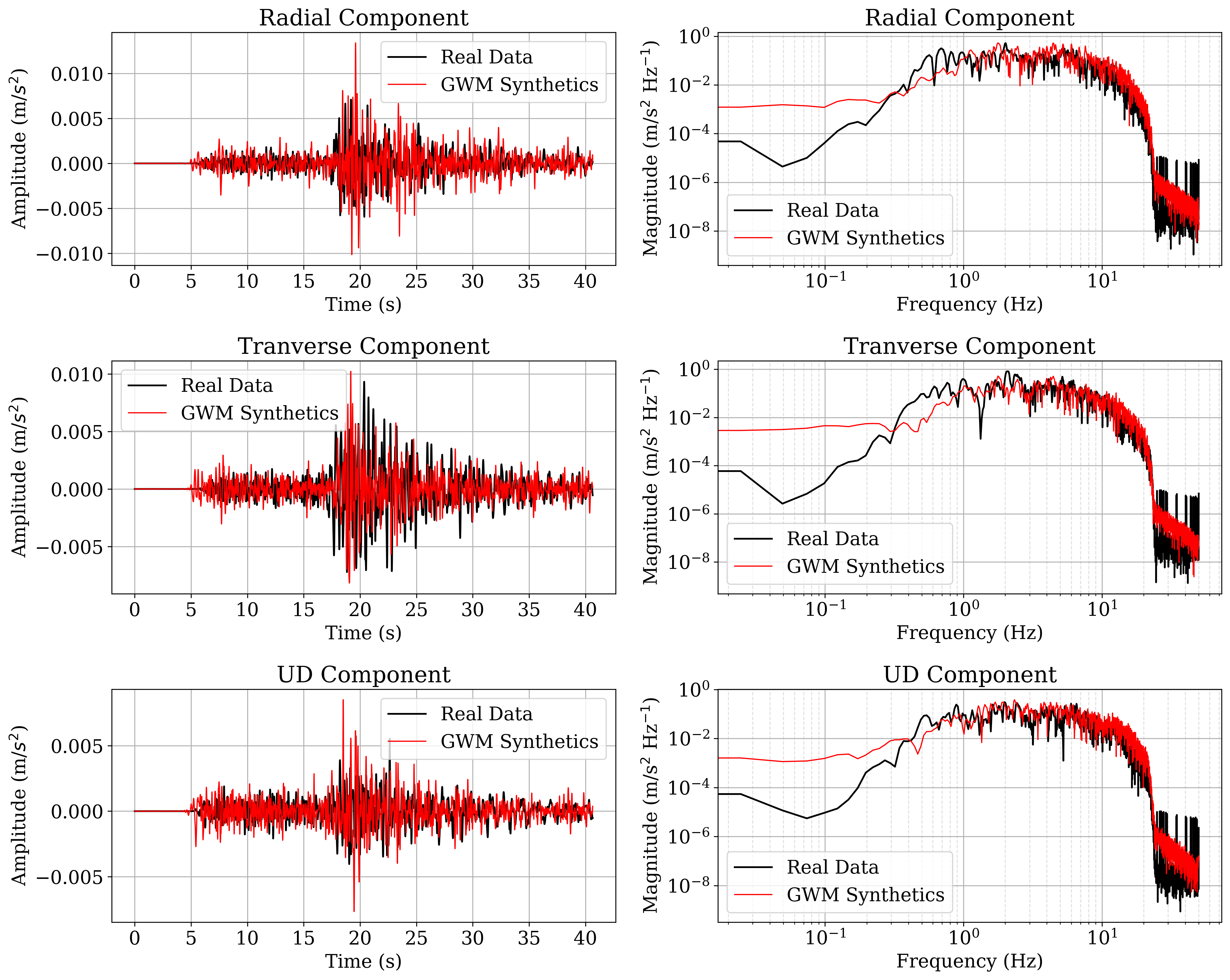}
    \caption{Fair comparison of temporal and frequency content between real data and generative waveform models.}
    \label{fig:waveform_stead_good}
\end{figure}

Figure \ref{fig:waveform_stead_good} presents the three-component waveform and its corresponding Fourier amplitude spectrum generated for a magnitude 4.9 event with a hypocentral distance of 105.28 km, hypocentral depth of 18.07 km, $V_{S30}=1250$ m/s, and an azimuthal gap of 31.22$^\circ$. The GWM reproduces the P-wave onset at roughly 5 s and matches the observed peak amplitude. It also clearly separates the P- and S-phase arrivals, reflecting the specified source–receiver distance. Although the model overestimates low-frequency energy, it captures spectral amplitudes above 0.8 Hz with good fidelity.

\begin{figure}[!htp]
    \centering
    \includegraphics[width=0.99\linewidth]{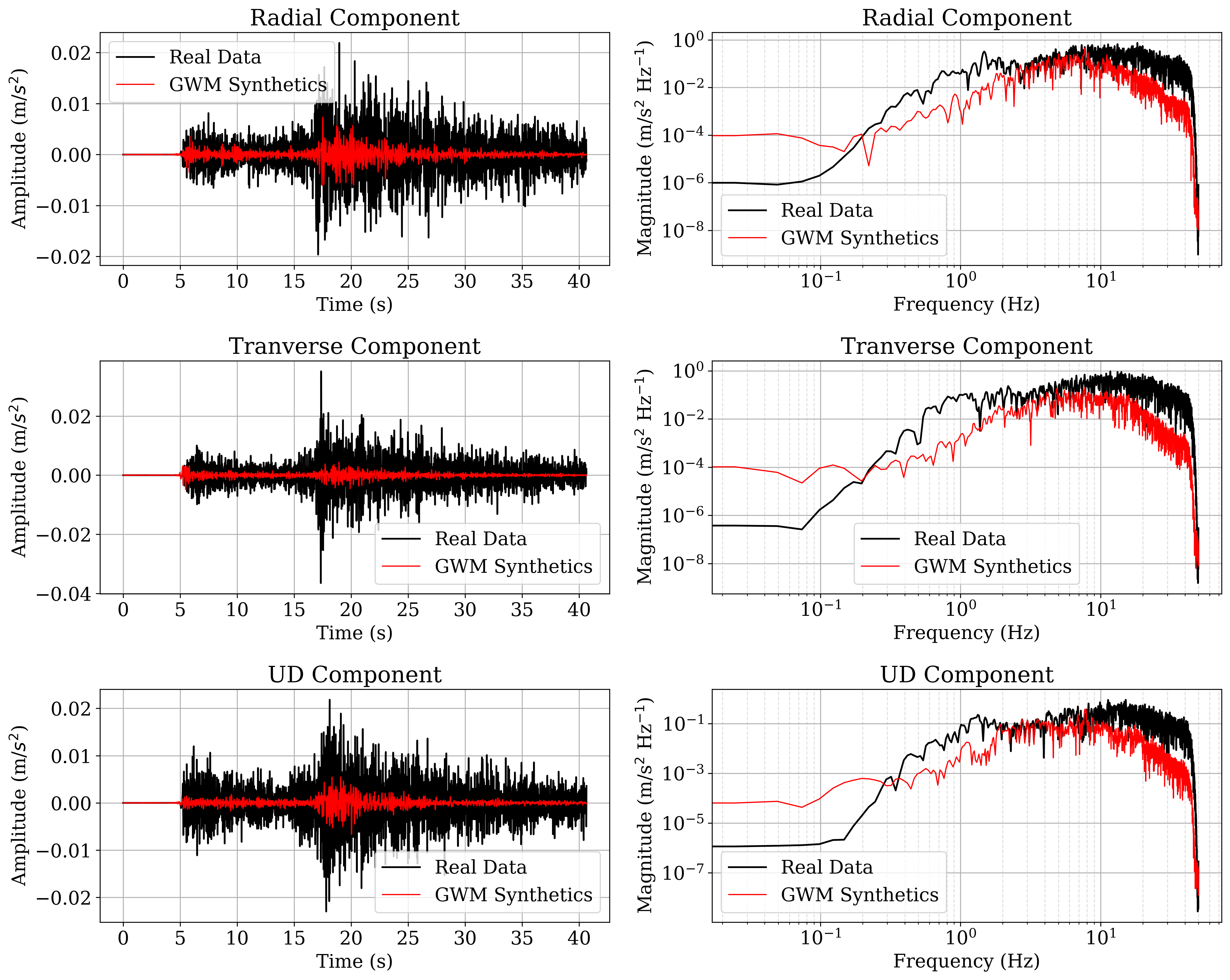}
    \caption{Poor comparison of temporal and frequency content between real data and generative waveform models.}
    \label{fig:waveform_stead_bad}
\end{figure}

Occasionally, the GWM fails to reproduce the observed waveform amplitudes, as illustrated in Figure \ref{fig:waveform_stead_bad}. The example corresponds to a magnitude 4.56 event with a hypocentral distance of 106.55 km, hypocentral depth of 57.31 km, $V_{S30}=964$ m/s, and an azimuthal gap of 140.91$^\circ$. As in Figure \ref{fig:waveform_stead_good}, the model correctly places the P-wave onset at $\approx$ 5 s and separates the P- and S-phase arrivals. However, it substantially underpredicts the overall time-domain and spectral amplitudes: low-frequency energy is overestimated, while amplitudes above 0.2 Hz are severely underestimated.

A training set of fewer than 3000 three-component waveforms enables the GWM to predict seismic traces with reasonable accuracy. Statistically, the model reproduces first-order intensity measures, such as the mean PGV and PGA, but it sometimes fails to match the amplitude and spectral content of individual records. This mixed performance likely stems from the broad variability within the training data (see main text). It requires at least 50 realizations to capture the mean of the amplitude of the real data. In addition, the random assignment of $V_{S30}$ values makes the experimental setup less than ideal for real-world applications.

\section*{Text S4: Comparing HighFEM with GANO}
We also compare the time-domain signal envelopes with the GANO model of \citeA{shi2024broadband} (Figures \ref{fig:supp_envelope_gano_0}–\ref{fig:supp_envelope_gano_2}). Their conditioning parameters are similar to those of the HighFEM GWM, except that i) the GANO uses finite rupture distance, whereas we use hypocentral distance, and ii) the 'faulting type' parameter, which GANO uses to distinguish between shallow crustal events and subduction zone events. Their signal lengths (60 s) also differ from ours (40 s). Furthermore, the HighFEM synthetics include 5 seconds of pre-P-wave noise, whereas the GANO synthetics begin directly at the P-wave onset. For all GANO comparisons we therefore compare signals only across the first 40 seconds that both models predict, we use faulting type 1 (shallow crustal) and hypocentral distances. Similar to our GWM, the GANO model also reproduces P- and S-phase arrivals as well as increasing ground-motion amplitudes with magnitude. 

Then, to compare with the GANO predictions, we calculate the mean model bias across distances $> 50$ km: 0.22 log-units for PGA and 0.14 log-units for PGV, corresponding to systematic underpredictions of $38\%$ for PGA and $65\%$ for PGV, both larger than those observed for the HighFEM GWM (Figure \ref{fig:bias_gano}). For distances $\leq 50$ km, GANO underpredicts PGV by $130\%$ (0.36 log-units) and PGA by approximately $75\%$ (0.24 log-units). However, this discrepancy at short hypocentral distances likely stems from the fact that GANO conditions on rupture distance rather than hypocentral distance. Therefore, this comparison is not strictly one-to-one. Since rupture distances can be substantially shorter than hypocentral distances, an underprediction of ground-motion amplitudes is expected. These systematic underpredictions at short hypocentral distances should therefore not be interpreted as poor performance of the GANO model.


\end{document}